\documentclass[a4paper]{article}
\usepackage[T1]{fontenc}
\usepackage[utf8]{inputenc}
\DeclareUnicodeCharacter{0301}{\'{e}}
\usepackage[brazilian]{babel}
\usepackage{graphicx}
\usepackage{hyperref}
\hypersetup{
  setpagesize  = false,
  colorlinks   = true,    
  urlcolor     = blue,    
  linkcolor    = black,   
  citecolor    = black    
}
\usepackage{subcaption}
\usepackage{authblk}
\usepackage{geometry}
\usepackage{setspace}
\usepackage{verbatim}
\usepackage[utf8]{inputenc}
\usepackage{amsmath}
\usepackage{amssymb}
\usepackage{gensymb}
\usepackage{verbatim} 
\usepackage{ragged2e} 
\usepackage{amsfonts}
\usepackage{cancel}
\usepackage{float}

\usepackage{booktabs}
\usepackage{multirow}
\usepackage{multicol}
\usepackage{soul}
\usepackage[dvipsnames]{xcolor}
\newcommand{\bra}[1]{\langle #1 |}
\newcommand{\ket}[1]{| #1 \rangle}
\newcommand{\bracket}[2]{\langle #1 | #2 \rangle}

\usepackage{fancyhdr}
\newcommand{\keywordsenglishname}{Keywords}

\renewenvironment{abstract}{%
        \begin{center}
	\begin{minipage}{14cm}
	{\textbf{\abstractname:}}
}{
        \end{minipage}
	\end{center}
}
\newenvironment{abstractinenglish}{
        \def\abstractname{\abstractinenglishname}
	\begin{abstract}
}{
        \end{abstract}
}
\newenvironment{keywords}{
        \def\abstractname{\emph{\keywordsportugues}}
	\begin{abstract}
}{
        \end{abstract}
}
\newenvironment{keywordsenglish}{
        \def\abstractname{\emph{\keywordsenglishname}}
	\begin{abstract}
}{
        \end{abstract}
}

\title{\textbf{Uma Introdução ao \textit{Variational Quantum Eigensolver} Aplicado à Química.}}
\author{Matheus da Silva Fonseca$^{1}$\thanks{Endereço de correspondência: msfonseca@estudante.ufscar.com}, Caio Moraes Porto$^2$,  Nicolás Armando Cabrera Carpio$^1$, Guilherme de Souza Tavares de Morais$^2$, Nelson Henrique Morgon$^2$, René Alfonso Nome$^2$, Celso Jorge Villas-Boas$^1$ }
\affil{\small $^1$ Universidade Federal de São Carlos, Departamento de Física, São Carlos, SP, Brasil. \\ 
       $^2$ Universidade Estadual de Campinas, Instituto de Química, Campinas, SP, Brasil.}
\date{}

\begin{document}

\maketitle
\vspace{6pt}

\begin{abstract}
A mecânica quântica introduziu um novo arcabouço teórico para o estudo de moléculas, possibilitando a previsão de propriedades e dinâmicas a partir da solução da equação de Schrödinger aplicada a esses sistemas. Contudo, a resolução dessa equação demanda um elevado custo computacional, o que levou ao desenvolvimento de diferentes formalismos matemáticos e métodos computacionais, concebidos para balancear os recursos disponíveis e a precisão desejada.
Em particular, os computadores quânticos surgem como uma tecnologia promissora, com o potencial de abordar esses problemas de forma mais eficiente nas próximas décadas, seja pela redução do consumo de memória, tempo e energia, \textit{i. e.}, a redução da complexidade computacional, ou pelo aumento da precisão. Essa linha de pesquisa é conhecida como Simulações Quânticas. Dada a limitação tecnológica atual dos computadores quânticos, os algoritmos quânticos variacionais (\textit{Variational Quantum Algorithms} - VQAs), especialmente o \textit{Variational Quantum Eigensolver} (VQE), emergem como uma abordagem viável para demonstrar vantagens em relação aos métodos clássicos no curto prazo. Essa viabilidade deve-se à menor demanda por portas lógicas e à profundidade reduzida dos circuitos necessários para sua implementação.
Neste trabalho, apresentamos a aplicação da mecânica quântica no estudo de moléculas, realizamos uma introdução aos fundamentos da computação quântica e exploramos a integração entre essas áreas por meio da utilização do VQE em simulações moleculares. Por fim, discutimos a complexidade espacial e temporal associada ao algoritmo, ressaltando suas implicações e desafios.

\end{abstract}

\begin{keywords}
Moléculas, Simulações Quânticas, VQE, Complexidade Computacional
\end{keywords}

\vspace{6pt}

\begin{abstractinenglish}
Quantum mechanics has introduced a new theoretical framework for the study of molecules, enabling the prediction of properties and dynamics through the solution of the Schrödinger equation applied to these systems. However, solving this equation is computationally expensive, which has led to the development of various mathematical frameworks and computational methods designed to balance the available resources with the desired level of accuracy.
In particular, quantum computers have emerged as a promising technology with the potential to address these problems more efficiently in the coming decades, whether through reductions in memory, time, and energy consumption, \textit{i. e.}, reductions in computational complexity or by enhancing precision. This research field is known as Quantum Simulation. Given the current technological limitations of quantum computers, Variational Quantum Algorithms (VQAs), especially the Variational Quantum Eigensolver (VQE), stand out as a feasible approach to demonstrating advantages over classical methods in the near term. This feasibility arises from their lower demand for quantum gates and the reduced depth of the circuits required for their implementation.
In this work, we present the application of quantum mechanics to the study of molecules, provide an introduction to the fundamentals of quantum computing, and explore the integration of these fields by employing the VQE in molecular simulations. Finally, we discuss the spatial and temporal complexity associated with the algorithm, highlighting its implications and challenges.
\end{abstractinenglish}

\begin{keywordsenglish}
Molecules, Quantum Simulations, VQE, Computational Complexity

\end{keywordsenglish}

\begin{multicols}{2}

\section{Introdução}

\noindent A computação quântica, um campo em rápido crescimento na ciência da computação e na física, utiliza os princípios fundamentais da mecânica quântica para processar informações de maneira revolucionária. Enquanto a computação clássica trabalha com bits, que representam informações em formato binário ($0$ ou $1$), a computação quântica opera com \textit{qubits}, sistemas quânticos que podem existir em superposição de estados, permitindo que representem simultaneamente $0$ e $1$~\cite{Steane1998,Valiev2005}. Essa capacidade singular possibilita que os computadores quânticos resolvam certos cálculos em um número de passos exponencialmente menor do que seus equivalentes clássicos~\cite{Zoller2005}.

A história da computação quântica remonta há décadas, com avanços notáveis que desafiaram as limitações da computação clássica. Um marco significativo surgiu em 1982, quando o físico Richard Feynman postulou que a simulação eficiente de sistemas quânticos deveria ser realizada por outros sistemas quânticos. Isso lançou as bases conceituais da computação quântica~\cite{Feynman1982,Preskill2021}. Posteriormente, em 1995, Cirac e Zoller~\cite{cirac1995quantum} propuseram o primeiro projeto tecnologicamente viável para a construção de um computador quântico usando íons aprisionados e no ano seguinte Lloyd~\cite{Lloyd1996} demonstrou como utilizar os computadores quânticos para simular de forma eficiente a dinâmica de sistemas físicos com interações locais.

Atualmente, a utilização de sistemas quânticos para realizar as mesmas tarefas que os computadores \emph{clássicos}, porém de forma a garantir alguma vantagem, seja por conta de cálculos mais precisos ou de forma mais eficiente, se expandiu para muitas outras atividades além da simulação de sistemas físicos (moléculas e materiais), como solução de problemas de otimização e de sistemas lineares, com diversas aplicações na indústria, mercado financeiro e previsões climáticas dentre outras. O próprio computador proposto por Cirac e Zoller é um exemplo desse tipo de sistema. Dessa forma, a busca pelo desenvolvimento de máquinas que resolvam tais problemas tornou-se um objetivo central para muitas empresas e Estados ao redor do mundo~\cite{quantummanifest,blattquantum},  dando a essas máquinas o nome de computadores \emph{quânticos} universais~\cite{nielsen2010}.
No entanto, essa busca tem enfrentado desafios técnicos gigantescos, dentre eles a decoerência dos \textit{qubits} devido a interações entre o sistema quântico e seu entorno, tornando a criação de computadores quânticos práticos uma tarefa árdua. Mesmo assim, avanços contínuos no desenvolvimento de hardware e de algoritmos estão nos aproximando cada vez mais dessa revolução da computação~\cite{Qing2023}.

A utilização de sistemas quânticas para simular outros sistemas quânticos é denominada atualmente de \emph{Simulações quânticas}, enquanto que para realizar tarefas de computação de propósito geral é chamada de \emph{Computação quântica}~\cite{quantummanifest,blattquantum}. Apesar de serem postas em categorias diferentes, um mesmo sistema pode ser utilizado para realizar algoritmos distintos para cada uma dessas tarefas, justificando o termo \emph{universal} frequentemente utilizado para descrevê-lo. 

Dentro da computação quântica universal, 
dois algoritmos destacam-se: O algoritmo de Shor~\cite{shor1997} e o de Grover~\cite{Grover1996}. O primeiro  pois eficientemente fatora números inteiros grandes em seus fatores primos, com implicações cruciais para a criptografia, já que poderia desafiar muitos sistemas criptográficos; O segundo destaca-se pois acelera a busca em bancos de dados não estruturados, com aplicações significativas em otimização e aprendizado de máquina.

Além dos mencionados algoritmos de Shor e Grover para a computação quântica, o \textit{Variational Quantum Eigensolver} (VQE)~\cite{Peruzzo2014} é um destaque na simulação quântica e desempenha um papel crucial na resolução de problemas complexos na área da química quântica. Este algoritmo apresenta uma abordagem híbrida quântico-clássica, especialmente projetada para determinar a energia do estado fundamental de sistemas físicos. O foco do VQE é encontrar o menor autovalor do Hamiltoniano molecular, que reflete a energia fundamental da molécula. Esse processo envolve a preparação de uma função de onda de teste, conhecida como `ansatz', em um computador quântico, seguida pela medição de sua energia. Em seguida, um algoritmo de otimização clássica é aplicado para ajustar os parâmetros da função de onda, otimizando-a iterativamente até que um conjunto ótimo de parâmetros seja encontrado, minimizando assim a energia da função de onda de teste~\cite{Li2019}.

A relevância do VQE na química quântica é incontestável, uma vez que permite a obtenção de informações fundamentais sobre as moléculas, como seus estados de energia, reatividade e propriedades químicas. Isso desempenha um papel fundamental na pesquisa de novos materiais, desenvolvimento de fármacos e na compreensão de reações químicas complexas. É possível adaptar o VQE para se obter a energia de estados além do fundamental, porém, tal abordagem foge do escopo deste trabalho~\cite{tilly2022variational}.

Apesar dos indícios teóricos e do esforço da comunidade científica no entendimento e melhora do VQE, o desenvolvimento de formalismos e métodos computacionais em computadores clássicos que buscam obter essas propriedades também são notórios, de forma que não é claro a partir da literatura se o estado da arte do VQE atualmente apresenta alguma vantagem com relação aos melhores algoritmos clássicos. De forma mais rigorosa, por vantagem nos referimos aos recursos necessários para realizar o algoritmo, tipicamente se referindo ao tempo de processamento ou memória; o termo complexidade computacional também pode ser empregado para se referir a essa quantidade de recursos de forma que algoritmos com maior complexidade demandam mais recursos e algoritmos de menor complexidade demandam menos recursos, assim, o cálculo de complexidade de algoritmos quânticos é essencial para verificar vantagens que eles podem possuir com relação aos algoritmos clássicos.  Neste trabalho iremos discutir a implementação do Variational Quantum Eigensolver e discutir sua complexidade dentro do cenário de problemas de otimização. 

O trabalho está estruturado da seguinte forma: Na Seção \ref{sec:fundamentacao} discute-se o formalismo e métodos utilizados pela física e química para o estudo de sistemas moleculares. Na Seção \ref{sec:compQeAlg} são discutidos os princípios básicos da computação clássica e quântica que são necessários para o algoritmo, seguida de uma breve discussão sobre análise de complexidade computacional. Já na Seção \ref{sec:AlgVar&SimFis} são apresentados os simuladores quânticos e porque os algoritmos variacionais são potenciais candidatos a essa tarefa no estágio de desenvolvimento atual de \textit{hardwares} quânticos. Na Seção \ref{sec:complexity} é feita uma discussão sobre o cálculo de complexidade do \textit{Variational Quantum Eigensolver}, utilizando um caso simples para exemplo. Por fim, na Seção \ref{sec:conclusoes} são apresentadas as conclusões desse trabalho.

\section{Fundamentação do problema}
\label{sec:fundamentacao}

Nesta Seção serão discutidos o arcabouço teórico e formalismo da mecânica quântica necessários para o estudo de formação de moléculas e reações químicas. Em especial, iremos nos focar nos métodos utilizados para se obter a energia do estado fundamental de Hamiltonianos utilizados para representar sistemas moleculares, essenciais para a obtenção de propriedades de interesse como estrutura eletrônica e reatividade~\cite{Frey2016,Bao2017,Levine2013-pg}. 

\subsection{O princípio variacional}

\noindent O princípio variacional, pilar da mecânica quântica e que guiará o VQE para a otimização da função de onda desejada, diz que, dado um sistema cujo operador Hamiltoniano $\hat{H}$ é independente do tempo e cujo autovalor de energia mais baixo da equação de Schrödinger é $E_{gr}$, e se $|\psi_0 \rangle$ é qualquer função bem-comportada, normalizada, das coordenadas das partículas do sistema que satisfaçam as condições de contorno do problema, então:
\begin{equation}
     \langle \psi_{0}|\hat{H}|\psi_{0} 
     \rangle = E_0 \geq  E_{gr},
    \label{eq:principioVariacional}
\end{equation}
\noindent onde $E_0$ representa a energia média associada a $\ket{\psi_0}$.

Como a equação de Schrödinger não pode ser resolvida exatamente exceto para casos muito simples como átomos monoeletrônicos~\cite{Levine2013-pg}, potenciais simples~\cite{Porto2020} ou partícula na caixa~\cite{Levine2013-pg}, a estratégia mais comum envolve o uso de funções de onda aproximadas $\Psi$, as quais consequentemente geram um resultado aproximado para a energia do sistema que, devido ao princípio variacional, apresentará um valor superior para a energia.  Dessa forma, se utilizarmos diferentes funções tentativa, aquela que apresentar menor energia será a que representa melhor o sistema. 

\subsection{Hamiltoniano molecular}
\label{sec:HamMol}

\noindent O Hamiltoniano de um sistema molecular pode ser dividido em duas partes: uma de energia cinética ($\hat{T}$) e outra energia potencial ($\hat{V}$), isto é,  
\begin{equation}
    \hat{H}_{\text{total}}=\hat{T}+\hat{V}.
    \label{eqn1p5}
\end{equation}

O potencial de interação, por sua vez, pode ser dividido em três partes: (a) potencial de repulsão entre elétrons; (b) potencial de repulsão entre núcleos; e (c) potencial de atração entre elétrons e núcleos, que podem ser expressos da seguinte forma:
\begin{equation}
    \begin{split}
    \hat{V}&= \sum_{i,j}^{Ne}\frac{e^2}{4\pi\varepsilon_0}\frac{1}{\left | \mathbf{\hat{r}}_i-\mathbf{\hat{r}}_j \right |} \\&+ \sum_{\alpha,\beta}^{Nn}\frac{Z_\alpha Z_\beta e^2}{4\pi\varepsilon_0}\frac{1}{\left | \mathbf{\hat{R}}_\alpha-\mathbf{\hat{R}}_\beta \right |} \\& - \sum_{i}^{Ne}\sum_{\alpha}^{Nn}\frac{Z_\alpha e^2}{4\pi\varepsilon_0}\frac{1}{\left | \mathbf{\hat{r}}_i-\mathbf{\hat{R}}_\alpha\right |},
    \end{split}
    \label{eqn1p6}
\end{equation}

\noindent onde $Z_\alpha$ é o número de prótons no $\alpha$-ésimo núcleo, $e$ é a carga eletrônica, $\mathbf{\hat{r}}_i$ são os operadores de posição dos elétrons e $\mathbf{\hat{R}}_{\alpha}$ os operadores de posição dos núcleos. $N_e$ e $N_n$ são os números totais de elétrons e de núcleos da molécula, respectivamente.

Na sequência, a energia cinética pode ser dividida em duas partes: (a) energia cinética eletrônica e (b) energia cinética nuclear atômica, de modo que o operador energia cinética total fica dado por
\begin{equation}
    \hat{T}=\sum_{\alpha}^{N_n}\frac{\hbar^2}{2m_\alpha}\nabla_\alpha^2 + \sum_{i}^{N_e}\frac{\hbar^2}{2m_e}\nabla_i^2,
    \label{eqn1p7}
\end{equation}
\noindent onde $m_\alpha$ é a massa do $\alpha$-ésimo núcleo,  $\nabla_\alpha^2$ é o operador de Laplace do $\alpha$-ésimo núcleo, $m_e$ é a massa de um elétron, $\nabla_i^2$ é o operador de Laplace do $i$-ésimo elétron.

Devido às massas dos núcleos serem muito maiores que a massa dos elétrons, podemos aproximar as posições dos núcleos para valores fixos. Essa aproximação recebe o nome de  ``Born-Oppenheimer'' e permite desconsiderarmos o termo de energia cinética e tomar como constante a energia de repulsão entre os núcleos ($V_{NN}$), além de permitir trabalharmos com funções que dependem apenas das coordenadas dos elétrons (e também do spin, apesar de não aparecerem explicitamente), sendo comumente utilizada na química quântica~\cite{Born1927}. Considerando essa aproximação e utilizando unidades atômicas ($m_e =1$, $\hbar =1$, etc.), podemos definir o Hamiltoniano eletrônico (para este caso, enquanto as coordenadas eletrônicas $\mathbf{\hat{r}}_i$ continuam sendo operadores, as coordenadas dos núcleos $\mathbf{R}_{\alpha}$ são números reais) :
\begin{equation}
    \begin{split}
    \hat{H}&= - \sum_{i}^{N_e}\frac{1}{2}\nabla_i^2 + \sum_{i,j}^{N_e}\frac{1}{\left | \mathbf{\hat{r}}_i-\mathbf{\hat{r}}_j \right |}  \\ &- \sum_{i}^{N_e}\sum_{\alpha}^{N_n}\frac{Z_\alpha}{\left | \mathbf{\hat{r}}_i-\mathbf{R}_\alpha\right |} +V_{NN}
    \end{split}
    \label{eq:n1p8}
\end{equation}
que permite definir a equação de Schrödinger eletrônica:
\begin{equation}
    \hat{H}(\{\mathbf{R}_\alpha\})\ket{\Psi(\{\mathbf{R}_\alpha\})}=E(\{\mathbf{R}_\alpha\})\ket{\Psi(\{\mathbf{R}_\alpha\})},
\end{equation}
onde deixamos escrito $(\{\mathbf{R}_\alpha\})$ para explicitar que, apesar do Hamiltoniano atuar apenas nas autofunções e autoenergias eletrônicas e dependerem de forma explícita apenas das coordenadas $\mathbf{r}_i$ dos elétrons (e, possivelmente, de seus spins), todos são parametrizados pelas coordenadas dos núcleos. Assim, devido ao custo computacional, em aplicações práticas busca-se obter soluções aproximadas da equação de Schrödinger eletrônica para diversas configuração dos núcleos em vez da solução da equação original em que os núcleos e elétrons são colocados no mesmo patamar. Dado isso, a não ser que se faça necessário explicitar, iremos omitir a dependência da parametrização utilizada das coordenadas nucleares ($\{\mathbf{R}_\alpha\}$).

\subsection{Funções de onda de $N_e$ elétrons}
\label{sec:FunNEletrons}

 Para resolver problemas de autovalores e autovetores torna-se necessário recorrermos a bases dos espaços de Hilbert para a realização de cálculos numéricos, sendo a escolha da base fundamental para melhorar a qualidade da solução e/ou reduzir o custo computacional. Uma forma simples e conveniente de criar funções de onda de $N_e$ elétrons, e, posteriormente, uma base para espaços de $N_e$ elétrons, é por meio do produto de funções de onda de $1$ elétron. Assim, dado um conjunto de $N_e$ funções de onda de $1$ elétron $\{\psi_a(\mathbf{r},m)\}$ que dependem das suas coordenadas espaciais ($\mathbf{r}$) e também de seu spin ($m$), o produto delas forma uma função de $N_e$ elétrons (por simplicidade, utilizaremos $\sigma=(\mathbf{r},m)$ para representar tanto a função das coordenadas espaciais quanto de spin): 
\[
\Psi(\sigma_1,\sigma_2,\cdots,\sigma_{N_e})=\psi_{a_1}(\sigma_1)\psi_{a_2}(\sigma_2)\cdots\psi_{a_{N_e}}(\sigma_{N_e}).
\]

Entretanto, como os elétrons são férmions, pelo princípio da exclusão de Pauli e por serem todos idênticos, as funções que representam estados físicos reais dos elétrons pertençam a um espaço de Hilbert anti-simétrico, \textit{i.e.}, $\Psi(\cdots \sigma_i,\cdots, \sigma_j \cdots) = -\Psi(\cdots \sigma_j,\cdots, \sigma_i \cdots)$~\cite{piza2003mecanica}. A forma mais simples de trabalharmos com esse espaço anti-simétrico é usando uma base de determinantes de Slater, que são funções de onda de $N_e$ elétrons produzidas a partir de determinantes de $N_e$ funções de $1$ elétron em vez de simples produtos. Esses determinantes já são, por construção, anti-simétricos. Assim, usando as mesmas funções de onda do exemplo de produtos simples, uma função de $N_e$ elétrons normalizada é:
\begin{equation}
\small
\Psi(\sigma_1,\cdots,\sigma_{N_e}) = \frac{1}{\sqrt{N_e!}}
    \begin{vmatrix}\psi_{b_{1}}(\sigma_{1}) &  \cdots & \psi_{b_{N_e}}(\sigma_{1})\\
    \psi_{b_{1}}(\sigma_{2}) &  \cdots & \psi_{b_{N_e}}(\sigma_{2})\\
    \vdots &  \ddots & \vdots\\
    \psi_{b_{1}}(\sigma_{N_e}) &  \cdots & \psi_{b_{N_e}}(\sigma_{N_e})
    \end{vmatrix}.
    \label{SlaterDeterminant}
\end{equation}
Observamos que essa função é caracterizada puramente pelas $N_e$ funções ${\psi_{b_j}}$ utilizadas e sua ordem no determinante, pois, como um determinante é anti-simétrico, ao inverter a ordem, invertemos o sinal da função.

A distinção entre as partes que dependem das coordenadas espaciais e a de spin não é necessária para a maior parte do presente trabalho, ainda assim é essencial um entendimento sobre esse tópico. Os estados dos elétrons são definidos por vetores ou funções que guardam informações tanto sobre sua posição $\mathbf{r}\in\mathbb{R}^3$ quanto seu spin $m\in\{-\frac{1}{2},\frac{1}{2}\}$. Para fazer essa representação, as funções que dependem de ambas as variáveis podem ser separadas como produto de duas funções, uma que depende da posição $\phi(\mathbf{r})$, chamadas de orbitais espaciais ou apenas orbitais, e uma que depende do spin $g(m)$. Como o spin dos elétrons pode estar apenas em dois estados, podemos representar o espaço de Hilbert dos spins a partir das funções $g_{-\frac{1}{2}}$ e $g_{\frac{1}{2}}$ definidas por
\[
g_s(m)=\delta_{s,m}.
\]
Fisicamente, a utilização das funções $g_s$ nos determinantes pode ser interpretada como a existência de um elétron com spin $s$ no sistema~\cite{piza2003mecanica}. 
Por outro lado, a escolha dos orbitais espaciais $\phi(\mathbf{r})$ varia para diferentes problemas, dependendo de propriedades do sistema, simetrias e recursos computacionais disponíveis~\cite{morgon2001funccoes}. Uma discussão mais aprofundada sobre a escolha dos orbitais será apresentada na Seção \ref{sec:basisFunction}. O produto de um orbital por uma função de spin é chamado de spin-orbital.

Por fim, bases muito úteis para trabalhar com o espaço anti-simétrico (ou sub-espaços dele, como veremos mais para a frente) de um sistema de $N_e$ elétrons são formadas por esses determinantes ao utilizar diferentes combinações entre as infinitas funções de base do espaço de $1$ elétron, sendo utilizadas nos métodos de Hartree-Fock e Coupled Cluster, respectivamente, Seções \ref{sec:HartreeFock} e \ref{sec:CoupledClusterTheory}.

\subsection{O método de Hartree-Fock e a energia de correlação}
\label{sec:HartreeFock}

Como já comentado no final da Seção \ref{sec:FunNEletrons}, os determinantes discutidos até agora formam uma base no espaço anti-simétrico das funções de $N_e$ elétrons. Porém, uma questão que surge é quais bases e quais determinantes são mais relevantes para o nosso problema e o método de Hartree-Fock~\cite{hartree1928wave,fock1930naherungsmethode,szabo1996modern} é capaz de produzir resultados para isso.

O método busca uma aproximação do estado fundamental com apenas um único determinante e, para isso, ele obtém o conjunto de $N_e$ funções de $1$ elétron que produzem o determinante com menor energia. 

Inicialmente é construído o funcional de energia
\[
E\left[\Psi\right]=\bra{\Psi}\hat{H}\ket{\Psi}
\]
onde $\ket{\Psi}$ é um determinante de Slater. A partir desse funcional, busca-se o conjunto de funções que o minimizam, sob a restrição de estarem ortonormalizadas, assim, esse conjunto deve satisfazer:
\[
\delta E\left[\Psi\right] - \delta\left[\sum_{j,k}\lambda_{jk}\left(\bracket{\psi_{j}}{\psi_{k}}-\delta_{jk}\right)\right]=0.
\]

É possível demonstrar que existe um conjunto de funções $\{\psi_{j}\}$ tal que $\lambda_{jk}=0$ se $j\ne k$ e cada uma, individualmente, deve satisfazer a equação de Hartree-Fock
\begin{equation}
    F(\sigma_i)\psi_j(\sigma_i)=\epsilon_j\psi_j(\sigma_i).
    \label{eq:FockEigen}
\end{equation}
onde $F(\sigma_i)$ é o operador de Fock que atua sobre o elétron $i$, definido como 
\[
    F(\sigma_i) = -\frac{1}{2}\nabla_i^2 - \sum_\alpha^{Nn}\frac{Z_\alpha}{\left|\mathbf{\hat{r}}_i-\mathbf{R}_\alpha\right|} + v^{HF}(\sigma_i),
\]
onde $v^{HF}(\sigma_i)$ é um potencial médio que o $i$-ésimo elétron sente devido à presença de outros elétrons.

O potencial $v^{HF}(\sigma_i)$ é o mesmo para todos os elétrons, assim, a variável de atuação é desnecessária, porém, na literatura é comum a substuição de $\sigma_i$ apenas por $i$. Além disso, $v^{HF}(\sigma_i)$ depende também da função $\psi_j(\sigma_i)$, ou seja, a equação \eqref{eq:FockEigen} é não linear, sendo resolvida habitualmente por métodos de campo autoconsistente (SCF - do inglês \textit{self-consistent-field}) nos quais escolhe-se uma solução tentativa $\psi_j^{(0)}(\sigma_i)$ e a partir dela é gerado um potencial $v_{(0)}^{HF}(\sigma_i)$. O potencial produz uma nova função, $\psi_j^{(1)}(\sigma_i)$, que, por sua vez, cria um novo potencial $v_{(1)}^{HF}(\sigma_i)$ e o processo é repetido até a convergência que ocorre quando o campo gerado por uma função final $\psi_j(\sigma_i)$ for igual ao campo que a produz. 

As autofunções do operador de Fock formam uma base completa no espaço de $1$ elétron, cada uma com autovalor (ou autoenergia) $\epsilon_j$. 
É possível demonstrar que o determinante que miniminiza o funcional de energia é formado pelas $N_e$ autofunções do operador de Fock com menores autovalores. Esse determinante é chamado de estado fundamental de Hartree-Fock e simbolizado por $\ket{\Psi_{HF}}$. 

Os spin-orbitais que estão presentes no estado fundamental de Hartree-Fock são chamados de ocupados, enquanto os demais são os virtuais. Observamos que apesar do elétron possuir spin, que está incluso na variável $\sigma$ utilizada, o operador de Fock não afeta o spin (assim como o Hamiltoniano $\hat{H}$), dessa forma, ele sempre produz duas autofunções de mesma energia que diferem entre si apenas pela parte de spin. 

A energia do estado de Hartree-Fock é chamada de energia de Hartree-Fock $E_{HF}$. A diferença entre ela e a energia do estado fundamental do Hamiltoniano exata ($E_{\text{gr}}$) é conhecida como energia de correlação $\mathcal{E}_{corr}$, pois, como o método de Hartree-Fock é um método de campo médio, ele não leva em consideração as correlações quânticas dos elétrons, ou seja, 
\begin{equation}
    \mathcal{E}_{corr} = E_{\text{gr}} - E_{HF}
\end{equation}
e pelo princípio variacional $E_{\text{gr}} < E_{HF}\Rightarrow\mathcal{E}_{corr} < 0 $.

\subsection{Bases no espaço de $N_e$ elétrons}
\label{sec:NdimBasis}
A partir do estado de Hartree-Fock podemos produzir as chamadas excitações, por meio da substituição nos spin-orbitais dos estados ocupados pelos estados virtuais. Por exemplo, se substituirmos um spin-orbital ocupado por um spin-orbital virtual, temos um determinante com uma excitação, e assim por diante. Habitualmente os estados ocupados são demarcados pelas letras $a,b,c\dots$ e os virtuais por $r,s,t,\dots$. Os determinantes onde ocorrem uma, duas, três, etc substituições nos estados ocupados são denotados, respectivamente, por $\ket{\Psi_a^r},\ket{\Psi_{ab}^{rs}},\ket{\Psi_{abs}^{rst}}$, etc, podendo chegar até $N_e$ excitações (todos os estados ocupados sendo substituídos por algum dos infinitos estados virtuais). 

O conjunto formado pelo estado de Hartree-Fock e todos as possíveis excitações formam uma base para o espaço de $N_e$ elétrons, dessa forma, qualquer estado pode ser escrito como uma combinação linear, normalizada, desses estados. Em especial, o estado fundamental é, a menos de um fator de normalização:
\begin{equation}
    \ket{\Psi_{\text{gr}}} = \ket{\Psi_{HF}} + \sum_{a,r} c_a^r\ket{\Psi_a^r} + \sum_{a,b,r,s} c_{ab}^{rs}\ket{\Psi_{ab}^{rs}}+\dots.
    \label{eq:FCIequation}
\end{equation}
Observa-se que apesar do nome ``excitações'', ainda utilizamos esses estados para construir o estado fundamental.

Essa base em especial é bastante vantajosa, pois, dado que $\ket{\Psi_{HF}}$ é, em algum sentido, uma boa aproximação para o estado fundamental de diversos sistemas moleculares, alguns métodos muito renomados na química quântica partem desse estado e tentam obter a energia do estado fundamental encontrando as ``contribuições'' dos determinantes excitados. Assim, a base formada pelos autoestados do operador de Fock é de grande relevância para a química quântica.

Por outro lado, é inviável trabalhar diretamente com uma base completa de determinantes, uma vez que o conjunto de determinantes formados por uma base infinita de funções de $1$ elétron também é infinito. Tal conjunto ultrapassa a capacidade de processamento de qualquer computador, seja clássico ou quântico. Por isso, na prática, é necessário restringir a análise a uma base finita de funções de $1$ elétron. Com essa base reduzida, a equação de Hartree-Fock é resolvida utilizando a forma matricial do operador de Fock.

O número de autoestados linearmente independentes gerados pelo método corresponde ao tamanho da base escolhida. Dentre esses autoestados, são selecionados aqueles com menor energia para formar o estado de Hartree-Fock \emph{na base utilizada}. Esse estado, no entanto, é apenas uma projeção do verdadeiro estado de Hartree-Fock (que seria obtido com uma base infinita) no subespaço formado pelos determinantes construídos a partir da base finita adotada. Consequentemente, as energias calculadas também são aproximações e estão limitadas pela escolha da base. Assim, ao tratar de propriedades computáveis, como a energia do estado fundamental, energia de Hartree-Fock e energia de correlação, essas são sempre entendidas como associadas à base específica utilizada, e não às quantidades ideais previstas pela teoria que assume uma base infinita.

Por fim, a quantidade de possíveis determinantes linearmente independentes para uma base com $N_f$ funções e $N_e$ elétrons é dada por ${N_f\choose N_e}$, isto é, o número de combinações possíveis de $N_e$ funções entre as $N_f$ disponíveis. Para sistemas com muitos elétrons, o aumento do tamanho da base, necessário para obter maior precisão, provoca um crescimento rápido no custo computacional do problema. Por isso, muitos métodos são desenvolvidos para operar com um conjunto limitado de determinantes. Um exemplo relevante é o de Configuração de Interação (CI - do inglês \textit{Configuration Interaction})~\cite{szabo1996modern}, que restringe a solução da equação de Schrödinger (equação \eqref{eq:FCIequation}) a um número fixo de excitações. Quando todas as excitações possíveis são consideradas, o método CI é chamado de \textit{Full Configuration Interaction} (FCI). Nesse caso, por englobar todos os determinantes possíveis, o FCI se torna equivalente à solução exata na base finita escolhida. Por isso, é comum referir-se à energia exata em uma base finita como a energia de FCI ($E_{\text{FCI}}$) associada a essa base e é usada frequentemente para testar novos métodos em problemas simples, \textit{i.e.}, \textit{toy models}.

\subsection{Construção dos circuitos para diferentes funções de base}
\label{sec:basisFunction}

As funções espaciais de $1$ elétron usadas para construir as funções $\Psi$ são chamadas de orbitais moleculares (OM)~\cite{Jensen2007}. No entanto, como não conhecemos a forma desses OM, são utilizadas combinações lineares de funções que conhecemos, os chamados orbitais atômicos (OA). Essa é a aproximação conhecida como combinação linear de orbitais atômicos (CLOA), onde cada OM $\phi$ é formado por:

\begin{equation}
  \phi = \sum_{i=1}^m c_i \chi_i  
\end{equation}

\noindent em que $m$ é o número de funções de base, $c_i$ são coeficientes da combinação linear e $\chi_i$ são os OA. Quanto maior o número de funções de base, melhor a representação do sistema. No limite, quando se utiliza um número infinito de funções de base, o conjunto é chamado de completo e a expansão de OA em OM deixa de ser uma aproximação. 

As funções de Slater, conhecidas como orbitais do tipo Slater (STO – do inglês \textit{Slater type orbitals}), foram uma das funções de base que primeiro conseguiu reduzir a complexidade computacional relacionada às integrais de energia. Elas tem a forma~\cite{Morgon2018}
\begin{equation}
    \chi_{n, l, m, \zeta}(r, \theta, \phi) = NY_{l,m}(\theta,\phi) r^{n-1} e^{-\zeta r},    
\end{equation}
\noindent em que $r$, $\theta$ e $\phi$ são coordenadas esféricas, $N$ é uma constante de normalização, $Y_{l,m}$ são esféricos harmônicos e $\zeta$ é chamado de expoente do orbital e é um parâmetro ajustado previamente. As vantagens das STO são que elas apresentam a cúspide que é encontrada no ponto onde o núcleo do átomo está localizado e ela apresenta um decaimento exponencial, similar ao decaimento encontrado nos orbitais exatos encontrados na resolução da equação de Schrödinger para o átomo de Hidrogênio. Essas vantagens fazem com que a convergência com o aumento do número de funções de base seja mais rápida que para funções que não apresentam essas características. No entanto, os cálculos de integrais de 3 e 4 centros e dois elétrons não podem ser resolvidos analiticamente (equações \eqref{eq:n1p10} e \eqref{eq:n1p12}), o que as torna computacionalmente pouco eficientes. 

\begin{figure}[H]
   \centering
   \includegraphics[width=0.53\textwidth]{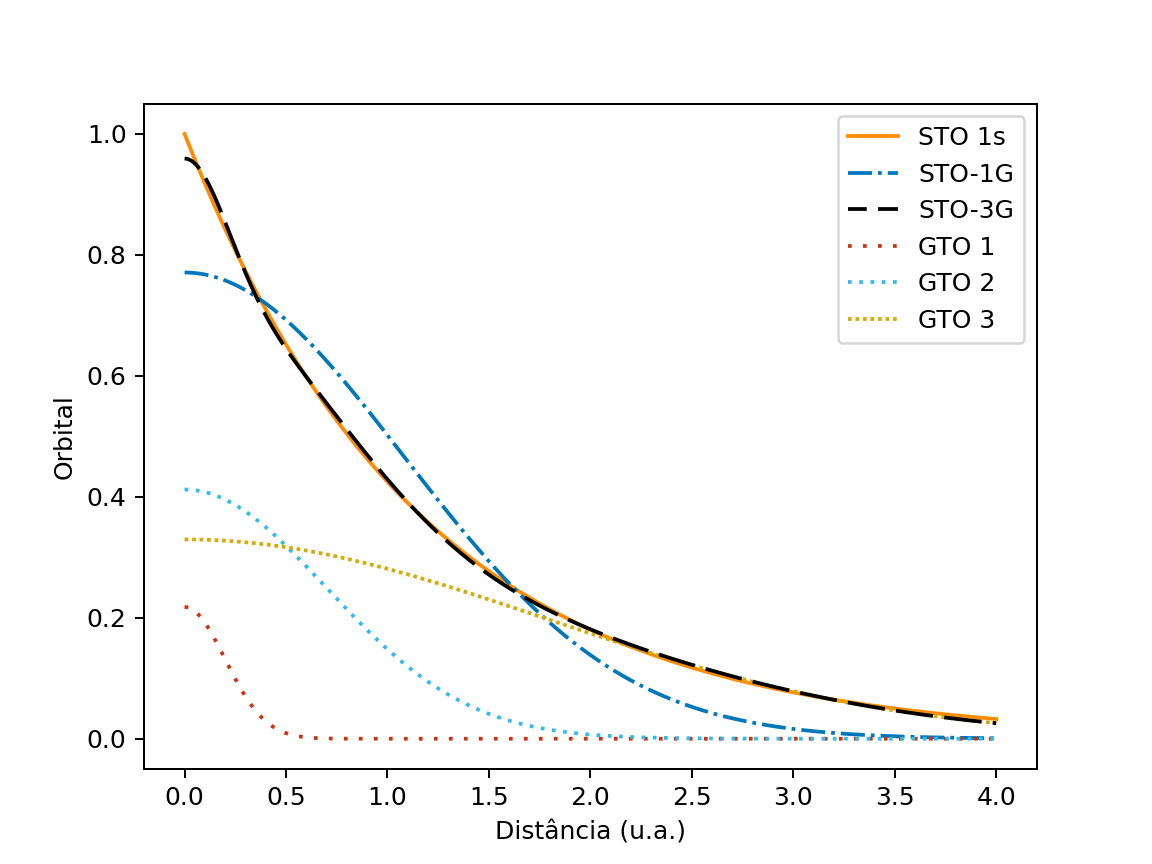}
   \caption{Curvas dos orbitais (funções) STO, STO-1G, STO-3G e GTO para o estado 1s. O uso de uma gaussiana (STO-1G) não reproduz adequadamente a curva STO para o orbital 1s, enquanto o uso de três gaussianas (STO-3G) o faz de forma muito próxima na região mostrada.
   }
   \label{fig:curva_sto_gto}
\end{figure}

Dessa forma, com objetivo de simplificar os cálculos e evitar essa limitação é feita a utilização de funções Gaussianas, também chamadas de orbitais do tipo Gauss (GTO – do inglês \textit{Gaussian type orbitals}). GTOs podem ser escritas em função de coordenadas cartesianas ou polares:
\begin{equation}
\chi_{n, l, m, \zeta}(r, \theta, \phi) = NY_{l,m}(\theta,\phi) r^{2n-2-l} e^{-\zeta r^2},
\end{equation}
\begin{equation}
\chi_{l_x, l_y, l_z, \zeta}(x, y, z) = N x^{l_x} y^{l_y} z^{l_z} e^{-\zeta r^2},
\end{equation}
\noindent em que $N$ é um fator de normalização, $l_x$, $l_y$ e $l_z$ são parâmetros que determinam o tipo e momento angular do orbital, e, por consequência, sua simetria. Através da somatória dos expoentes $l_x$, $l_y$ e $l_z$, pode-se saber qual orbital se trata: $l_x$ + $l_y$ + $l_z$ = 0 tem um orbital do tipo $s$,  $l_x$ + $l_y$ + $l_z$ = 1 temos um orbital do tipo $p$, e dai em diante. As desvantagens do uso de GTOs vem de que a função apresenta derivada zero no núcleo, ao invés de uma cúspide, e não apresenta decaimento exponencial. Por isso, as GTOs não conseguem representar satisfatoriamente nem a região próxima nem a região afastada do núcleo, como pode ser visto na Fig. \ref{fig:curva_sto_gto}. Para corrigir essa limitação é utilizada uma combinação linear de GTOs para descrever uma única STO. Essas funções de base são conhecidas como STO-\textit{n}G, em que \textit{n} representa o número de GTOs que são utilizadas para cada STO e G representa que são funções Gaussianas~\cite{Hehre1969}. Enquanto a curva STO-1G, formada por apenas uma gaussiana, não consegue representar nem o início nem o fim da função STO, nota-se que a função STO-3G, formada pela combinação linear das funções GTO 1, 2 e 3, representa tanto valores próximos quanto afastados do núcleo. Mostrou-se que a utilização de mais de 3 GTOs não ocasionava grandes mudanças, então a mais utilizada é a STO-3G~\cite{Jensen2007}.

Funções de base como STO-nG são chamadas de funções mínimas, ou de zeta único (SZ - do inglês \textit{single zeta}). Essas têm apenas o número de funções de base necessárias para representar os elétrons e orbitais de um átomo neutro. Dessa forma, os átomos do primeiro período, a primeira linha, da tabela periódica terão apenas uma única função de base para cada, já que esses apresentam apenas um orbital s. Isso pode ser visto na Tabela \ref{tb:STO-3G_moleculas_diatomicas}, que apresenta o número de funções de base, de orbitais e de qubits para diversas moléculas diatômicas utilizando STO-3G. Tanto H$_2$ quanto He$_2$ apresentam apenas 2 orbitais, o que explica porque é impossível fazer um cálculo de Coupled Cluster (CC) (explicado na Seção \ref{sec:CoupledClusterTheory}) com STO-nG, ou outra função de base mínima, para He$_2$: não existem orbitais virtuais para os quais os elétrons possam ser excitados. O mesmo ocorre para Ne$_2$, que apresenta 10 orbitais sendo todos ocupados.

\begin{table*}[!htb]
\begin{center}
\caption{Número de elétrons, orbitais ocupados e virtuais, funções de base e de qubits para moléculas diatômicas e H$_2$O.}\label{tb:STO-3G_moleculas_diatomicas}
\begin{tabular}{lccccc}
\toprule%
\multirow{2}{*}{Molécula} & \multirow{2}{*}{Elétrons} & \multicolumn{2}{c}{Orbitais}  & Funções & \multirow{2}{*}{qubits}       \\\cmidrule{3-4}%
                          &                           & ocupados & virtuais           & de base & \\
\midrule
H$_2$    &   2   & 1  & 1    &      2   & 4 \\
He$_2$   &   4   & 2  & 0    &      2   & 4 \\
O$_2$    &   16  & 8  & 2    &      10  & 20 \\
Ne$_2$   &   20  & 10 & 0    &      10  & 20 \\
H$_2$O   &   10  & 5  & 2    &      7   & 14 \\    
HLi      &   4   & 2  & 4    &      6   & 12 \\
HBe      &   5   & 3  & 3    &      6   & 12 \\
HB       &   6   & 3  & 3    &      6   & 12 \\
HC       &   7   & 4  & 2    &      6   & 12 \\
HN       &   8   & 4  & 2    &      6   & 12 \\
HO       &   9   & 5  & 1    &      6   & 12 \\
HF       &   10  & 5  & 1    &      6   & 12 \\
HNe      &   11  & 6  & 0    &      6   & 12 \\
\bottomrule
\end{tabular}
\end{center}
\end{table*}

Podemos usar a molécula de O$_2$ como exemplo de como se calcular o número de funções de base e de qubits utilizados com STO-3G. Observa-se que o número de qubits é igual ao número de spin-orbitais e a razão disso será explicado na Seção \ref{sec:VQEsteps}. O átomo de oxigênio apresenta uma distribuição eletrônica 1s$^2$2s$^2$2p$^4$, então será utilizada uma função de base para o orbital 1s, uma para o orbital 2s e três para os orbitais 2p$_x$, 2p$_y$ e 2p$_z$, totalizando cinco funções de base. Seguindo então o método CLOA, a molécula de O$_2$ será descrita por cinco orbitais de cada oxigênio gerando dez funções de base e 20 qubits para representar os spin-orbitais que cada um dos elétrons pode ocupar. Já para molécula de H$_2$O, teremos cinco orbitais do oxigênio e mais dois de cada hidrogênio, totalizando 7 orbitais e funções de base. A Tabela \ref{tb:STO-3G_moleculas_diatomicas} também apresenta uma série de moléculas formadas por um hidrogênio e um átomo da segunda linha da tabela periódica. Algumas dessas moléculas não são estáveis, mas são utilizadas apenas como exemplo de preenchimento dos orbitais e da função de base STO-3G. Nessa série vemos que o número de elétrons aumenta e os orbitais vão sendo ocupados com um ou dois elétrons. No entanto, o número de funções de base e de qubits não é alterado.

\begin{figure*}[!htb]
\captionsetup[subfigure]{justification=centering}
     \centering
     \begin{subfigure}[b]{0.48\linewidth}
         \centering
         \includegraphics[width=\linewidth]{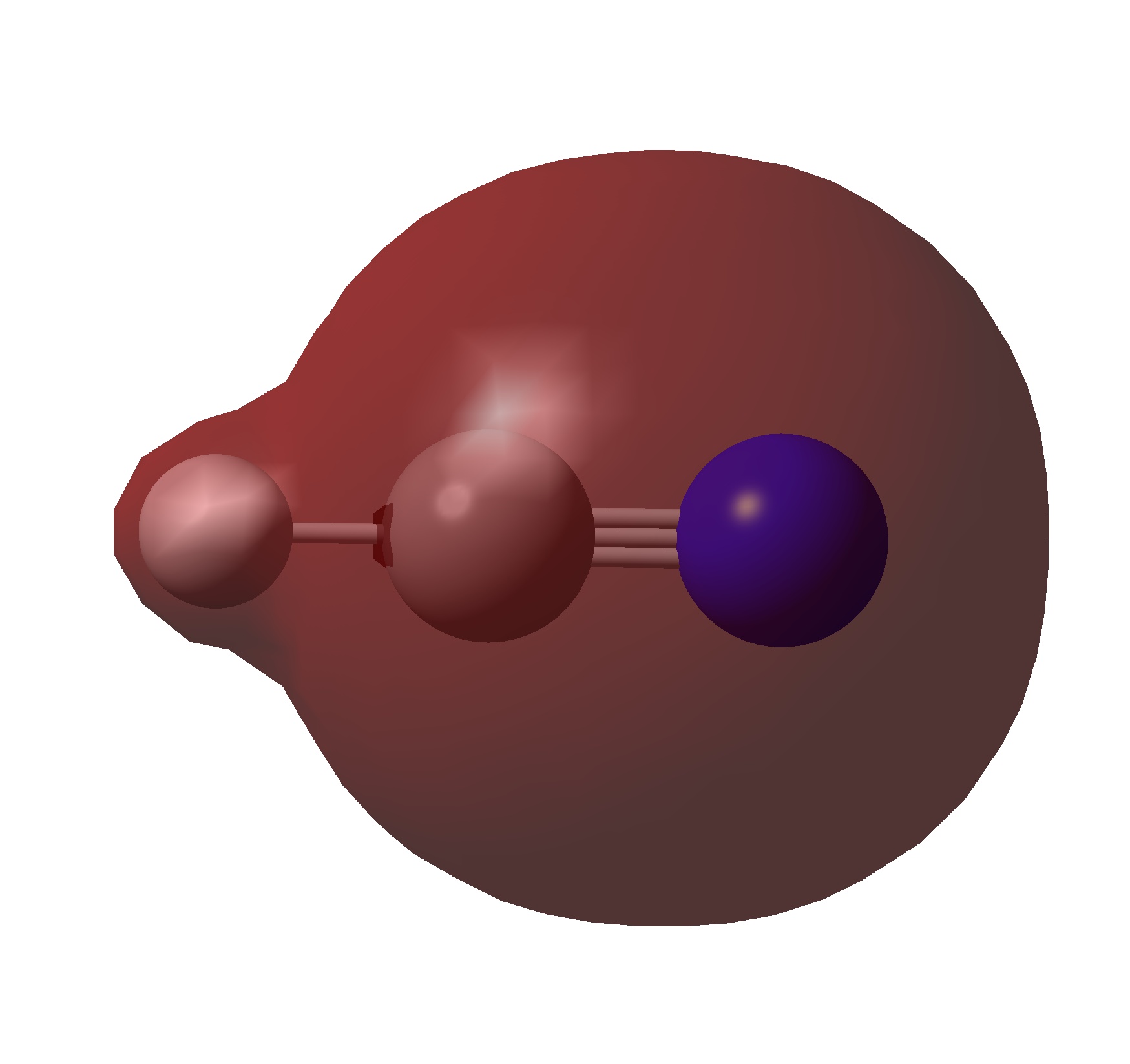}%
          \caption{}
     \end{subfigure}%
     \hfill
     \begin{subfigure}[b]{0.48\linewidth}
         \centering
         \includegraphics[width=\linewidth]{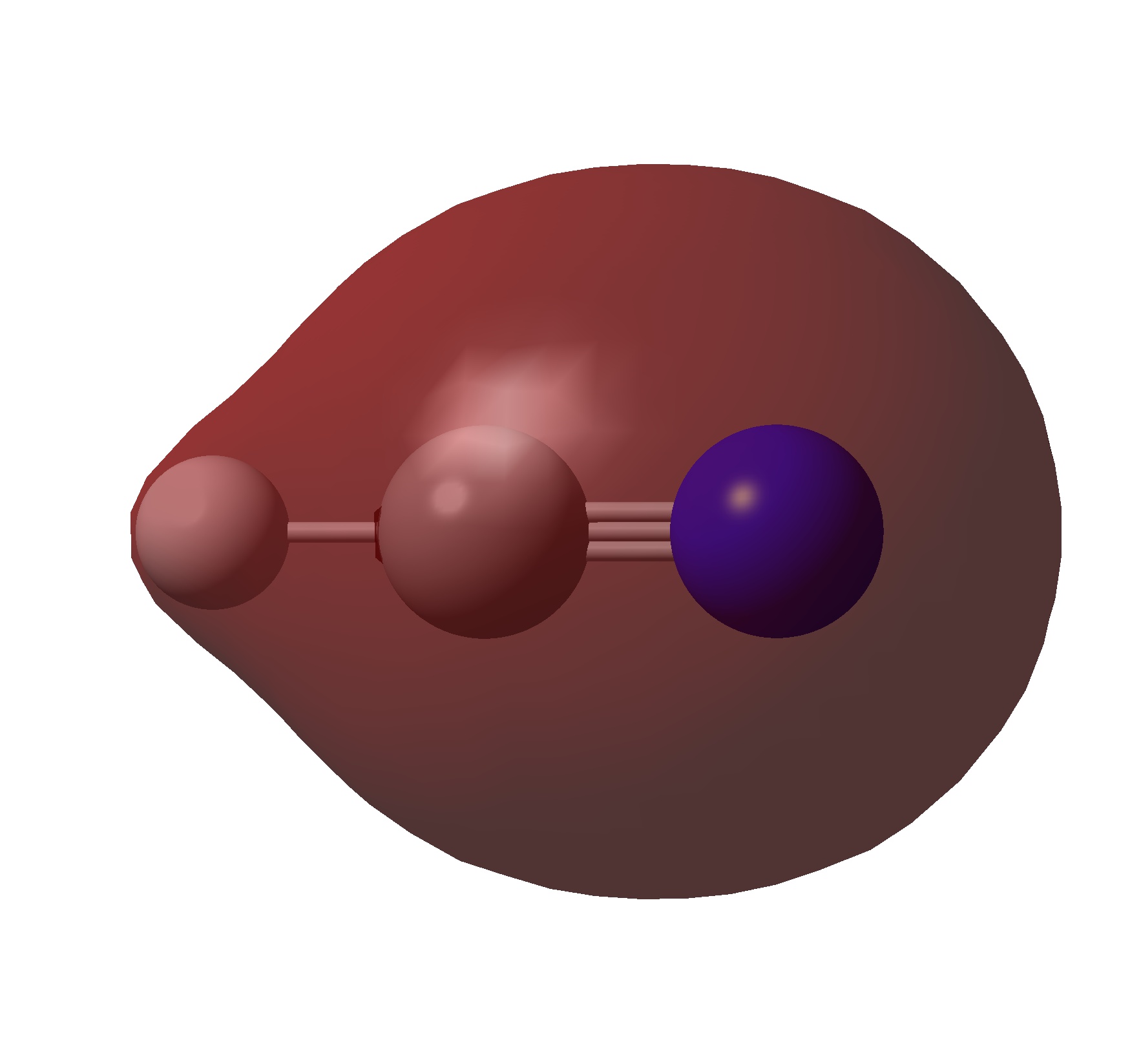}%
        \caption{}
     \end{subfigure}%
        \caption{Orbitais moleculares calculado para a molécula de HCN utilizando função de base a) STO-3G e b) 6-31G(d,p). Os átomos são representados por esfera cinza para carbono, azul para nitrogênio e branco para hidrogênio.}
        \label{fig:orbitais_SZ_DZ}
\end{figure*}

Uma primeira forma de melhorar funções de base é utilizando mais funções para descrever os mesmos orbitais. Quando se dobra o número de funções de base elas são chamadas de funções de base de zeta duplo (DZ - do inglês \textit{double zeta}). O termo zeta vem historicamente por esta ser a letra grega normalmente utilizada nos expoentes das funções de base. Dessa forma, uma função de base DZ irá usar duas funções de base do tipo 1s para o átomo de hidrogênio. Para o átomo de oxigênio, e outros átomos do segundo período, serão usadas duas funções de base para os orbitais 1s e duas para 2s, bem como dois grupos de orbitais 2p, ou seja, dois de cada um dos 2p$_x$, 2p$_y$ e 2p$_z$. Para átomos isolados, as funções SZ podem ser capazes de descrever bem os orbitais atômicos, no entanto, como as ligações que um determinado átomo irá fazer podem ser diferentes no ambiente molecular, a adição de mais funções de base permite que os diferentes expoentes possam melhor descrever os diversos tipos de ligações em diferentes direções às quais os átomos estão sujeitos. Por exemplo, um carbono pode estar ligado ao mesmo tempo a um outro átomo de carbono e a um hidrogênio, o que torna os expoentes ideais para a primeira ligação diferentes dos expoentes da segunda ligação. Tal questão pode ser vista na Fig. \ref{fig:orbitais_SZ_DZ}, onde se nota, para a molécula de HCN com função STO-3G, que enquanto o átomo de carbono interage com o de nitrogênio, o orbital 1s do hidrogênio se mantém relativamente isolado do resto da molécula, enquanto que para uma função de base DZ há uma maior interação.

Seguindo o mesmo raciocínio da base DZ, pode-se usar funções de base de zeta triplo (TZ - do inglês \textit{triple zeta}), zeta quadruplo (QZ - do inglês \textit{quadruple zeta}), e daí em diante. Como é de se esperar, funções de maior zeta são capazes de descrever e aumentar a precisão dos resultados, bem como também aumentam os custos computacionais.

Como as ligações químicas ocorrem essencialmente entre elétrons da camada mais externa, chamada de camada de valência, muitas vezes só duplica, triplica, etc, os orbitais que compõem essa camada, sendo chamado de função de base de valência separada. Estas são utilizadas na vasta maioria dos cálculos de estrutura eletrônica, já que pode-se economizar os custos computacionais dos elétrons do caroço (\textit{i. e.} elétrons que não são os de valência) que têm participação reduzida em reações químicas. Com isso em mente, Pople desenvolveu um grupo de funções de base e com uma notação específica {$a$-$bcdG$}~\cite{Hehre1969,Binkley1980,Ditchfield1971,Gordon1982} em que $a$ representaria o número de Gaussianas utilizadas para descrever os orbitais de caroço, seguido de um traço e de dois ou três números, que representam o número de Gaussianas que são utilizadas em cada grupo, terminado pela letra \textit{G}, para indicar que se tratam de Gaussianas. Por exemplo, a notação {3-21G} corresponde a usar uma função de base formada por três Gaussianas para os orbitais de caroço e duas funções de base contendo duas (2) e uma (1) Gaussiana, para os orbitais de valência, o que a torna uma função de base DZ. Já a função de base 6-311G~\cite{Krishnan1980} usará uma função de base formada por seis Gaussianas para os orbitais de caroço, e três funções de base para cada orbital da camada de valência, formados por três, uma e uma Gaussiana, para 311 respectivamente.

A Tabela \ref{tb:funções_base_exemplos}, mostra exemplos do número de orbitais para moléculas pequenas com as funções de base de Pople. Para a função de base 3-21G~\cite{Binkley1980} como H$_2$ não tem orbitais de caroço, temos apenas a duplicação dos orbitais de valência para cada hidrogênio, totalizando quatro funções de base e oito qubits. O mesmo vale para He$_2$, que agora apresenta dois orbitais virtuais e pode ser calculado. Na molécula de O$_2$, o oxigênio terá uma função de base para os orbitais de caroço 1s, e duas para os orbitais de valência 2s, 2p$_x$, 2p$_y$ e 2p$_z$, num total de nove funções de base. Por isso, a molécula terá dezoito funções de base e usará 36 qubits.

\begin{table*}[!htb]
\begin{center}
\caption{Número de elétrons, orbitais ocupados e virtuais, funções de base e de qubits para moléculas diatômicas e H$_2$O.}\label{tb:funções_base_exemplos}
\begin{tabular}{llccccc}
\toprule%
\multirow{2}{*}{Molécula} & Função & \multirow{2}{*}{Elétrons} & \multicolumn{2}{c}{Orbitais}  & N.º funções & \multirow{2}{*}{qubits}       \\\cmidrule{4-5}%
                          &  de base &  & ocupados & virtuais           & de base & \\
\midrule
H$_2$    &   3-21G    &  2     &   1   &   3    &  4    &  8  \\
He$_2$   &   3-21G    &  4     &   2   &   2    &  4    &  8  \\
O$_2$    &   3-21G    &  16    &   8   &   10   &  18   &  36  \\
H$_2$O   &   3-21G    &  10    &   5   &   8    &  13   &  26  \\
H$_2$O   &   6-31G    &  10    &   5   &   8    &  13   &  26  \\
H$_2$O   &   6-311G   &  10    &   5   &   14   &  19   &  38  \\
\bottomrule
\end{tabular}
\end{center}
\end{table*}

Para o cálculo da energia eletrônica em computadores quânticos, vários fatores que são importantes em computadores clássicos, como distribuição das Gaussianas, sua otimização, balanceamento, expoentes e tipos, não têm nenhuma influência. Nesse sentido, as funções de base 3-21G e 6-31G~\cite{Hehre_1972}  produzem circuitos idênticos, mesmo que os resultados de cálculos utilizando computadores clássicos sejam significativamente diferentes. Isso porque ambas utilizam uma função de base para o caroço e duas para os orbitais de valência, mesmo que elas sejam formadas por diferentes números de Gaussianas, como pode ser visto na Tabela \ref{tb:funções_base_exemplos} para a molécula de H$_2$O. No caso da função de base TZ 6-311G ~\cite{Krishnan1980} serão seis funções de base para cada orbital de valência 1s dos hidrogênios, uma função de base para o orbital de caroço 1s do oxigênio, e três funções de base para cada orbital de valência 2s, 2p$_x$, 2p$_y$ e 2p$_z$.

\begin{figure*}[!htb]
\captionsetup[subfigure]{justification=centering}
     \centering
     \begin{subfigure}[b]{0.48\linewidth}
         \centering
         \includegraphics[width=\linewidth]{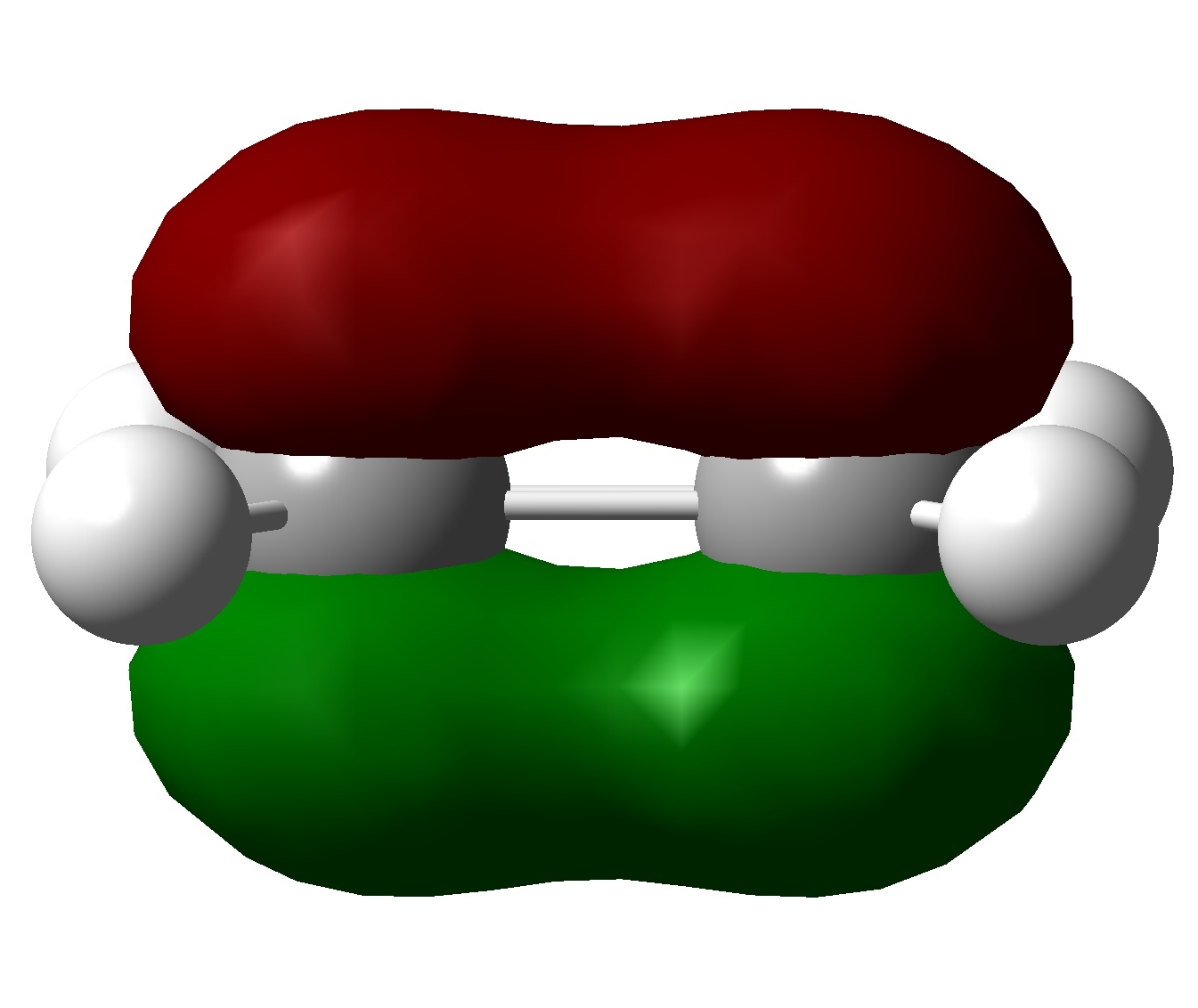}%
          \caption{}
     \end{subfigure}
     \hfill
     \begin{subfigure}[b]{0.48\linewidth}
         \centering
         \includegraphics[width=\linewidth]{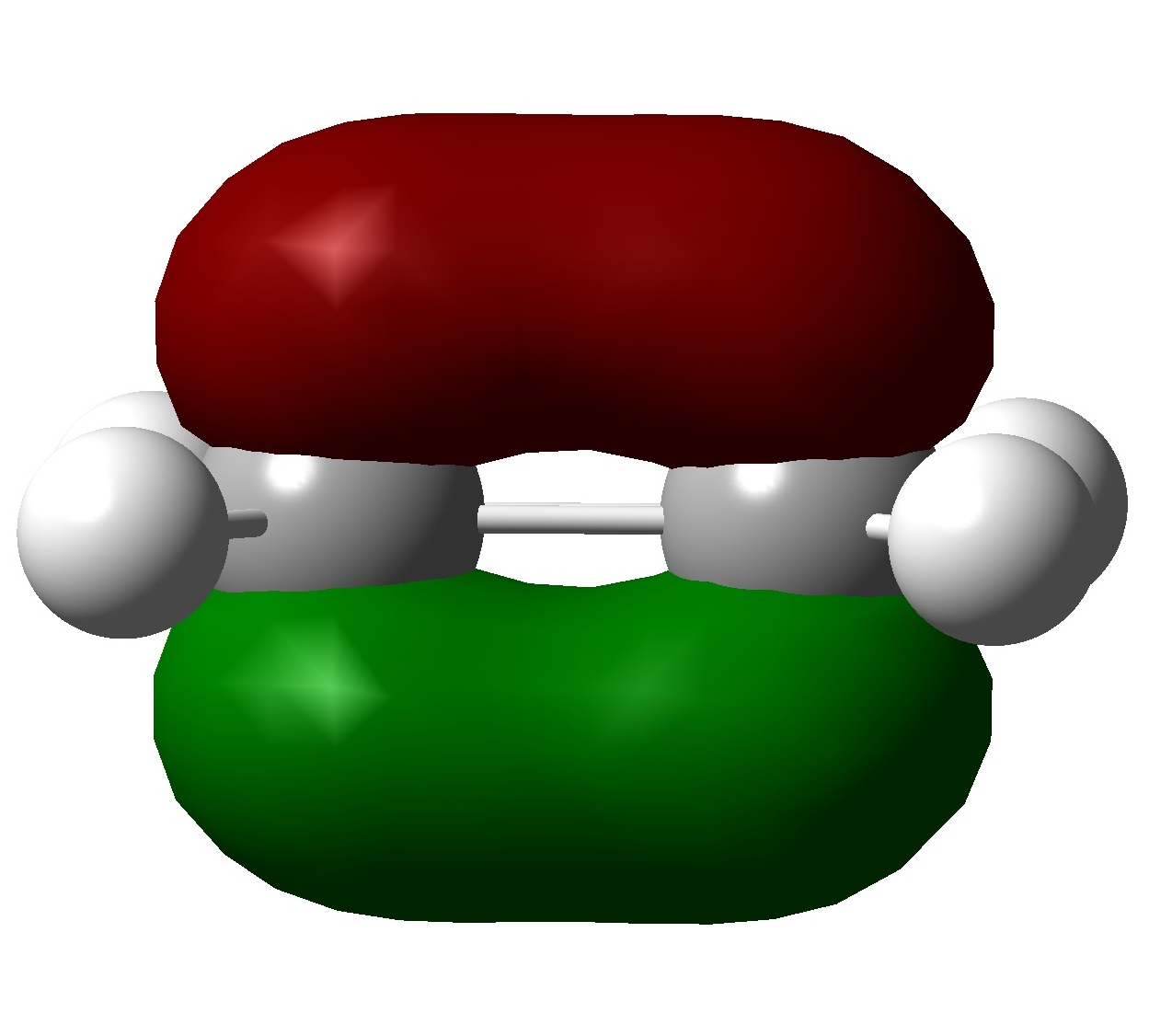}%
        \caption{}
     \end{subfigure}
        \caption{Orbitais moleculares calculados para a molécula de C$_2$H$_4$ utilizando função de base a) STO-3G e b) 6-31G(d,p). Os átomos são representados por esfera cinza para carbono e branco para hidrogênio.}
        \label{fig:orbitais_polarization}
\end{figure*}

O número de funções de base tem efeito essencialmente na parte radial do orbital. No entanto, em ambiente molecular, os orbitais estarão sujeitos a distorções angulares, como por exemplo distorcer um orbital puramente esférico 1s em uma forma elipsoide em um dos eixos devido a uma ligação com outro átomo. Tal efeito é obtido adicionando à função de base original outras funções de base de maior momento angular. No exemplo, se adicionarmos uma função 2p$_x$ a uma função de base 1s esférica, poderemos obter um orbital que tem flexibilidade para se distorcer na direção $x$ da forma que melhor representar a ligação em questão. Outro exemplo são ligações duplas entre átomos de carbono, que podem ser vistas na Fig. \ref{fig:orbitais_polarization}, nas quais a primeira ligação ocorre no eixo formado pelos núcleos dos átomos, e a segunda ocorre perpendicularmente à primeira. Como esses orbitais 2p se distorcem fora do eixo, mas na direção desse para melhor interação entre eles, são necessárias funções 3d para que essa distorção possa ser melhor descrita.

Para as funções de base de Pople, a notação de funções de polarização é {$a$-$bcdG(e,f)$} em que é adicionado um parênteses com os tipos de funções de polarização após $G$. O primeiro número dentro dos parâmetros, $e$, representam as funções adicionadas a átomos diferentes do hidrogênio, e o segundo, $f$, aos átomos de hidrogênio. Essa separação é interessante pois átomos de hidrogênio são frequentemente mais numerosos que átomos mais pesados, e são átomos que na maior parte dos casos estão envolvidos em apenas uma ligação e não fazem parte das ligações em que as reações estão acontecendo. Dessa forma, pode-se reduzir consideravelmente o custo computacional com um erro relativamente pequeno. No entanto, caso o hidrogênio tenha um papel importante no sistema que está sendo estudado, pode ser importante também adicionar funções de polarização a esses átomos presentes na molécula.

O número de orbitais virtuais, de funções e qubits para a molécula de H$_2$O utilizando diferentes funções de base são apresentados na Tabela \ref{tb:H2O_dif_func_base}. A diferença do número de funções de base utilizadas de 6-31G para 6-31G(d) são as cinco funções 3d adicionadas ao átomo de oxigênio. Em seguida, pode-se ver que a função de base 6-31G(d,p) tem 6 funções a mais que a 6-31G(d). Essas funções vem dos três orbitais 2p adicionados a cada hidrogênio. Isso nos mostra o quanto as funções de polarização em átomos de hidrogênio podem impactar o custo computacional, aumentando nesse caso o número de funções em 33\%.

\begin{table*}[!htb]
\begin{center}
\caption{Número de orbitais virtuais, de funções e qubits para a molécula de H$_2$O, que possui 10 elétrons e 5 orbitais preenchidos.}\label{tb:H2O_dif_func_base}
\begin{tabular}{llccccc}
\toprule%
 Função  &  Orbitais & N.º funções & \multirow{2}{*}{Qubits}       \\
 de base &  virtuais &  de base    &                               \\
\midrule
6-31G         &   8    &   13   &   26     \\
6-31G(d)      &   13   &   18   &   36     \\
6-31G(d,p)    &   19   &   24   &   48     \\
6-31+G        &   12   &   17   &   34     \\
6-31++G       &   14   &   19   &   38     \\
def2-SVP      &   19   &   24   &   48     \\
def2-TZVP     &   38   &   43   &   86     \\
def2-TZVPP    &   54   &   59   &   118    \\
cc-pVDZ       &   19   &   24   &   48     \\
cc-pVTZ       &   53   &   58   &   116    \\
aug-cc-pVDZ   &   36   &   41   &   82     \\
aug-cc-pVTZ   &   87   &   92   &   184    \\
\bottomrule
\end{tabular}
\end{center}
\end{table*}

As funções GTO decaem rapidamente em relação às funções STO, e, por isso, não descrevem bem essa região, que por ser uma parte da função de onda longe do núcleo tem pouco impacto na energia final. No entanto, para alguns sistemas como ânions~\cite{Nagy2017} estados excitados~\cite{Jensen2007} ou interações não-covalentes~\cite{Johnson_2013}, a descrição da função de onda a maiores distâncias pode ser muito importante. A solução normalmente utilizada é adicionar uma função difusa, com expoente menor, às funções de base utilizadas. Para as funções de base de Pople, podem ser adicionados um ou dois ``+'', que, como as funções de polarização, representam a adição de funções difusas em átomos diferentes de hidrogênio e átomos de hidrogênio. Os argumentos para adicionar apenas aos átomos pesados é o mesmo do das funções de polarização. Na Tabela \ref{tb:H2O_dif_func_base}, podemos ver que para 6-31+G são adicionados quatro funções à função de base 6-31G, que correspondem a uma função 2s e três funções 2p, e para 6-31++G temos, além das funções adicionadas ao oxigênio, também a adição de uma função 1s para cada hidrogênio.

Um grupo de funções mais atual, considerado como um melhor custo benefício~\cite{kirschner2020performance} e que tem resultados mais consistentes~\cite{Pitman2023} são as funções de base de Ahlrichs~\cite{Weigend2005} que tem qualidade DZ, TZ e QZ. As do tipo DZ são chamadas de def2-SVP, pois são bases padrão (def - do inglês \textit{default}) de valência separada e polarizada (SVP - do ingês \textit{single valence polarized}). Elas apresentam um esquema de 2s1p, que significa dois orbitais do tipo s e um orbital do tipo p, para o átomo de hidrogênio e hélio, e um esquema 3s2p1d para átomos do segundo período. Então para a molécula de água, o hidrogênio terá dois orbitais do tipo s mais um orbital do tipo p, que é formado pelos orbitais 2p$_x$, 2p$_y$ e 2p$_z$, totalizando 5 funções de base, e o oxigênio terá três orbitais do tipo s, dois do tipo p e um do tipo d, que é formado por cinco orbitais, totalizando quatorze funções de base, que faz com que a água possua vinte e quatro, cinco de cada hidrogênio mais quatorze do oxigênio, funções de base (Tabela \ref{tb:H2O_dif_func_base}).

As funções de base do tipo TZ são chamadas de def2-TZVP e def2-TZVPP, em que TZV quer dizer valência de zeta triplo (TZV - do inglês \textit{triple zeta valence}) e P e PP representam conjuntos de polarização menores ou maiores, respectivamente. Elas possuem esquemas 3s1p e 3s2p1d para o átomo de hidrogênio, para TZVP e TZVPP respectivamente, e 5s3p2d1f para os átomos do segundo período. Pode-se ver que nessa região há apenas o aumento de polarização para átomos de hidrogênio. No entanto, existem diferenças no número de polarizações para átomos mais pesados. Para a molécula de água, teremos então três orbitais tipo s e um do tipo p para os hidrogênios ($3+3\cdot1=6$) e cinco orbitais s, três p, dois d e um f para o oxigênio ($5+3\cdot3+2\cdot5+7=31$), totalizando quarenta e três funções para a função de base def2-TZVP. Para a função de base def2-TZVPP há apenas o aumento de 1p e 1d para os átomos de hidrogênio, que geram 8 funções de base e adicionam 16 funções de base a mais do que para a molécula de água com def2-TZVP.

As funções de base chamadas de correlação consistente (cc - do inglês \textit{correlation consistent} escrito no minúsculo para não confundir com CC) foram desenvolvidas por Dunning\cite{Dunning1989} para serem capazes de recuperar a correlação de elétrons da camada de valência. São populares por atingirem 99\% da correlação de outras bases do mesmo tipo com dois terços do número de funções Gaussianas e metade do número de funções de polarização. São escritas através do esquema cc-pVXZ, onde X é a qualidade zeta da função. A função de base cc-pVDZ tem o mesmo esquema que a função de base DZ def2-SVP, de forma que o número de funções é também 24, enquanto a função de base cc-pVTZ utiliza um esquema 3s2p1d para hidrogênios e 4s3p2d1f para oxigênios, totalizando 58 funções de base. As funções de base cc, assim como as funções de Pople, podem ser aumentadas com funções difusas, que são adicionadas como uma função a mais para cada momento angular. Então aug-cc-pVDZ tem esquemas 3s2p1d e 4s3p2d e aug-cc-pVTZ tem 4s3p2d e 5s4p3d2f, para hidrogênios e átomos do segundo período respectivamente.

De forma geral, é difícil se falar de recomendações de funções de base pois a função de base mais acurada e a mais eficiente podem não ser as mesmas, e dependendo da estrutura do problema estudado diferentes funções de base podem se encaixar no parâmetro de ``mais adequada''. Além disso, várias funções de base foram desenvolvidas especificamente para determinados tipos de cálculos ou sistemas, como cálculos de espectroscopia ou extrapolação ao limite da função de base completa. Ao mesmo tempo, pode-se dizer que, em cálculos utilizando computadores clássicos, é frequentemente necessário o uso de pelo menos bases do tipo TZ para se obter uma acurácia adequada~\cite{Nagy2017,Jensen2007}, sendo que em alguns sistemas funções de bases maiores são necessárias.

\subsection{Segunda quantização para sistemas de $N_e$ elétrons \label{segunda_quantizacao}}

 A segunda quantização é um formalismo bastante útil para o estudo de sistemas de muitas partículas quânticas, sendo o mais utilizado em trabalhos que implementam o VQE~\cite{tilly2022variational}. Neste formalismo, usa-se o espaço de Fock~\cite{piza2003mecanica,szabo1996modern} que é um espaço de Hilbert onde as propriedades de anti-simetria são passadas para operadores de criação e aniquilação das funções de base de 1 elétron, e os estados são objetos abstratos que representam quais funções de onda de $1$ elétron estão ocupadas, sem fazer indicação a nenhum elétron específico pois são todos idênticos. Assim, o determinante da equação \eqref{SlaterDeterminant} é representado por
\begin{equation} 
\small \ket{\psi_{b_{1}},\cdots,\psi_{b_{N_e}}}
    \to
    \frac{1}{\sqrt{N_e!}}
    \begin{vmatrix}\psi_{b_{1}}(\sigma_{1}) &  \cdots & \psi_{b_{N_e}}(\sigma_{1})\\
    \psi_{b_{1}}(\sigma_{2}) &  \cdots & \psi_{b_{N_e}}(\sigma_{2})\\
    \vdots & \ddots & \vdots\\
    \psi_{b_{1}}(\sigma_{N_e}) &  \cdots & \psi_{b_{N_e}}(\sigma_{N_e})
    \end{vmatrix},
\end{equation}
com ${\psi_j}$ sendo uma base de funções de $1$ elétron e $b_j\in\{0,1,2,\dots\}$.
Para cada função de $1$ elétron pertencente à base associamos dois operadores $\hat{a}_j$ e $\hat{a}_j^\dagger$ com as propriedades:
\begin{equation}
    \{\hat{a}_j,\hat{a}_{j'}^{\dagger}\}=\delta_{j,j'} 
    \label{eq:antComut1}
\end{equation}
e
\begin{equation}
    \{\hat{a}_j,\hat{a}_{j'}\}= 0,
    \label{eq:antComut2}
\end{equation}
com $\{\cdot , \cdot\}$ sendo o anti-comutador definido para qualquer par $\hat{A}$ e $\hat{B}$ de operadores como
\begin{equation*}
    \{\hat{A},\hat{B}\}=\hat{A}\hat{B} + \hat{B}\hat{A}.
\end{equation*}

Os operadores $\hat{a}_j^{\dagger}$ e $\hat{a}_j$ são responsáveis por, respectivamente, ``criar e aniquilar'' elétrons nos estados aos quais são associados. Assim, se temos um sistema no estado $\ket{\psi_{b_{1}},\cdots,\psi_{b_{N_e}}}$ e o $\psi_j$ está desocupado (não há elétron nele), então:
\begin{equation}
    \hat{a}_j^\dagger\ket{\psi_{b_{1}},\cdots,\psi_{b_{N_e}}} = \ket{\psi_j,\psi_{b_{1}},\cdots,\psi_{b_{N_e}}},
    \label{eq:a_daggerEffect}
\end{equation}
ao mesmo tempo, se $\psi_j$ estiver ocupado (há elétron nele), então o operador $a_j$ atua da seguinte forma:
\begin{equation}
    \hat{a}_j\ket{\psi_j,\psi_{b_{1}},\cdots,\psi_{b_{N_e}}}= \ket{\psi_{b_{1}},\cdots,\psi_{b_{N_e}}} ,
    \label{eq:aEffect}
\end{equation}
caso o estado esteja ocupado (respectivamente desocupado), o operador $\hat{a}^\dagger$ (respectivamente $\hat{a}$) gera o vetor nulo.

Reparemos que as funções criadas ou aniquiladas nas equações \eqref{eq:a_daggerEffect} e \eqref{eq:aEffect} estão sempre à esquerda no vetor de estado (o \textit{ket}), indicando que a função nova deve ser inserida à ou retirada da esquerda do determinante representado pelo vetor de estado, por exemplo, 
\begin{equation*}
    \begin{split}
    \hat{a}^{\dagger}_{j}\ket{\psi_{b_2},\psi_{b_5}} =& \ket{\psi_j\psi_{b_2},\psi_{b_5}} \\ \to &
    \frac{1}{\sqrt{3}}\begin{vmatrix}\psi_{j}(\sigma_{1}) & \psi_{b_{2}}(\sigma_{1}) & \psi_{b_{5}}(\sigma_{1})\\
    \psi_{j}(\sigma_{2}) & \psi_{b_{2}}(\sigma_{2}) & \psi_{b_{5}}(\sigma_{2})\\
    \psi_{j}(\sigma_{3}) & \psi_{b_{2}}(\sigma_{3}) & \psi_{b_{5}}(\sigma_{3})
    \end{vmatrix},
    \end{split}
\end{equation*}
que é distinto, por exemplo, de
\[
\frac{1}{\sqrt{3}}\begin{vmatrix}\psi_{b_2}(\sigma_{1}) & \psi_{j}(\sigma_{1}) & \psi_{b_{5}}(\sigma_{1})\\
\psi_{b_2}(\sigma_{2}) & \psi_{j}(\sigma_{2}) & \psi_{b_{5}}(\sigma_{2})\\
\psi_{b_2}(\sigma_{3}) & \psi_{j}(\sigma_{3}) & \psi_{b_{5}}(\sigma_{3})
\end{vmatrix}.
\]

A posição no determinante onde a função de onda aparece é consequência da relação de anti-comutação dos operadores e uma dedução rigorosa é apresentada em ~\cite{piza2003mecanica}. Para os propósitos desse artigo essa informação é relevante porque na representação desses estados no espaço de qubits, que será apresentada posteriormente, não existe nenhuma menção à ordem dos determinantes. Uma forma de lidar com isso é trabalharmos sempre com estados que estão \emph{ordenados} de acordo com o índice, por exemplo, um estado que contém as funções $\psi_1,\psi_3$ e $\psi_5$ sempre será representado como  $\ket{\psi_1\psi_3\psi_5}$ e nunca como $\ket{\psi_3\psi_1\psi_5}$ ou $\ket{\psi_1\psi_5\psi_3}$. 

Como a troca de posição de duas funções dentro do ket é equivalente a permutar duas colunas do determinante, a diferença entre dois estados que possuem duas funções trocadas é apenas um fator de sinal negativo. Por exemplo, $\ket{\psi_1\psi_3\psi_5} = -\ket{\psi_3\psi_1\psi_5}$.
Essa propriedade permite obter um estado ordenado a partir de um estado com uma única função fora de ordem mais à esquerda, realizando trocas sucessivas com as funções imediatamente à sua direita e multiplicando o estado por \((-1)\) a cada troca realizada. Exemplificando novamente, para relacionar o estado $\ket{\psi_1\psi_3\psi_4\psi_5}$ a partir do estado $\ket{\psi_4\psi_1\psi_3\psi_5}$ fazemos:
\[
\ket{\psi_4\psi_1\psi_3\psi_5}=-\ket{\psi_1\psi_4\psi_3\psi_5}=\ket{\psi_1\psi_3\psi_4\psi_5}
\]
o que pode ser apresentado de forma mais geral como:
\[
 \ket{\psi_j,\psi_{b_{1}},\cdots,\psi_{b_{N_e}}} =  (-1)^{s_j}\ket{\psi_{b_{1}},\cdots,\psi_j,\cdots\psi_{b_{N_e}}},
\]
onde $s_j$ é a quantidade de funções com índice menor que $j$ presentes no estado e o estado do lado direito da equação está ordenado. 

Com isso, a atuação dos operadores $\hat{a}_j^{\dagger}$ e $\hat{a}_j$ pode ser descrita usando sempre estados ordenados como:
\begin{equation}
    \hat{a}_j^\dagger\ket{\psi_{b_{1}},\cdots,\psi_{b_{N_e}}} = (-1)^{s_j}\ket{\psi_{b_{1}},\cdots\psi_j,\cdots,\psi_{b_{N_e}}},
    \label{eq:a_daggerEffect_v2}
\end{equation}
\begin{equation}
    \hat{a}_j\ket{\psi_{b_{1}},\cdots,\psi_j,\cdots\psi_{b_{N_e}}}= (-1)^{s_j}\ket{\psi_{b_{1}},\cdots,\psi_{b_{N_e}}}.
    \label{eq:aEffect_v2}
\end{equation}

\subsubsection{Observáveis na segunda quantização}
\label{sec:ObsSecQuant}

Tendo os estados no espaço de Fock, é necessário construir os operadores que representam os observáveis, de forma que tragam os mesmos resultados físicos dos operadores no espaço de funções. Para obter essa equivalência é necessário que dada uma base qualquer no espaço de funções de $N_e$ elétrons, construídos a partir de uma base de funções de $1$ elétron $(\psi_{i}(\sigma))$, e os vetores equivalentes no espaço de Fock, os elementos de matriz do operador $O_{\text{1st}}$ no espaço de funções sejam iguais aos elementos de matriz do operador que o represente no espaço de Fock, $O_{\text{2nd}}$  (aqui, utilizamos os sob escritos 1st e 2nd para diferenciar os operadores na primeira e na segunda quantização), ou seja,
\end{multicols}
\begin{equation}
    \bra{\Psi(\sigma_1,\sigma_2,\cdots,\sigma_{N_e})}O_{\text{1st}}\ket{\Psi'(\sigma_1,\sigma_2,\cdots,\sigma_{N_e})} =\bra{\psi_{b_{1}},\psi_{b_{2}},\cdots,\psi_{b_{N_e}}}O_{\text{2nd}}\ket{\psi_{b'_{1}},\psi_{b'_{2}},\cdots,\psi_{b'_{N_e}}},
    \label{eq:1st=2nd}
\end{equation}
\begin{multicols}{2}
\noindent onde $\Psi(\sigma_1,\sigma_2,\cdots,\sigma_{N_e})$ é o determinante formado pelas funções $\psi_{b_1},\psi_{b_2},\dots\psi_{b_{N_e}}$ com as colunas nessa ordem, e o determinante $\Psi'(\sigma_1,\sigma_2,\cdots,\sigma_N)$ é definido de forma análoga.

Os operadores no espaço de Fock que satisfazem a igualdade \eqref{eq:1st=2nd} possuem formas distintas dependendo da quantidade de elétrons sobre os quais os seus correspondentes operadores no espaço de funções atuam. Por exemplo, o operador energia cinética dos elétrons e o operador de interação elétron-núcleo, para o caso que consideramos apenas o Hamiltoniano eletrônico, são operadores que atuam apenas nas coordenadas de um único elétron por vez, chamado de \emph{operador de 1 corpo}. Por outro lado, os termos de interação coulombiana entre os elétrons atuam nas coordenadas de dois elétrons por vez, sendo \emph{operadores de 2 corpos}. Dessa maneira, é possível demonstrar que, na segunda quantização, os operadores de 1 corpo atuando em funções de 1 elétron (demarcados pelo super índice 1) que satisfazem a equação \eqref{eq:1st=2nd} são
\begin{equation}
    O^{(1)}_{\text{2nd}} = \sum_{p,q} \bra{\psi_p(\sigma)}O^{(1)}_{\text{1st}}\ket{\psi_{q}(\sigma)}\hat{a}_p^\dagger \hat{a}_{q}
    \label{2nd-1corpo}
\end{equation}
e os operadores de 2 corpos atuando sobre funções de 2 elétrons (demarcados pelo super índice 2) são
\end{multicols}
\begin{equation}
    O^{(2)}_{\text{2nd}} = \frac{1}{2}\sum_{p,p',q,q'} \bra{\psi_p(\sigma)\psi_{p'}(\sigma')}O^{(2)}_{\text{1st}}\ket{\psi_{q}(\sigma)\psi_{q'}(\sigma')}\hat{a}_p^\dagger \hat{a}_{p'}^\dagger \hat{a}_{q'} \hat{a}_{q},
    \label{2nd-2corpo}
\end{equation}
\begin{multicols}{2}
\noindent onde as somas são sobre todas as funções de base. A generalização para operadores de mais corpos é direta, porém, não é necessária para os nossos problemas dado que o Hamiltoniano tratado só possui termos de 1 ou 2 corpos. Dessa forma, toda informação sobre o Hamiltoniano passa para os produtos internos das equações \eqref{2nd-1corpo} e \eqref{2nd-2corpo}. Esses produtos são integrais sobre as funções de base utilizadas. De maneira explícita:
\end{multicols}
\begin{equation}
    \bra{\psi_p(\sigma)}O^{(1)}_{\text{1st}}\ket{\psi_{q}(\sigma)} = \int d\sigma \psi^*_p(\sigma) O^{(1)}_{\text{1st}} \psi_{q}(\sigma) \equiv \bra{p}O^{(1)}\ket{q},
    \label{eq:integral1Corpo}
\end{equation}
\begin{equation}
    \bra{\psi_p(\sigma)\psi_{p'}(\sigma')}O^{(2)}_{\text{1st}}\ket{\psi_{q}(\sigma)\psi_{q'}(\sigma')} =
    \int \int d\sigma d\sigma ' \psi^*_p(\sigma) \psi^*_{p'}(\sigma') O^{(2)}_{\text{1st}} \psi_{q}(\sigma)\psi_{q'}(\sigma') 
    \equiv \bracket{pp'}{qq'}.
    \label{eq:integral2Corpo}
\end{equation}
\begin{multicols}{2}
A notação utilizada para essas integrais é bastante comum em textos de química quântica.

Uma vez definida a base de trabalho, podemos representar o Hamiltoniano da equação \eqref{eq:n1p8} a partir dos resultados dessas integrais. Para isso, as integrais \eqref{eq:integral1Corpo} e \eqref{eq:integral2Corpo} devem ser aplicadas considerando essa base e os operadores 
\[
-\frac{1}{2} \nabla_i^2 + \sum_\alpha \frac{Z_\alpha}{\left | \mathbf{\hat{r}}_i - \mathbf{R}_\alpha\right |}
\]
e 
\[
 \frac{1}{\left | \mathbf{\hat{r}}_j-\mathbf{\hat{r}}_j \right |}
\]
de forma que o Hamiltoniano seja dado pela expressão
\begin{equation}
    \hat{H} = \sum_{p,q} h_{q}^{p} \hat{a}_p^\dagger \hat{a}_q +\frac{1}{2} \sum_{p,p',q,q'} h_{qq'}^{pp'} \hat{a}_p^\dagger \hat{a}_{p'}^\dagger \hat{a}_{q'} \hat{a}_q + h_0,
    \label{eq:n1p13}
\end{equation}
onde 
\end{multicols}
\begin{equation}
    h_{q}^{p} =\bra{\psi_p}\left [ -\frac{1}{2} \nabla_i^2 + \sum_\alpha \frac{Z_\alpha}{\left | \mathbf{\hat{r}}_i -\mathbf{R}_\alpha\right |} \right ]\ket{\psi_q}= \int_{-\infty }^{\infty } \psi_p^* \left ( \mathbf{r}_i \right ) \left [ -\frac{1}{2} \nabla_i^2 + \sum_\alpha \frac{Z_\alpha}{\left | \mathbf{r}_i -\mathbf{R}_\alpha\right |} \right ] \psi_q\left ( \mathbf{r}_i \right ) \mathrm{d}\mathbf{r}_i,
    \label{eq:n1p10}
\end{equation}
\begin{equation}
     h_{qq'}^{pp'}  =\bra{\psi_p\psi_{p'}}\frac{1}{\left | \mathbf{\hat{r}}_j-\mathbf{\hat{r}}_j \right |}\ket{\psi_q\psi_{q'}}= \int_{-\infty }^{\infty } \int_{-\infty }^{\infty } \psi_p^* \left ( \mathbf{r}_i \right ) \psi_{p'}^* \left ( \mathbf{r}_j \right ) \frac{1}{\left | \mathbf{r}_j-\mathbf{r}_j \right |} \psi_q\left ( \mathbf{r}_i \right ) \psi_{q'}\left ( \mathbf{r}_j \right )\mathrm{d}\mathbf{r}_i \mathrm{d}\mathbf{r}_j 
    \label{eq:n1p12}
\end{equation}
\begin{multicols}{2}
\noindent e $h_0$ é a constante de energia potencial de repulsão núcleo-núcleo, dentro da aproximação de Born-Oppenheimer.

A segunda quantização é um bom método para depois codificarmos nosso sistema no espaço de qubits, pois precisamos apenas representar quais estados de 1 elétron estão ocupados e sermos capazes de codificar os operadores de criação e aniquilação em operadores de qubits.

\subsection{Resolução do problema por Coupled Cluster}
\label{sec:CoupledClusterTheory}

Existem diversos métodos de se obter a energia de correlação, no entanto o custo computacional destes pode ser proibitivo dependendo do sistema estudado. Nesse sentido, o método \textit{Coupled Cluster} (CC) é um dos mais bem sucedidos~\cite{Crawford2000} juntando tanto a precisão quanto o custo relativamente mais baixo~\cite{Bartlett2011}. Além disso, existem outras vantagens como a consistência em relação ao tamanho do sistema, onde a energia calculada se mantém consistente mesmo com a separação e redução da interação entre os subsistemas. Um exemplo prático é que a energia de duas moléculas calculadas separadamente é igual à energia calculada com as duas moléculas a uma grande distância, mas no mesmo sistema~\cite{Bartlett2007,Lyakh2011}. Tal propriedade não ocorre em outros métodos de obtenção da correlação.

Devido às aproximações das funções de base de $1$ elétron e da forma que é construída, normalmente a matriz de HF é formada por orbitais ocupados e virtuais. Podemos escrever a energia de correlação como a soma das contribuições de cada um dos pares de spin-orbitais ($\mathcal{E}_{\text{corr}}=\sum_{p>q}{e_{pq}}$), o que nos sugere que podemos utilizar a função \textit{cluster} de duas partículas para correlacionar dois elétrons num par de orbitais ocupados:

\begin{equation}\label{eq:funcao_cluster}
    f_{ab}(\sigma_1, \sigma_2)=\sum_{r>s}t^{rs}_{ab}\psi_a(\sigma_1)\psi_b(\sigma_2),
\end{equation}

\noindent em que $a, b, ...$ representam orbitais ocupados, $r, s, ...$ representam orbitais virtuais, $\sigma_1$ ($\sigma_2$ ) é o vetor das coordenadas e spin do elétron 1 (2), $t^{rs}_{ab}$ são os coeficientes de cluster. Inserindo a equação \eqref{eq:funcao_cluster} na função de onda obtida através do determinante de Slater:
\begin{equation}
    \ket{\Psi} = \ket{[\psi_a(\sigma_1)\psi_b(\sigma_2) + f_{ab}(\sigma_1, \sigma_2)] \psi_c(\sigma_3)\psi_d(\sigma_4)},
\end{equation}
e rearranjando a equação, temos que:
\begin{equation}\label{eq:psi+funcaocluster}
    \ket{\Psi} = \ket{\Psi_0} + \sum_{r>s}t^{rs}_{ab}\ket{\psi_r(\sigma_1)\psi_s(\sigma_2)\psi_c(\sigma_3)\psi_d(\sigma_4)}.
\end{equation}

\noindent Pode-se perceber que essa nova função de onda, que é uma melhor aproximação à função de onda exata, representa trocar os orbitais $\psi_a$ e $\psi_b$, que são orbitais ocupados, pelos orbitais $\psi_r$ e $\psi_s$, que são orbitais virtuais. É importante perceber que, apesar de estar tratando de apenas uma única mudança entre dois elétrons, como o determinante de Slater gera uma combinação linear dos orbitais presentes no sistema, tem-se que  na equação \eqref{eq:psi+funcaocluster} a correlação é gerada para todos os elétrons presentes nos orbitais $\psi_a$ e $\psi_b$. Ao mesmo tempo, já que não há nada de especial nos orbitais que foram selecionados acima, uma melhor representação do sistema seria gerar a correlação entre todos os pares de orbitais presentes.

A troca de orbitais preenchidos por orbitais virtuais corresponde a excitar o elétron a um orbital de maior energia e descreve exatamente o que foi discutido na Seção sobre segunda quantização. Seguindo o que foi discutido sobre o assunto, o operador de cluster pode ser definido como:
\begin{equation}\label{eq:clusteroperator}
    \hat{t}_{ab}=\sum_{r>s}t^{rs}_{ab}\hat{a}^{\dagger}_r \hat{a}^{\dagger}_s \hat{a}_b \hat{a}_a.
\end{equation}

Assim como foi definido o operador de cluster para uma troca dupla, o operador pode corresponder a uma troca arbitrária de um até $N$ orbitais, sendo $N$ o número total de orbitais no sistema. Quando se inclui todos os operadores na função de onda, obtém-se o resultado exato para esta no espaço definido pelas funções de base $\psi$.

Importante notar que tanto o determinante de Slater quanto a forma da equação \eqref{eq:psi+funcaocluster} tem como resultado que os coeficientes do cluster apresentam uma antissimetria em relação à permutação de qualquer um dos orbitais, ou seja, $t^{rs}_{ab}=-t^{rs}_{ba}=-t^{sr}_{ab}$.

A partir dos operadores de um e dois orbitais, pode-se escrever a função de onda para um sistema de quatro elétrons como:
\begin{equation}\label{eq:psi_cc_expandida}
\begin{split}
    \ket{\Psi} = \biggl ( 1 + \sum_a \hat{t}_a + \frac{1}{2} \sum_{ab} \hat{t}_a \hat{t}_b + \frac{1}{6} \sum_{abc} \hat{t}_a \hat{t}_b \hat{t}_c \\+  
    \frac{1}{2} \sum_{ab} \hat{t}_{ab} + \frac{1}{8} \sum_{abcd} \hat{t}_{ab} \hat{t}_{cd} + \frac{1}{24} \sum_{abcd} \hat{t}_a \hat{t}_b \hat{t}_c \hat{t}_d \\+ 
    \frac{1}{2} \sum_{abc} \hat{t}_{ab} \hat{t}_c + \frac{1}{4} \sum_{abcd} \hat{t}_{ab} \hat{t}_c \hat{t}_d \biggr ) \ket{\Psi_0}.
\end{split}
\end{equation}

Para que se possa perceber mais claramente a forma da equação \eqref{eq:psi_cc_expandida}, pode-se simplificá-la definindo operadores de cluster de um e dois orbitais da forma:
\begin{align}
    \hat{T}_1 = \sum_a \hat{t}_a, \\
    \hat{T}_2 = \sum_{ab} \hat{t}_{ab}.
\end{align}

\noindent Substituindo esses operadores na equação \eqref{eq:psi_cc_expandida}, esta se torna:
\begin{equation}\label{eq:psi_cc_contraida}
\begin{split}
    \ket{\Psi} = \biggl ( 1 + \hat{T}_1 + \frac{1}{2!} \hat{T}^2_1 + \frac{1}{3!} \hat{T}^3_1 + \hat{T}_2 + \frac{1}{2!} \hat{T}^2_2\\ + 
    \frac{1}{4!} \hat{T}^4_1 + \hat{T}_2 \hat{T}_1 +  \frac{1}{2!} \hat{T}_2 \hat{T}^2_1\biggr ) \ket{\Psi_0}.
\end{split}
\end{equation}

A expressão da equação \eqref{eq:psi_cc_contraida} tem o mesmo formato da expansão de séries de potências de uma função exponencial, de forma que pode-se escrever essa como:
\begin{equation}
    \ket{\Psi} = e^{\hat{T_1}+\hat{T_2}}\Psi_0 = e^{\hat{T}}\ket{\Psi_0},
    \label{eq:CoupleCluster}
\end{equation}
\noindent em que $\hat{T}_n = \sum^n_i \hat{T}_i$. Como o exemplo didático utilizado apresenta apenas 4 elétrons, os termos de ordem mais alta, como $\hat{T}_3$ não aparecem. No entanto, eles podem fazer parte do sistema e podem ser importantes para a precisão do resultado. Por isso, quando se tem um grande número de elétrons, e devido ao aumento significativo do custo computacional quando se adiciona operadores de cluster de mais orbitais, pode-se truncar o operador de cluster num determinado número de trocas. Os níveis presentes na expansão geram as siglas usuais para os métodos de CC, como CCSD (do inglês \textit{singles and doubles}) que apresenta os operadores $\hat{T}_1$ e $\hat{T}_2$, CCSDT (do inglês \textit{singles, doubles and triples}) que além de $\hat{T}_1$ e $\hat{T}_2$ também inclui o operador $\hat{T}_3$, e daí em diante.

\section{Computação quântica e o algoritmo VQE} 
\label{sec:compQeAlg}
Nesta Seção, abordaremos os fundamentos da computação clássica e quântica. Em seguida, discutiremos brevemente como calcular os recursos necessários para a execução de um algoritmo, contextualizando essa análise no âmbito da otimização de algoritmos.
\subsection{Fundamentos de Computação Clássica}\label{sec:cc}

A \emph{computação clássica} é a base dos sistemas computacionais modernos e opera manipulando unidades fundamentais de informação conhecidas como \emph{bits}. Um bit pode assumir apenas dois estados, representados por 0 e 1. Esses dois valores binários são a chave para todas as operações realizadas pelos computadores. A manipulação desses bits é feita por meio de \emph{portas lógicas}, que são circuitos eletrônicos projetados para realizar operações lógicas elementares, como \emph{E (AND)}, \emph{OU (OR)} e \emph{NÃO (NOT)}. Essas operações seguem as regras da \emph{álgebra booleana}, criada por George Boole no século XIX, que estabelece as bases matemáticas para o tratamento binário de informações ~\cite{tanenbaum2015}.

As portas lógicas são combinadas em circuitos mais complexos que podem realizar tarefas mais sofisticadas, como adição e multiplicação de números binários, controle de fluxos de informação e tomada de decisões lógicas dentro dos processadores. A álgebra booleana simplifica a maneira como essas operações são realizadas, permitindo que computadores manipulem dados de forma eficiente e escalável. Por exemplo, somadores binários, que são blocos de construção fundamentais para a aritmética dos computadores, são formados a partir da combinação de múltiplas portas lógicas ~\cite{patterson2017}.

\subsubsection{Processamento e Limitações da Computação Clássica}

Os computadores clássicos realizam suas operações de maneira \emph{sequencial} ou \emph{paralela}, sempre respeitando as regras de processamento de informações clássicas. No processamento \emph{sequencial}, as instruções são executadas uma após a outra, o que simplifica o controle do fluxo de dados, mas pode limitar o desempenho quando há múltiplas tarefas que podem ser realizadas de forma simultanêa. Já o processamento \emph{paralelo} envolve a execução de várias operações ao mesmo tempo, distribuindo a carga de trabalho entre vários núcleos de processamento, o que melhora significativamente o desempenho em tarefas intensivas, como simulações científicas e renderização gráfica ~\cite{stallings2016}.

Entretanto, a eficiência dos computadores clássicos é limitada por fatores físicos. A \emph{Lei de Moore}, observada por Gordon Moore em 1965, previu que o número de transistores em um chip dobraria aproximadamente a cada dois anos, aumentando o desempenho dos processadores~\cite{moore1965}. Por décadas, essa previsão se manteve verdadeira, impulsionando o crescimento exponencial do poder computacional. Contudo, à medida que os transistores se tornam menores, o processo de miniaturização enfrenta limitações físicas significativas, como o aumento da \emph{dissipação de calor} e problemas relacionados à interferência quântica em escalas nanométricas~\cite{tanenbaum2015}.

A \emph{memória} também desempenha um papel crucial no desempenho da computação clássica. A capacidade de armazenar grandes quantidades de dados e acessá-los rapidamente é essencial para a eficiência de qualquer sistema computacional. A \emph{hierarquia de memória}---composta por registradores, memória cache, RAM (memória de acesso aleatório) e armazenamento secundário (como discos rígidos ou SSDs)---foi desenvolvida para otimizar a relação entre velocidade e custo, permitindo que os sistemas operem de maneira mais eficiente~\cite{patterson2017}.

À medida que problemas computacionais se tornam mais complexos, como aqueles encontrados em áreas como inteligência artificial, criptografia e simulação de sistemas físicos, a demanda por \emph{poder computacional} continua a crescer. Esse aumento de demanda levanta questões sobre o futuro da computação clássica, uma vez que os avanços na miniaturização de transistores e na eficiência energética dos sistemas enfrentam limitações físicas e tecnológicas importantes~\cite{stallings2016}. Tais restrições impõem desafios à continuidade do aumento de desempenho esperado pelos usuários. No contexto dessas limitações, surgem novas abordagens computacionais, como a computação quântica, que tem sido apontada como uma alternativa promissora para expandir as fronteiras do processamento e armazenamento de dados.

\subsection{O Qubit e a Superposição}\label{sec:qubits}

O qubit, ou bit quântico, é a unidade fundamental de informação na computação quântica. Diferentemente do bit clássico, que só pode assumir os valores de \(0\) ou \(1\), o qubit pode existir em uma \emph{superposição quântica} de ambos os estados simultaneamente. Em termos matemáticos, o estado de um qubit é representado por um vetor de estado em um espaço vetorial complexo bidimensional, \textit{i.e.}, um espaço de Hilbert, expresso da seguinte forma:
\[
|\psi\rangle = \alpha |0\rangle + \beta |1\rangle,
\]
onde $\alpha$ e $\beta$ são números complexos que representam as \emph{amplitudes de probabilidade} associadas aos estados $\ket{0}$ e $\ket{1}$, respectivamente. Essas amplitudes, quando elevadas ao quadrado em módulo, $|\alpha|^2$ e $|\beta|^2$, correspondem às \emph{probabilidades} de encontrar o sistema nos estados $\ket{0}$ e $\ket{1}$, garantindo que a soma das probabilidades satisfaça a condição de normalização $|\alpha|^2 + |\beta|^2 = 1$
~\cite{nielsen2010}. 

Os estados base \( |0\rangle \) e \( |1\rangle \) são, na verdade, vetores ortonormais em um espaço vetorial de dimensão 2, que podem ser explicitamente representados como:
\[
|0\rangle = \begin{pmatrix} 1 \\ 0 \end{pmatrix}, \quad |1\rangle = \begin{pmatrix} 0 \\ 1 \end{pmatrix}.
\]
Esses vetores formam uma base completa para o espaço de estados de um qubit, permitindo que qualquer estado quântico seja descrito como uma combinação linear desses vetores de base. A capacidade de um qubit de existir em uma superposição desses estados é o que o diferencia fundamentalmente do bit clássico e é um dos pilares da \emph{computação quântica}.

Além disso, essa propriedade de superposição permite que o qubit processe múltiplas possibilidades ao mesmo tempo. Por exemplo, um sistema de \( n \) qubits pode estar em uma superposição de \( 2^n \) estados diferentes, explorando todas essas possibilidades simultaneamente~\cite{ladd2010}. Isso oferece uma vantagem teórica significativa em comparação com os sistemas clássicos, especialmente em problemas que envolvem busca e otimização em grandes espaços de soluções.

\subsubsection{Medição de Qubits e a Esfera de Bloch}\label{sec:medqubits}

Uma das características mais intrigantes dos qubits é o processo de medição. Diferentemente da computação clássica, onde o estado de um bit pode ser lido de maneira determinística, a medição de um qubit é probabilística. Quando medimos um qubit que está em um estado de superposição, ele colapsa para um dos estados base \( |0\rangle \) ou \( |1\rangle \), com probabilidades dadas por \( |\alpha|^2 \) e \( |\beta|^2 \), respectivamente~\cite{preskill2018}. Por exemplo, se o qubit estiver no estado de superposição \( |\psi\rangle = \frac{1}{\sqrt{2}} |0\rangle + \frac{1}{\sqrt{2}} |1\rangle \), a medição deste qubit resultará no estado \( |0\rangle \) com probabilidade de 50\% e no estado \( |1\rangle \) também com 50\% de chance. Após a medição, o estado de superposição original desaparece, e o qubit assume o estado correspondente ao resultado da medição~\cite{shor1997}.

Esse comportamento probabilístico é uma manifestação direta dos princípios da \emph{mecânica quântica}, onde a medição pode alterar o estado de um sistema, influenciando os resultados das medidas subsequentes. Tal comportamento é crucial para diversas aplicações quânticas, incluindo a \emph{criptografia quântica}, onde a incerteza inerente pode ser explorada para gerar números verdadeiramente aleatórios~\cite{bennett1984}.

Além disso, em sistemas com múltiplos qubits (que serão explicados na Seção \ref{sec:2qubits}), o ato de medir um qubit pode impactar os estados de outros qubits, dependendo de como esses qubits estão \emph{emaranhados}, entretanto só é possível verificar essa influência ao confrontar os resultados das medidas nos qubits correlacionados. Esse emaranhamento quântico adiciona uma camada adicional de complexidade e potencial à computação quântica, permitindo a execução de algoritmos quânticos com eficiências inatingíveis por sistemas clássicos~\cite{james2001measurement}.

\subsubsection{A Esfera de Bloch}
\label{sec:EsferaBloch}

\begin{figure}[H]
    \centering
    \vspace{-0.5cm}
    \hspace{-0.2cm}\includegraphics[width=0.5\columnwidth]{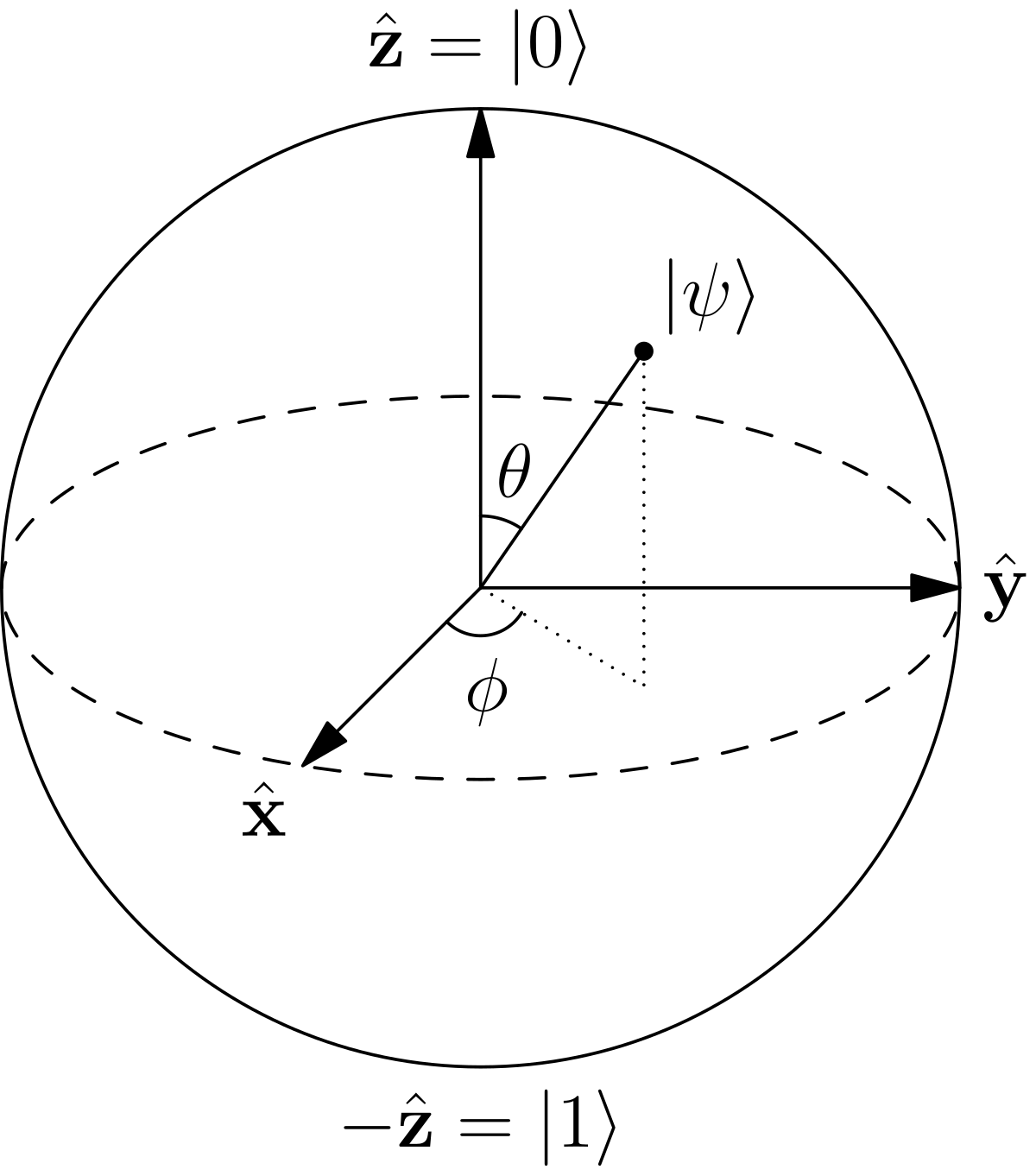} 
    \vspace{-0.2cm}
    \caption{Representação da esfera de Bloch, utilizada para descrever o estado de um qubit em um espaço tridimensional. Os eixos \(X\), \(Y\) e \(Z\) indicam as direções principais associadas aos operadores de Pauli, enquanto o vetor aponta para a posição correspondente ao estado do qubit no espaço de Hilbert projetado. Nessa representação, um estado normalizado arbitrário $\ket{\psi}$ pode então ser completamente descrito pelos ângulos polar ($\theta$) e azimutal ($\phi$).}
    \label{fig:esfera-bloch}
\end{figure}

A representação gráfica mais comum para um qubit é a \emph{esfera de Bloch}, uma esfera tridimensional que visualiza qualquer estado puro de um qubit como um ponto em sua superfície. Neste modelo, os estados \( |0\rangle \) e \( |1\rangle \) correspondem a pontos nos polos norte e sul da esfera, respectivamente. Qualquer outro ponto na superfície da esfera representa uma superposição de \( |0\rangle \) e \( |1\rangle \).

O estado geral de um qubit pode ser escrito como:
\[
|\psi\rangle = \cos\left(\frac{\theta}{2}\right)|0\rangle + e^{i\phi}\sin\left(\frac{\theta}{2}\right)|1\rangle,
\]
onde \( \theta \) e \( \phi \) são ângulos esféricos que definem a posição do vetor de estado na esfera de Bloch. A amplitude \( \cos\left(\frac{\theta}{2}\right) \) corresponde à projeção do estado do qubit no estado da base computacional \( |0\rangle \), enquanto \( e^{i\phi}\sin\left(\frac{\theta}{2}\right) \) está associado ao estado \( |1\rangle \). Aqui, \( \theta \) controla a magnitude da superposição entre \( |0\rangle \) e \( |1\rangle \), enquanto \( \phi \) introduz uma fase relativa entre os dois estados~\cite{nielsen2010}.

\subsubsection{Medição na Base Computacional e a Esfera de Bloch}

A medição na base computacional corresponde a medir o estado do qubit nos estados base \( |0\rangle \) e \( |1\rangle \). Esses estados são representados na esfera de Bloch como os polos norte (\( |0\rangle \)) e sul (\( |1\rangle \)). 
Se o vetor estiver no equador da esfera (correspondendo ao estado de superposição máxima), as probabilidades de medir \( |0\rangle \) ou \( |1\rangle \) serão iguais (50\% cada). Esta correspondência se reflete nas amplitudes \( \alpha \) e \( \beta \), uma vez que \( |\alpha|^2 = \cos^2\left(\frac{\theta}{2}\right) \) e \( |\beta|^2 = \sin^2\left(\frac{\theta}{2}\right) \).

Além disso, a fase \( \phi \), que está associada ao ângulo ao redor do eixo z da esfera de Bloch, não afeta diretamente a probabilidade de medir \( |0\rangle \) ou \( |1\rangle \), mas influencia as interferências quânticas quando o qubit interage com outros qubits ou passa por portas quânticas antes da medição~\cite{preskill2018}.

\subsection{2 qubits e correlações quânticas}
\label{sec:2qubits}

A Seção \ref{sec:qubits} apresentou os conceitos e formalismo básicos para o entendimento de um único qubit. A partir de agora, será discutido um sistema de dois qubits, o que permite explorar uma nova propriedade do mundo quântico chamada emaranhamento, capaz de abrir novas possibilidades para a computação. 

Um par de bits clássicos pode estar em quatro diferentes combinações: $00$, $01$, $10$ e $11$. Dessa forma, a transposição para um par de qubits é bem semelhante ao caso de um qubit, ou seja, um par de qubits pode estar em qualquer estado de superposição dentre os quatro casos, sendo representado matematicamente como
\[
\ket{\psi} = \alpha_{00}\ket{00} + \alpha_{01}\ket{01} + \alpha_{10}\ket{10} + \alpha_{11}\ket{11}.
\]
Ao realizar uma medida no par de qubits pode-se obter qualquer uma das combinações com probabilidade $\vert\alpha_{q_1q_2}\vert^2$ ($q_1,q_2\in \{0,1\}$) e, após a medida, o sistema ``colapsa'' para o respectivo estado $\ket{q_1q_2}$. Os coeficientes devem satisfazer a condição de normalização $\vert\alpha_{00}\vert^2 + \vert\alpha_{01}\vert^2 + \vert\alpha_{10}\vert^2 +\vert\alpha_{11}\vert^2 = 1 $ para representarem probabilidades.

Para discutir sobre as possibilidades que surgem em sistemas de dois qubits, consideremos como exemplo o estado $\ket{\Psi^+}$ dado por
\[
\ket{\Psi^+} = \frac{\ket{01}+\ket{10}}{\sqrt{2}}.
\]
Vamos então assumir que, uma vez preparados nesse estado, um dos qubits é mantido em um laboratório e o outro é enviado para outro laboratório distante do primeiro, sem que o estado seja alterado. Dessa forma, cada laboratório só tem acesso e consegue medir apenas um dos qubits. Caso uma pessoa em qualquer um dos laboratórios meça o seu qubit, as probabilidades de se obter o estado $\ket{0}$ ou o estado $\ket{1}$, devido à forma do estado $\ket{\Psi^+}$, são ambas iguais a $\frac{1}{2}$. Supondo que o primeiro laboratório meça $\ket{0}$, então a única possibilidade para o estado do sistema todo, após a medida, é colapsar para $\ket{01}$. Porém,  dado esse estado global, o estado do segundo qubit fica automaticamente determinado, ou seja, só pode-se medi-lo no estado $\ket{1}$. Por outro lado, se no primeiro qubit tivesse sido medido o estado $\ket{1}$, então o estado do sistema após a medida seria $\ket{10}$, e qualquer medida posterior no segundo qubit resultaria em $\ket{0}$. O argumento também vale se o segundo laboratório medisse seu qubit primeiro. Essa correlação ainda não manifesta os aspectos quânticos desse sistema, uma vez que correlações desse tipo poderiam estar em sistemas clássicos, por exemplo, se tivessemos duas caixas e soubessemos que em uma há uma bola branca e na outra uma preta. Porém, nos sistemas quânticos ainda é possível fazer mudanças de base de medida, de forma que as correlações continuam presentes em diferentes bases, algo que só é possível devido aos aspectos não-clássicos da teoria.

Em outras palavras, se existe um par de qubits no estado $\ket{\Psi^+}$, o resultado da medida em apenas \emph{um} dos qubits determina o resultado que se obtém em uma medida posterior do outro qubit, o que demonstra a existência de uma \emph{correlação} entre os resultados da medida de um dos qubits com o resultado da medida no outro, lembrando que para verificar correlações precisamos sempre observar as informações de ambos os qubits. Um par de qubits nesse estado é dito emaranhado. 

Estados emaranhados entre dois (ou mais qubits/sistemas quânticos) apresentam diversas propriedades interessantes. Em especial, chama atenção que, após emaranhados, a correlação entre os resultados das medidas se mantém mesmo que os qubits estejam extremamente distantes entre si quando as medidas forem realizadas, de modo que uma medida não pode interferir na outra e o resultado é uma consequência apenas da interação entre eles antes de serem separados, (\textit{i.e.}, as medidas não possuem relação causal entre si)~\cite{einstein1935can,nielsen2010}. Além disso, as correlações produzidas por esses estados seriam impossíveis de se obter  a partir de uma teoria clássica, como demonstrado por John Bell em 1964~\cite{bell1964einstein}. Em outras palavras: Emaranhamento é uma consequência da teoria quântica cujos resultados estão em total confronto com qualquer teoria clássica causal.

Com relação à computação, a superposição e o emaranhamento permitem desenvolver algoritmos que seriam impossíveis apenas com bits em estados $0$ e $1$ e, por isso, espera-se que a computação quântica seja capaz de trazer vantagens com relação à computação clássica, principalmente na realização de tarefas com menor gasto de tempo. Alguns dos algoritmos mais famosos são o algoritmo de Deutsch–Jozsa~\cite{deutsch1992rapid}, que pode verificar se uma função binária é constante ou balanceada, o algoritmo de busca de Grover~\cite{Grover1996}, o algoritmo de fatoração de Shor~\cite{shor1997}, o algoritmo de teletransporte quântico~\cite{bennett1993teleporting}, o algoritmo de distribuição de chaves quânticas BB84~\cite{bennett1984}, dentre outros.

Ao generalizar esses conceitos para o caso de $n$ qubits, um estado genérico é uma superposição de todas as $2^n$ possíveis combinações de $n$ bits 
\begin{equation}
    \ket{\psi}_{\text{n-qubits}} = \sum_{q_1,\cdots q_n\in\{0,1\}}\alpha_{q_1\cdots q_n} \ket{q_1\cdots q_n},
    \label{eq:state_vector_nqubits}
\end{equation}
onde também aparecer estados emaranhados, mas agora entre múltiplas partículas.

\subsection{Operações em computadores quânticos}

 Nesta Seção serão apresentadas quais operações são possíveis em qubits e como elas se relacionam com processamento de informação em computadores quânticos. De forma intuitiva, um computador aplica funções em um conjunto de valores para produzir um resultado, sendo necessário fazer o mapeamento dos valores de entrada e saída e das funções em uma forma que o computador consiga operar. Por exemplo, a  soma dos números $2$ e $3$ obtendo o número $5$ é possível mapeando $2$ e $3$ em um conjunto de $n$ bits nos quais é aplicado um circuito lógico que representa a soma, produzindo ao final um conjunto de $n$ bits em que aplicamos o mapeamento inverso para obter o número $5$ como resultado.  

O processo é semelhante na computação quântica, porém, com o mapeamento dos valores de entrada em um vetor de estado de $n$ qubits como o da equação \eqref{eq:state_vector_nqubits}, a representação da função (operação matemática) por um operador linear unitário no espaço de Hilbert do sistema, e os resultados de medidas sobre o vetor de estado final sendo relacionados com a resposta do problema abordado. Destaca-se que, como as medidas na mecânica quântica produzem resultados aleatórios, podem ser necessárias diversas medidas para se obter a resposta desejada, implicando em diversas execuções do algoritmo.

É possível demonstrar que a computação quântica realizada por operadores unitários é capaz de reproduzir qualquer resultado da computação clássica, ou seja, é possível construir computadores usando sistemas quânticos e clássicos que são capazes de resolver os mesmos tipos de problema, porém, espera-se que computadores quânticos sejam capazes de resolver os problemas de forma mais rápida e eficiente que os clássicos~\cite{nielsen2010}. 

Neste trabalho será utilizado um modelo de computação quântica denominado computação por portas lógicas~\cite{deutsch1989quantum}, análogo à computação digital implementada em computadores clássicos, conforme discutido na Seção \ref{sec:cc}, que também recebe esse nome. Neste modelo, são realizadas operações unitárias nos qubits por meio de dispositivos de controle. Um exemplo é a utilização de laser (dispositivo de controle) para a manipulação de íons (qubits) aprisionados em armadilhas eletromagnéticas. Essas operações são chamadas de portas lógicas. A partir delas, é possível construir operações mais gerais, que são necessárias em algoritmos complexos, o que pode ser feito via decomposição em sequências discretas de portas lógicas fundamentais. Como exemplo, a porta lógica $Y$ pode ser realizada por meio da aplicação seguida das portas lógicas $Z$, $X$ e $e^{i\frac{\pi}{2}}$ (O significado dessas portas será discutido em mais detalhes no decorrer do texto). Além disso, apesar das portas lógicas serem operadores, não iremos utilizar o símbolo ``\^{}'' nelas por simplicidade de notação.

É possível decompor, com precisão arbitrária, qualquer operação unitária que atua em qualquer quantidade de qubits em uma sequência de aplicações de portas de um ou dois qubits de um conjunto pequeno de portas lógicas~\cite{barenco1995elementary}. Tal conjunto é dito universal e um dispositivo físico capaz de realizar tal conjunto de operações, com um sistema de medição do estado dos qubits ao final do processo, pode executar qualquer algoritmo.

Existem diversos tipos de sistemas físicos que podem ser controlados e manipulados, de modo que se apresentam como candidatos ao hardware de um computador quântico. Os exemplos mais comuns são os qubits supercondutores~\cite{acharya2024quantum}, íons aprisionados~\cite{fernandes2022ions}, sistemas fotônicos~\cite{aghaee2025scaling}, sistemas baseados em ressonância magnética nuclear~\cite{jones2001nmr}, e átomos neutros~\cite{bloch2008quantum}. Cada um deles possui seu próprio conjunto de portas lógicas universais dado que esses conjuntos não são únicos.

Entre as portas lógicas de 1 qubit mais comuns estão a porta $X$ (ou porta NOT), que inverte o valor do qubit, a porta $Z$ que gera uma fase de $e^{i\pi}$ no estado $\ket{1}$ e a porta Hadamard $H$ que gera uma superposição quando aplicado aos estados da base computacional. A representação de seus efeitos nos qubits da base computacional, \textit{i. e.}, a base $\{\ket{0},\ket{1}\}$, é
\begin{align*}
 X\ket{0}	&=\ket{1},  	& Z\ket{0}	&=\ket{0}, 	& H\ket{0}	&=\frac{\ket{0}+\ket{1}}{\sqrt{2}},\\
 X\ket{1}	&=\ket{0}, & Z\ket{1}	&=-\ket{1}, &	H\ket{1}	&=\frac{\ket{0}-\ket{1}}{\sqrt{2}}. 
\end{align*}

Com relação às portas de 2 qubits, tem-se o exemplo da porta \textit{Controlled-Not} (C-NOT ou CX). Um dos qubits que a porta atua é chamado de qubit de controle enquanto o outro de qubit alvo e a função dela é inverter o estado do qubit alvo se o qubit de controle estiver no estado $\ket{1}$, ou seja, ela realiza o efeito da porta $X$ no qubit alvo com base no estado do qubit de controle, assim:
\begin{equation*}
    \begin{split}
        \text{C-NOT}\ket{00} = \ket{00}, \quad \text{C-NOT}\ket{01} = \ket{01}, \\
        \text{C-NOT}\ket{10} = \ket{11}, \quad \text{C-NOT}\ket{11} = \ket{10} .
    \end{split}
\end{equation*}

Em teoria, podemos construir versões controladas para qualquer porta $U$ de 1 qubit, denominada de forma geral de \textit{Controlled} $U$ ($CU$). Outra exemplo bastante comum na literatura é a \textit{Controlled} $Z$ ($CZ$).

Existem ainda as portas parametrizáveis que recebem este nome por possuírem parâmetros livres cujo efeito dessas portas em um ou mais qubits depende do valor dos parâmetros. Entre as portas parametrizáveis mais comuns estão as operações de rotação sobre os eixos $\hat{\mathbf{x}}$, $\hat{\mathbf{y}}$ e $\hat{\mathbf{z}}$ na esfera de Bloch, todas dependentes de um parâmetro $\theta$ e definidas por:
\begin{equation}
    R_x(\theta) = e^{-i\frac{\theta}{2}X}, \quad R_y(\theta) = e^{-i\frac{\theta}{2}Y}, \quad R_z(\theta) = e^{-i\frac{\theta}{2}Z}.
    \label{eq:rotacoes_eixos}
\end{equation}

\subsection{Cálculo de recursos para um algoritmo}
\label{sec:recursos}

Para diversas aplicações práticas, como problemas que trabalham com grandes valores de entradas ou simulações de sistemas físicos de alta dimensionalidade, é essencial saber quantos recursos são necessários para executar um algoritmo, como tempo e memória. Ao fazer isso, é possível comparar os algoritmos e verificar qual é mais indicado para sua aplicação. Por exemplo, você pode ter um algoritmo que gasta menos tempo mas consome mais memória do que o seu computador possui, dessa forma é melhor optar por um algoritmo mais lento mas que seja executável na máquina. A quantidade de recursos demandada por um algoritmo também é chamada de ``Complexidade'', onde a quantidade máxima de memória necessária para rodar o algoritmo ao longo da sua execução é sua complexidade espacial, enquanto o tempo necessário é a complexidade temporal~\cite{cormen2022introduction}

O cálculo de complexidade depende do modelo do dispositivo em que um algoritmo é implementado e, usualmente, considera-se um processador que realiza instruções uma seguida da outra (sem paralelismo), com mesma duração, e com memória finita.  Outras hipóteses também precisam ser consideradas~\cite{cormen2022introduction}, porém, não são necessárias para os propósitos desta Seção. Dessa forma, a quantidade de instruções se torna uma medida da complexidade temporal que, apesar de depender do modelo do sistema, não depende da máquina específica em que será implementado. Por exemplo, somar todos os números de $1$ até $1000$ um a um demanda $999$ operações sequenciais no modelo apresentado, porém, realizar essa tarefa em um computador moderno deve ser mais rápido do que em um computador dos anos 2000. Por ser independente de máquina, a quantidade de instruções se torna uma forma mais rigorosa para definir a complexidade temporal do algoritmo. Ademais, essa complexidade é usualmente quantificada em termos do tamanho das entradas.

Explicaremos o cálculo de complexidade de um algoritmo por meio de um exemplo simplificado de busca em uma lista com $N$ itens. Imagine que uma pessoa está interessada em saber se algum produto está catalogado em uma lista salva em um computador e qual o seu preço. Cada produto da lista possui uma região da memória onde estão salvas as informações sobre ele, como seu nome e preço. O que a pessoa pode fazer é falar para o computador checar cada item da lista, comparar o nome do item com o nome do produto e se encontrar o produto de interesse, verificar o preço e informá-lo. Dessa forma, $N$ é o tamanho da entrada e será necessário, em média, realizar $N/2$ comparações e no máximo $N$ (a utilização de pior caso ou caso médio depende do tipo de análise considerada), implicando que a complexidade temporal seja proporcional a $N$. A quantidade de memória necessária seria $N\cdot \alpha + \beta$, onde $\alpha$ é a quantidade de memória para guardar um item da lista com todas as suas informações e $\beta$ é a memória necessária para guardar o nome do item que está sendo procurado.

Reparem que essa análise desconsidera uma série de fatores, como o tempo para leitura do nome que será procurado, a busca pelo preço na região da memória com informações do produtos, o tempo para o preço ser escrito na tela do computador e outras variáveis de memória auxiliares como a variável de interação do \textit{loop}. Porém, esses exemplos só iriam adicionar termos constantes no tempo de busca de forma que a informação que eles trazem não compensam pelo esforço necessário para quantificá-los e poderia, inclusive, tornar o resultado final mais difícil de interpretar. Por essas razões, a complexidade computacional normalmente é quantificada considerando apenas o fator que cresce mais rapidamente com o aumento da entrada e desconsiderando termos que o multiplicam, o que é chamado de análise assintótica~\cite{cormen2022introduction}. Então, se tivéssemos um algoritmo com complexidade temporal de $5N^2 + 30N$, diríamos que a complexidade é de $N^2$. Para o exemplo previamente apresentado, a complexidade temporal e espacial seriam apenas $N$. Os fatores constantes são descartados pois, quando consideramos problemas de tamanho real, eles normalmente são desprezíveis para efeitos de comparação, como no caso de $5N^2+50N$, em que um valor de apenas $N=100$ já tornaria o termo $5N^2$ dez vezes maior que o $50N$. A partir deste ponto iremos utilizar a notação ``big-$O$'' que torna simples a análise assintótica. Com essa notação, em vez de falarmos de forma explícita que ``a complexidade do algoritmo é superiormente limitada por $N$'', falamos apenas que ele tem complexidade de $O(N)$.

Toda essa discussão foi feita para o caso de um computador clássico com instruções sequenciais, porém, podemos fazer uma transposição simples para a Computação Quântica que trabalha com o modelo de circuitos. Para o modelo de circuitos, a memória é análoga aos qubits necessários para o algoritmo. Assim, o número de qubits é a complexidade espacial. Já para complexidade temporal precisamos considerar a quantidade de portas lógicas não sequenciais implementadas no circuito, semelhante ao modelo de computação clássica apresentado. Porém, por escolhermos especificamente quais qubits cada porta lógica atua, é possível considerar parte delas operando em paralelo, desde que atuem em qubits diferentes. Um conjunto de portas lógicas que atuam em paralelo é chamado de camada do circuito. Assim, a complexidade temporal é proporcional ao número de camadas, que por sua vez é chamada de profundidade do circuito. Além da quantidade de camadas, devido a natureza probabilística da mecânica quântica, existem algoritmos em que consegue-se a resposta correta apenas com uma dada probabilidade. Devido a isso, diversas execuções do circuito quântico devem ser realizadas para se obter a resposta certa com uma probabilidade desejada. Isso implica que a complexidade temporal também é proporcional a esse número de execuções.

\begin{figure}[H]
    \centering
    \vspace{-0.2cm}
    \hspace{-0.2cm}\includegraphics[width=0.7\columnwidth]{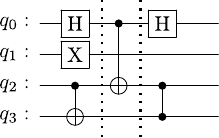} 
    \vspace{-0.2cm}
    \caption{Circuito simples para exemplificar o cálculo da profundidade de um algoritmo. Nele podemos ver três camadas de portas lógicas, com a primeira tendo as portas Hadamard ($H$), $X$ e C-NOT, a segunda apenas a porta C-NOT e a terceira com uma porta $H$ e uma porta CZ. As linhas tracejadas são ilustrativas, apenas para facilitar a visualização das camadas.}
    \label{fig:profundidade}
\end{figure}

Vamos tomar o circuito da Figura \ref{fig:profundidade} para exemplificar. Ele possui três portas de um qubit e três portas de dois qubits. Apesar disso, pode-se ver que nele existem três camadas de circuito, separadas na Figura \ref{fig:profundidade} por linhas tracejadas, em que as respectivas portas em cada uma delas podem ser realizadas paralelamente, ou seja, sua profundidade é igual à $3$. A saber, na primeira camada são aplicadas as portas $X$, Hadamard e C-NOT, na segunda apenas a porta C-NOT, e na última uma porta Hadamard e uma CZ. Agora, vamos supor que esse circuito seja útil para resolver algum problema em particular e que sabemos da teoria que ao rodá-lo quatro vezes conseguimos a solução com uma probabilidade suficientemente alta para pararmos o algoritmo. Então a complexidade temporal será proporcional ao produto da profundidade do circuito pela quantidade de vezes que precisamos rodar o circuito, totalizando uma complexidade temporal de $4\cdot 3 = 12$. 

Se o circuito para resolver o problema crescesse linearmente com o tamanho da entrada, mas a quantidade de medidas se mantivesse constante, podemos dizer que o algoritmo quântico possui complexidade $O(N)$, que seria a mesma se o número de medidas necessárias crescesse linearmente com a entrada enquanto a profundidade se mantivesse constante. Porém, se tanto a quantidade de medidas quanto a profundidade do circuito crescessem linearmente, tal algoritmo teria complexidade total $O(N^2)$.

O algoritmo que iremos apresentar se enquadra como um algoritmo de otimização. Isso significa que conseguimos a solução para o problema a partir de um ponto ótimo de uma função (um mínimo da função nesse caso), chamada \emph{função objetivo} ou \emph{função custo} $C(\boldsymbol{\theta})$ em que $\boldsymbol{\theta}$ são os parâmetros ou entradas da função. Devido à complexidade das funções objetivo e à sua dependência de múltiplos parâmetros, os métodos de otimização enfrentam diversas dificuldades, como a presença de mínimos locais que podem impedir os algoritmos de convergir para soluções ótimas. Em muitas aplicações, a obtenção de soluções satisfatórias, ainda que não ótimas, já é considerada suficiente; contudo, até mesmo essas soluções podem ser desafiadoras de se alcançar. Além disso, uma limitação intrínseca dos métodos de otimização é a impossibilidade de prever o tempo necessário para encontrar uma solução ótima ou próxima da ótima, uma vez que tal estimativa exigiria conhecimento completo da função objetivo, o que, na prática, equivaleria a conhecer previamente a própria solução.

Os algoritmos de otimização usualmente se baseiam no cálculo da função objetivo para um estado (conjunto de parâmetros) de entrada, que pode ser chamado da solução atual, e em outros estados que são relacionados por alguma regra com o estado de entrada. A partir desses resultados é produzido um novo estado, que pode ser algum dos estados relacionados ou um estado completamente diferente dependendo do método. Em seguida, o novo estado passa a ser a solução atual e o processo, chamado de ``passo'', é repetido até o resultado ótimo ser encontrado. O cálculo de um estado novo a partir de um estado de entrada são realizados por \emph{otimizadores}. Um otimizador bastante simples é o \emph{Gradiente descendente}, em que o novo estado é a subtração do estado atual com o gradiente da função custo no estado atual multiplicado por um fator positivo, ou seja,
\begin{equation}
    \boldsymbol{\theta}_{\text{novo}} = \boldsymbol{\theta}_{\text{antigo}} - \eta \nabla_{\boldsymbol{\theta}}C(\boldsymbol{\theta}).
    \label{eq:grad}
\end{equation}
Por depender do gradiente, esse método usualmente precisa calcular a função custo em diversos pontos próximos a $\boldsymbol{\theta}$.

Diferentes otimizadores calculam a função custo para quantidades diferentes de estados antes de gerar um estado novo e demandam diferentes quantidade de atualizações da solução atual para chegar ao ponto ótimo. Essas quantidades variam bastante de otimizador para otimizador e, como dito, saber quantos passos são necessários para cada um demandaria um conhecimento completo da função que buscamos otimizar.

Nos algoritmos quânticos que iremos apresentar a partir da Seção \ref{sec:CompRuidAlgVar}, o processo de otimização é realizado em um computador clássico e o trabalho do computador quântico é apenas fornecer um método para o cálculo da função custo nos diversos estados de forma eficiente, sendo independente do otimizador. Para um mesmo problema, diferentes circuitos ou funções custo podem ser propostas~\cite{cerezo2021variational}, com a complexidade dessa etapa podendo ser calculado da forma como apresentada para os algoritmos que não são de otimização, \textit{i. e.}, obtendo a profundidade do circuito e a quantidade de medidas necessárias para se calcular a função custo. Dessa maneira, obter a complexidade do cálculo de uma função custo é essencial para saber qual o melhor circuito para a resolução do problema. 

\section{Algoritmos Quânticos Variacionais para a simulação de sistemas físicos}
\label{sec:AlgVar&SimFis}

Nesta Seção, exploraremos os simuladores quânticos e os desafios que enfrentam no contexto dos hardwares quânticos disponíveis atualmente. Além disso, apresentaremos o \textit{Variational Quantum Eigensolver}, detalhando suas principais etapas e o papel que desempenha no avanço da computação quântica.

\subsection{Simuladores quânticos}

Como comentado na Seção \ref{sec:2qubits}, as propriedades únicas de sistemas quânticos como superposição e emaranhamento não podem ser reproduzidas em sistemas clássicos, porém, elas podem ser simuladas por métodos numéricos. Hoje em dia, simulações numéricas de propriedades quânticas são necessárias em quase todos os campos da física, mas infelizmente essas simulações são extremamente custosas computacionalmente. 

Nesse contexto, Feynman traz em 1982 uma questão fundamental~\cite{Feynman1982}: é possível utilizar um sistema quântico, no qual exista uma alta capacidade tecnológica de controle sobre suas interações e evolução, de modo a permitir simular um outro sistema de interesse e obter informações sobre suas propriedades? A esses sistemas com alto controle foi atribuído o termo \emph{simuladores quânticos}. 

Em princípio, simuladores quânticos e computadores quânticos são conceitualmente diferentes, porém, como os computadores quânticos são dispositivos que necessitam de alto controle, se torna natural questionar se eles podem servir como simuladores quânticos. A resposta para essa questão veio com S. Lloyd em 1996~\cite{Lloyd1996} no seu trabalho ``Universal Quantum Simulators'' onde ele demonstra como utilizar computadores quânticos baseados em portas lógicas para simular de forma eficiente a evolução de sistemas com interações locais e obter suas propriedades.

Em especial, as autoenergia de Hamiltonianos $\hat{H}$ são de grande interesse e desde a proposta de Lloyd alguns algoritmos foram desenvolvidos para obtê-las usando computadores quânticos. Um dos mais conhecidos é o \textit{quantum phase estimation} capaz de obter de forma eficiente as fases $\phi$ dos autovalores $e^{i2\pi\phi}$ de um operador Unitário $\hat{U}$ (observa-se que essa é a forma genérica para os autovalores de qualquer operação unitária) e, para o caso particular que $\hat{U}=e^{i2\pi \hat{H}}$, as fases são autovalores do Hamiltoniano $\hat{H}$.

\subsection{Computadores ruidosos e algoritmos variacionais}
\label{sec:CompRuidAlgVar}

A realização do \textit{quantum phase estimation}, e diversos outros algoritmos como o de P. Shor para fatoração de números, são impraticáveis para problemas reais em computadores quânticos atuais. Isso se deve à alta sensibilidade de sistemas quânticos a ruídos oriundos do ambiente e que produzem alterações imprevisíveis no sistema. O ruído está presente tanto na física clássica quanto na quântica, sendo crucial para nosso entendimento do universo e desenvolvimento tecnológico~\cite{cohen2005history}. 

Nos computadores clássicos, o desgaste dos dispositivos eletrônicos e condutores ao longo do tempo, variações de campos eletromagnéticos, aquecimento dos circuitos e muitos outros fenômenos produzem ruídos inconvenientes que podem afetar o processamento de informação. Porém, esses dispositivos \emph{clássicos} e \emph{macroscópicos} são bastante resilientes e podem executar muitas operações de maneira correta apesar desses efeitos e, mesmo que erros ocorram, existem protocolos eficientes para corrigi-los~\cite{shannon1948mathematical,hamming1950error}. Por outro lado, na computação quântica, os sistemas físicos que realizam o processamento de informação são formados por poucos átomos ou sistemas microscópicos extremamente sensíveis aos ruídos oriundos do meio ambiente em que se encontram. Desse modo, como é impossível isolá-los completamente do universo e de suas interações eletromagnéticas (mesmo no vácuo) ou com partículas próximas, a superposição e o emaranhamento podem ser facilmente destruídas, introduzindo assim erros no algoritmo. Dessa forma, reduzir os efeitos do ruído e desenvolver métodos \emph{quânticos} de correção/mitigação de erros~\cite{krinner2022realizing,postler2022demonstration,ai2024quantum} é essencial para que os computadores quânticos possam ser aplicados para resolver problemas reais.

Para quantificar o quanto o ruído interfere no computador quântico são utilizados dois conceitos. O primeiro é o tempo de coerência que se refere ao tempo máximo que é possível evoluir um conjunto de qubits utilizando o circuito sem que os efeitos do ruído alterem o estado o suficiente para que as informações obtidas ao medi-lo não sejam mais confiáveis. O segundo é a fidelidade das portas lógicas, pois as interações do sistema com os dispositivos que geram os efeitos das portas, como lasers e campos magnéticos, também estão sujeitas a imperfeições e ruídos e, por isso, os estados de fato produzidos por uma porta lógica diferem dos estados que deveriam ser produzidos na teoria, inserindo novamente erros no algoritmo. Normalmente, portas lógicas distintas possuem diferentes fidelidades, com as de um qubit podendo chegar a fidelidades acima de $99,9\%$ e as de dois qubits acima de $99\%$~\cite{frisch2024trapped}. Esses números podem impressionar, mas ainda são baixos, pois, na execução de algoritmos de aplicação prática, com mais de milhares de operações, a propagação de erros faz com que o estado final dos qubits difira totalmente do desejado.

O tempo de coerência e as fidelidades das portas lógicas definem o controle sobre os sistemas quânticos utilizados para a computação e, devido às capacidades tecnológicas atuais, atribui-se ao período que estamos o termo \textit{noisy intermediate-scale quantum} (NISQ) \textit{era}~\cite{preskill2018}. Como forma de tentar obter alguma vantagem utilizando dispositivos da era NISQ na resolução de problemas e simulações de interesse prático (industrial/comercial) com relação aos computadores clássicos, foram desenvolvidos os Algoritmos Quânticos Variacionais (VQAs - do inglês \textit{Variatonal Quantum Algorithms})~\cite{Peruzzo2014, cerezo2021variational}. 

\begin{figure*}[!htb]
    \centering
    \includegraphics[width=0.9\textwidth]{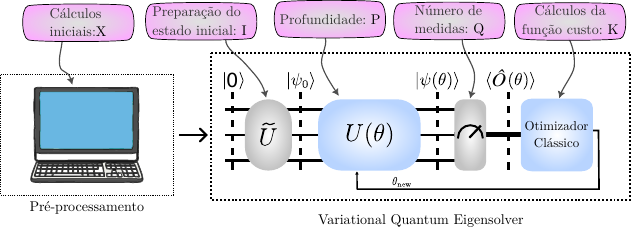} 
    \caption{Representação esquemática de um \textit{Variational Quantum Algorithm}. A primeira etapa consiste no pré-processamento de dados em um computador clássico, cuja complexidade temporal é $X$. O restante do algoritmo é dividido em quatro partes: A preparação do estado inicial dos qubits, o circuito variacional, o número de medidas e o otimizador (que define o número de cálculos da função custo para cada passo de otimização, bem como o número de passos para a otimização), cujas complexidades temporais são, respectivamente, $I$, $P$, $Q$ e $K$. Enquanto os valores de $I$, $P$ e $Q$ podem ser determinados por uma análise teórica, o valor de $K$ depende de uma ampla gama de fatores e só pode ser obtido empiricamente ou se já conhecêssemos a solução do problema.}
    \label{fig:VQA-diagram}
\end{figure*}

Os VQAs são uma classe de algoritmos híbridos que fazem uso de um circuito quântico parametrizável, ou seja, que possui portas lógicas que dependem de parâmetros reais $(\boldsymbol{\theta} )$ que podem ser ajustados, como as rotações da equação \eqref{eq:rotacoes_eixos}, e um otimizador clássico. Dessa maneira, o estado final é dependente dos parâmetros, \textit{i. e.}, $\ket{\psi_{\text{final}}}=\ket{\psi_{\text{final}}(\boldsymbol{\theta} )}$. Nesses casos, a função custo do problema de interesse é obtida a partir do estado final, por exemplo, por meio do cálculo do valor esperado de algum observável, \textit{i. e.}, $C(\boldsymbol{\theta}) = \bra{\psi_{\text{final}}(\boldsymbol{\theta})}\hat{O}\ket{\psi_{\text{final}}(\boldsymbol{\theta} )}$. Dessa forma, o computador quântico calcula de forma eficiente a função custo e o computador clássico utiliza esse cálculo para atualizar os parâmetros da função. Assim, o cerne dos VQAs consiste na construção de um estado inicial e circuito que permitam obter uma solução adequada para o problema por meio de um conjunto de parâmetros que precisam ser descobertos. Um diagrama para representar os VQAs onde a função custo é o valor esperado de um observável pode ser observado na Figura \ref{fig:VQA-diagram}

A utilização dos VQAs na era NISQ ocorre devido à capacidade de ``adaptação'' dos circuitos, uma vez que possuem parâmetros livres, o que permite que eles sejam menos profundos e, portanto, possam ser executados sem a necessidade de computadores com longos tempos de coerência e fidelidades de portas extremamente altas.  
Em muitos casos, os VQAs se baseiam no princípio variacional, equação \eqref{eq:principioVariacional}, que pode ser estendido para qualquer observável $O$ além do Hamiltoniano, onde dado um estado $\ket{\psi}$ normalizado, o menor autovalor de $O$, $O_{\text{min}}$, é limitado superiormente por
\begin{equation}
    O_{\text{min}} \le \bra{\psi}O\ket{\psi},\forall \ket{\psi},
\end{equation}
onde a igualdade ocorre se $\ket{\psi}$ for o autoestado $\ket{\psi}_{\text{min}(O)}$ de $O$ relacionado ao seu menor autovalor $O_{\text{min}}$. Assim, constrói-se um observável de forma que seu autoestado de menor autovalor codifique a solução do problema de interesse. Apesar da generalização, usualmente $O$ realmente representa um Hamiltoniano $H$ e, por consequência, $\ket{\psi}_{\text{min}(O)}$ é o estado fundamental do sistema $\ket{\psi}_{\text{gr}}$ e $O_{\text{min}}$  é a energia fundamental $E_{\text{gr}}$. 

Quando o menor autovalor é a quantidade que se busca descobrir por meio de um VQA, ele é chamado de \textit{Variational Quantum Eigensolver} (VQE), como é o caso de uma das aplicações mais comuns desses algoritmos, em sistemas moleculares, onde objetiva-se obter a energia do estado fundamental e que será discutida no restante deste trabalho.

\subsection{Etapas do Variational Quantum Eigensolver}
\label{sec:VQEsteps}

A primeira etapa do VQE para se obter a energia fundamental de um sistema molecular é o mapeamento do espaço de Hilbert $\mathcal{H}_S$ do sistema de interesse no espaço de Hilbert dos qubits $\mathcal{H}_Q=\mathbb{C}^{2^n}$, onde $n$ é o número de qubits utilizados. 
Fundamentalmente, essa correspondência entre os dois sistemas é o que permite as simulações quânticas. 

As matrizes de Pauli $\{I,X,Y,Z\}$ são bastante úteis nesta tarefa já que grande parte das operações em computação quântica são definidas a partir dessas matrizes, suas exponenciais e portas controladas como $CX$ e $CZ$. Além disso, os processos de medidas são usualmente realizados nas bases dos observáveis do conjunto $\{I,X,Y,Z\}^{\otimes n}$, formado por todos os possíveis produtos tensoriais de $n$ matrizes de Pauli, chamados de \textit{Pauli strings}. Por exemplo, em um espaço de 4 qubits, a \textit{Pauli string} $I_0 Z_1 I_2  X_3$ representa uma medida em $Z$ no qubit $1$ e em $X$ no qubit $3$, enquanto não é feita nenhuma medida nos qubits $0$ e $2$.

Como o conjunto $\{I,X,Y,Z\}^{\otimes n}$ é uma base no espaço $\mathcal{L}\left(\mathbb{C}^{2^n}\right)$ (\textit{i. e.} o espaço dos operadores lineares que atuam nos vetores de $\mathbb{C}^{2^n}$), é possível medir qualquer observável  $O$ a partir da soma ponderada de medidas das \textit{Pauli strings} que formam $O$ via combinação linear, com os pesos sendo os coeficientes da combinação linear, por exemplo, o valor esperado do observável 
\[
O=2\cdot I_0 Z_1 I_2  X_3 + 4\cdot I_0 Z_1 Z_2  Y_3
\]
 satisfaz a igualdade
\[
\langle O\rangle = 2\cdot \langle I_0 Z_1 I_2  X_3\rangle  + 4 \cdot \langle I_0 Z_1 Z_2  Y_3\rangle .
\]

Dessa forma, representar os estados de $\mathcal{H}_S$ por meio de qubits de forma lógica e unívoca, e operadores unitários e observáveis utilizando as \textit{Pauli strings} de forma consistente é como realizamos o mapeamento. De forma mais rigorosa, dados quaisquer dois estados $\ket{\psi}_S,\ket{\varphi}_S\in\mathcal{H}_S$, suas representações em qubits, $\ket{\psi}_Q$ e $\ket{\varphi}_Q$, e um operador $U_S$ que realiza $U_S\ket{\psi}_S=\ket{\varphi}_S$, então o operador $U_Q$ que atua no espaço de qubits e representa $U_S$ deve realizar $U_Q\ket{\psi}_Q=\ket{\varphi}_Q$. Para deixar mais claro, será exemplificado esse processo utilizando o mapeamento de Jordan-Wigner (JW)~\cite{jordan1928paulische}, que é o mais simples e comum para relacionar um subespaço do espaço de Fock fermiônico gerado por um conjunto finito com $N_f$ funções de base/spin-orbitais e o espaço de $N_f$ qubits. Nesse mapeamento, cada spin-orbital do espaço de Fock é representado por um qubit de tal forma que se algum elétron estiver ocupando o spin-orbital, o qubit é igual a $\ket{1}$. Se não houver elétron, ele é igual a $\ket{0}$ e a superposição de spin-orbitais são obtidas a partir da superposição de qubits. 

Como foi apresentado na Seção \ref{sec:ObsSecQuant}, os operadores do espaço de Fock são construídos a partir de produtos de operadores de criação e aniquilação de spin-orbitais ($\hat{a}_{j}$ e $\hat{a}^{\dagger}_{j}$). Desse modo, para conseguir representar qualquer operador do espaço de Fock, basta encontrar uma forma consistente de representar $\hat{a}_{j}$ e $\hat{a}^{\dagger}_{j}$ no espaço de qubits. 

Como o spin-orbital $j$ está ocupado se o qubit $j$ for $\ket{1}$ e não ocupado se o qubit for $\ket{0}$, os operadores $\hat{a}_{j}$ e $\hat{a}^{\dagger}_{j}$ devem levar, respectivamente, $\left( \ket{0}_j\to 0,   \ket{1}_j \to \ket{0}_j\right)$ e $\left( \ket{0}_j\to \ket{1}_j,   \ket{1}_j\to 0 \right)$ e para isso é possível relacionar $\hat{a}_{j}\to \sigma_j^-=\ket{0}_j\bra{1}_j= \frac{X+iY}{2}$ e $\hat{a}_{j}^{\dagger}\to \sigma_j^+=\ket{1}_j\bra{0}_j=\frac{X-iY}{2}$. Porém, para garantir os efeitos das equações \eqref{eq:a_daggerEffect_v2} e \eqref{eq:aEffect_v2}, consequente das relações de anti-comutação dos operadores (equações \eqref{eq:antComut1} e \eqref{eq:antComut2}), é preciso aplicar portas $Z$'s nos spin-orbitais de índice menor que $j$, pois, ela realizará uma multiplicação de $-1$ nos qubits que estão em $\ket{1}$. Assim, a representação de $\hat{a}_{j}$ e $\hat{a}^{\dagger}_{j}$ são:
\begin{equation*}
    \begin{split}
        \hat{a}_{j} \to Z_0 Z_1 Z_2 \cdots \sigma_j^-,\\
        \hat{a}_{j}^{\dagger} \to Z_0Z_1  Z_2 \cdots \sigma_j^+,
    \end{split}
\end{equation*}
 e o subíndice indica em qual qubit o operador atua. Ressalta-se que estão sendo omitidas as operações identidades aplicadas nos qubits de índice maior que $j$, porém, para alguns casos, elas serão escritas explicitamente para facilitar o entendimento. Com isso, qualquer operador pode ser produzido a partir de produtos de $\hat{a}_{j}$ e $\hat{a}_{j}^{\dagger}$. A partir deste ponto, por simplicidade, iremos nos referir tanto aos operadores do espaço de Fock quanto do espaço de qubits apenas como operadores ou circuitos quânticos. 

Outros dois mapeamentos comumente apresentados e discutidos na literatura são os de paridade e Bravyi-Kitaev (BK)~\cite{seeley2012bravyi}. No primeiro desses, o qubit $j$ está em $\ket{0}$ ou $\ket{1}$ se a quantidade de spin-orbitais ocupados com índice menor ou igual a $j$ for, respectivamente, par ou ímpar. Enquanto o mapeamento de BK combina os mapeamentos de JW e de paridade. Para as discussões deste artigo, será utilizado apenas o mapeamento de JW que está sumarizado na Tabela \ref{tab:Fock-qbits}.

\begin{table*}
    \centering
    \begin{tabular}{|c|c|c|}
         \hline
         {\small Sistema} & Eletrônico & Qubits\\
         \hline
         {\small Espaço vetorial} & Fock $\mathcal{F}$ &  $\mathbb{C}^{2^{n}}$\\
         \hline
         {\small Vetores de estado} &$\ket{\psi_0,\psi_1,\psi_4}$ & $\ket{11001\cdots0}$\\
         \hline
         {\small Operadores de criação e aniquilação} &$\hat{a}^{\dagger}_j,\hat{a}_j$ & $Z^{\otimes < j}\frac{X_j-iY_j}{2},Z^{\otimes < j}\frac{X_j+iY_j}{2}  $\\
         \hline
        {\small Operadores gerais} &$\hat{H} = \sum h_p^{q} \hat{a}^{\dagger}_p \hat{a}_{q} + \frac{1}{2}\sum h_{qq'}^{pp'} \hat{a}^{\dagger}_p \hat{a}^{\dagger}_{p'} \hat{a}_{q'} \hat{a}_{q}$ & $\hat{H} = \sum_a P_a$\\
         \hline
    \end{tabular}
    \caption{Exemplificação da correspondência entre o espaço de sistemas eletrônicos e um sistema de qubits usando a codificação de Jordan-Wigner. $P_a$ se refere a uma \textit{Pauli string} genérica. $Z^{\otimes < j}$ indica a aplicação da matriz de Pauli $Z$ nos qubits de índice menor que $j$. $\hat{H}$ definido pela equação \eqref{eq:n1p13} com o termo constante $h_0$ omitido.}
    \label{tab:Fock-qbits}
\end{table*}

O mapeamento afeta diretamente alguns aspectos da simulação, sendo dois bastante relevantes: o número de qubits e a quantidade de \textit{Pauli strings} necessárias para decompor o observável que representa o Hamiltoniano eletrônico no espaço de qubits. Assim como o mapeamento de Jordan-Wigner, o de paridade e Bravyi-Kitaev também utilizam $N_f$ qubits.

É importante notar que o tamanho da dimensão de um sistema de $N_f$ qubits é $2^{N_f}$, independente do número de elétrons. Por outro lado, o subespaço do espaço de Fock que representa $N_e$ elétrons com $N_f$ funções de base tem dimensão $\binom{N_f}{N_e} < 2^{N_f}$, ou seja, o espaço vetorial do simulador é maior que o subespaço de interesse. Isso fica claro se considerarmos, por exemplo, o estado 
\[\ket{0}^{\otimes N_f}=\underset{N_f\text{ termos}}{\underbrace{\ket{0\cdots 0}}}\] 
que, para o mapeamento de Jordan-Wigner, representa um sistema sem nenhum spin-orbital ocupado, ou seja, sem elétrons. Da mesma forma, o espaço de qubits com essa codificação pode representar sistemas com qualquer quantidade de elétrons desde $0$ até $N_f$ enquanto nosso espaço de interesse é apenas aquele que representa um sistema de $N_e$ elétrons. Isso significa que é necessário tomar cuidados durante a execução do algoritmo para descartar soluções que possam surgir e não representem um sistema com exatamente $N_e$ elétrons, sendo uma forma de garantir isso a utilização de um estado inicial que já possua $N_e$ elétrons e um circuito que represente uma unitária que conserve o número de partículas no espaço de Fock.

No que se refere à quantidade de \textit{Pauli strings} necessárias para representar o Hamiltoniano eletrônico no espaço de qubits, todas as codificações que citamos produzem uma quantidade de \textit{Pauli strings} de $O\left(N^4\right)$. Uma discussão sobre como chegar nesse valor para a codificação de Jordan-Wigner será apresentada na Seção \ref{sec:complexity}.

Assim, os três mapeamentos apresentados não introduzem mudanças significativas no número de qubits necessários nem na quantidade de \textit{Pauli strings} que precisam ser medidas. A principal diferença reside no mapeamento de Bravyi-Kitaev em comparação aos outros dois, especificamente em uma característica denominada \textit{Pauli weight}. Esse termo refere-se ao número de matrizes de Pauli distintas da identidade presentes em uma \textit{Pauli string}. Mapeamentos que geram \textit{Pauli strings} de baixo \textit{Pauli weight} influenciam significativamente a complexidade e a qualidade da solução obtida pelo VQE, variando de acordo com a codificação utilizada.

Uma das razões para \textit{Pauli weights} baixos trazerem vantagens para o algoritmo é o aparecimento frequente de exponenciais de \textit{Pauli strings} nos operadores dos circuitos do VQE, \textit{e. g.} $e^{-i\frac{\theta}{2}X_0 I_1 I_2 Z_3 X_4}$. Esses são operadores de múltiplos qubits que precisam ser decompostos em portas de 1 e 2 qubits que afetam os qubits relacionados aos operadores diferentes da identidade no expoente. 
Dessa forma, $e^{-i\frac{\theta}{2}X_0 I_1 I_2 Z_3 X_4}$ é um operador de três qubits que afeta os qubits 0, 3 e 4. 

Usualmente, a decomposição dessas exponenciais em portas de 1 e 2 qubits gera um termo no circuito de profundidade proporcional ao \textit{Pauli weight} da \textit{Pauli string} do expoente. Dessa maneira, \textit{Pauli strings} com menores \textit{Pauli weights} permitem construir circuitos menos profundos. Similarmente, \textit{Pauli strings} que atuam em operadores diferentes podem ser medidas em paralelo, diminuindo o custo temporal dessa etapa do VQE, além de garantir maior resistência a erro de medidas, dado que a quantidade de qubits medidos ao final do circuito é menor. Uma explicação mais detalhada, com exemplos sobre essas duas questões, será apresentada na Seção \ref{sec:complexity}.

Por fim, foi demonstrado que \textit{Pauli strings} com \textit{Pauli weights} mais baixos geram vantagens no processo de otimização, fornecendo mais resistência ao problema do \textit{Barren plateau} (BP)~\cite{mcclean2018barren, larocca2025barren} que será explicado na Seção \ref{sec:measurement}. O valor máximo de \textit{Pauli weight} das \textit{Pauli strings} geradas pelos mapeamentos de Jordan-Wigner e de paridade são da ordem do número de qubits $N_f$, enquanto na de Bravyi-Kitaev é da ordem de $\log_{2}(N_f)$.

Partindo para a escolha do circuito que será utilizado, ele deve manter um balanceamento entre duas características: expressividade e treinabilidade. A primeira se refere à capacidade do circuito de gerar estados diferentes e de forma uniforme, que é importante para garantir a existência de um conjunto de parâmetros capaz de produzir o estado fundamental de interesse a partir do estado inicial. Já o segundo se refere à dificuldade da obtenção desse conjunto de parâmetros pelo processo de otimização, pois um conjunto grande de parâmetros e/ou que gere uma superfície de energia muito rugosa com vários mínimos locais pode tornar o treinamento extremamente custoso e demorado. 

Um simples exemplo pode ser feito com circuitos de um qubit. Tomemos dois circuitos 
\[U_1(\alpha,\beta,\gamma,\delta)=e^{i\alpha}R_z(\beta)R_y(\gamma)R_z(\delta)\] 
e
\[U_2(\theta)=R_x(\theta)\]
com $\alpha,\beta,\gamma,\delta$ e $\theta$ parâmetros reais, e o estado inicial do algoritmo variacional sendo $\ket{0}$. É possível demonstrar que qualquer porta lógica $U$ de um qubit pode ser produzida com a escolha adequada de $\alpha,\beta,\gamma$ e $\delta$, chamada decomposição $Z-Y$ de $U$~\cite{nielsen2010}. Isso implica que qualquer estado de um qubit pode ser produzido por $U_1(\alpha,\beta,\gamma,\delta)$, por outro lado, é possível verificar que $U_2(\theta)\ket{0}=\cos\left(\frac{\theta}{2}\right)\ket{0}-i\sin\left(\frac{\theta}{2}\right)\ket{1}=\cos\left(\frac{\theta}{2}\right)\ket{0}+e^{i\pi}\sin\left(\frac{\theta}{2}\right)\ket{1}$ que não é estado na forma mais geral (Seção \ref{sec:EsferaBloch}): $\cos\left(\frac{\theta}{2}\right)\ket{0}+e^{i\phi}\sin\left(\frac{\theta}{2}\right)\ket{1}$, $\phi\in\mathbb{R}$. Dessa forma, enquanto os parâmetros de $U_1$ podem ser otimizados para se obter qualquer estado, incluindo o estado fundamental de interesse (alta expressividade), o mesmo não ocorre para $U_2$ (baixa expressividade). 

Por outro lado, existem quatro parâmetros para otimizar em $U_1$, podendo tornar seu treinamento mais difícil (baixa treinabilidade) que o de $U_2$, que possui apenas um (alta treinabilidade). Então, caso haja alguma informação prévia informando um estado inicial capaz de ser levado a um estado próximo do fundamental por $U_2$, para uma precisão desejada, ele se torna mais eficiente. 

O número de parâmetros de um circuito capaz de levar um estado qualquer em outro estado qualquer, como o caso de $U_1$ para um qubit, escala com a exponencial do número de qubits, tornando os treinamentos impraticáveis. Assim, para produzir um ansatz com um balanceamento entre expressividade, para garantir que seja possível gerar um estado suficientemente próximo do fundamental, e treinabilidade, para que a sua obtenção ocorra em um tempo razoável, é essencial conhecer profundamente o sistema simulado e os possíveis hardwares disponíveis como simuladores.  

Neste contexto, um candidato proeminente é a versão unitária do \textit{Coupled Cluster}. O método \textit{Unitary Coupled Cluster} (UCC)~\cite{taube2006new} consiste em obter o estado gerado pela aplicação do operador da equação \eqref{eq:CoupleCluster} no estado de referência, mas substituindo o operador de \textit{cluster} $\hat{T}$ por $\hat{T}-\hat{T}^{\dagger}$, que é anti-hermitiano. Ou seja, o estado final fica dado por 
\begin{equation}
    \ket{\Psi_{\text{final}}} = e^{\hat{T}-\hat{T}^{\dagger}}\ket{\Psi_0}.
    \label{eq:UCC}
\end{equation}

 O UCC foi discutido na literatura antes do desenvolvimento dos algoritmos variacionais quânticos, pois, por utilizar um operador unitário, ele satisfaz o teorema variacional, garantindo que os resultados obtidos sejam um limite superior para a energia real do sistema. Isso faz com que o UCC apresente uma vantagem natural com relação ao CC. Destaca-se a existência de cenários onde o método usual diverge para menos infinito por não ser variacional. Existem  também indícios de que o UCC produz resultados mais acurados do que o CC usual~\cite{kutzelnigg2010unconventional}. Por outro lado, a realização do UCC em computadores clássicos é extremamente custosa e demorada devido ao comutador de $\hat{T}$ e $\hat{T}^{\dagger}$ que aparece nas equações do método, tornando-o pouco utilizado nessa forma  de computação~\cite{taube2006new}.
 
Porém,  no caso de computadores quânticos, é possível construir o estado final a partir de um circuito quântico parametrizável que realize a operação $e^{\hat{T}-\hat{T}^{\dagger}}$, com os valores $\{t\}$ a serem otimizados no operador de \textit{cluster} sendo os parâmetros do circuito.

Dessa maneira, o UCC é uma adaptação do CC usual que pode ser realizada em um computador quântico. Como já se sabe da teoria que existe um conjunto de parâmetros capaz de gerar uma boa aproximação do estado fundamental, o UCC cumpre todos os requisitos para ser o ansatz do VQE em problemas de estrutura eletrônica.

Realizar o UCC com todas as excitações é impraticável, mesmo em um computador quântico, tornando necessário selecionar quais excitações serão consideradas ao aplicar o algoritmo. Algumas versões muito utilizadas na literatura são o \textit{Unitary Coupled Cluster Single and Double} (UCCSD), o \textit{Unitary Coupled Cluster Generalized Single and Double} (UCCGSD) e o \textit{k- Unitary Pair Coupled Cluster Generalized Singles and Doubles} (k-UpCCGSD). Todos esses consideram apenas excitações simples e duplas e por isso, por brevidade de notação, iremos omitir as letras S e D nas siglas a partir deste ponto. 

O UCC é a versão unitária do CC. O termo \textit{generalized} em UCCG significa que ele considera todas as possíveis transições entre spin-orbitais, \textit{i.e.}, ocupados para virtuais, virtuais para ocupados , virtuais para virtuais e ocupados para ocupados, sendo inspirado na versão generalizada, mas não unitária, do \textit{coupled cluster} proposta por Nooijen~\cite{nooijen2000can}. Em seu trabalho, Nooijen conjectura ser possível produzir o autoestado \emph{exato} do sistema utilizando \emph{apenas as excitações duplas e simples generalizadas}. Apresentando resultados iniciais promissores, sua conjectura foi posteriormente provada \emph{incorreta}, porém, ainda sendo capaz de produzir resultados acurados, conforme demonstrado por Mukherjee e Kutzelnigg~\cite{mukherjee2004some}.

Por último, o k-UpCCG considera também as transições entre todos os spin-orbitais, porém, são realizadas entre as transições duplas apenas aquelas entre pares de funções com spins diferentes mas mesmo orbital. Por exemplo, dados três orbitais $\phi_p,\phi_q, \ \text{e} \ \phi_r$, o operador de \textit{cluster} não contém o termo $\hat{a}_{p,\uparrow}^{\dagger}\hat{a}_{p,\downarrow}^{\dagger}\hat{a}_{q,\uparrow}\hat{a}_{r,\uparrow}$ -- o símbolo $\uparrow$ e $\downarrow$ representam, respectivamente, os spins $\frac{1}{2}$ e $-\frac{1}{2}$ -- porque ele transiciona dois orbitais \emph{distintos} para um mesmo orbital, independente da parte de spin. Porém, ele conteria o termo  $\hat{a}_{p,\uparrow}^{\dagger}\hat{a}_{p,\downarrow}^{\dagger}\hat{a}_{q,\uparrow}\hat{a}_{q,\downarrow}$, que transiciona os dois elétrons do orbital $\phi_q$ para o orbital $\phi_p$. Além disso, a letra $k$ indica que a exponencial do operador de \textit{cluster} será aplicada $k$ vezes de forma sequencial com parâmetros independentes.

\begin{figure*}[!htp]
    \centering
    \includegraphics[width=0.5\textwidth]{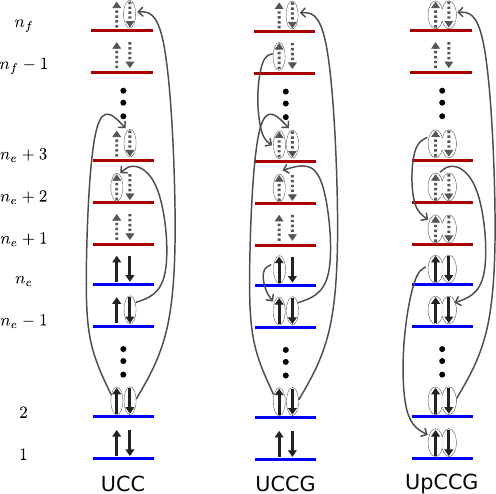} 
    \caption{Esquema ilustrativo dos operadores de excitação presentes em cada um dos ansatz apresentados. O UCC possui apenas operadores que levam spin-orbitais ocupados em estados virtuais. O UCCG possui operadores que permitem a transição entre spin-orbitais quaisquer. O UpCCG é semelhante, porém possui operadores que permitem a transição apenas entre pares de orbitais em que cada par possui mesma parte espacial com um spin $\frac{1}{2}$ e um $-\frac{1}{2}$. Consideramos um caso que o número de elétrons são pares, com $n_e$ e $n_f$ os números de orbitais ocupados e virtuais, respecticamente.}
    \label{fig:transicoes}
\end{figure*}

O k-UpCCG foi introduzido na literatura em 2019 por Lee \textit{et al.}~\cite{lee2018generalized}. Neste trabalho é feita a comparação entre os três métodos, UCC, UCCG e k-UpCCG, comparando as energias obtidas por cada um deles com a energia de FCI, definida na Seção \ref{sec:NdimBasis} e equivalente ao resultado exato obtido ao diagonalizar o Hamiltoiano, para pequenas moléculas. Seus resultados indicam que a utilização das excitações generalizadas apresenta resultados mais acurados, mostrando vantagem na utilização do UCCG em comparação ao UCC. Eles informam que a principal razão para a criação do k-UpCCG é a geração de circuitos com menor profundidade ($O(kN)$), quando comparado aos demais métodos ($O(N^3)$), além de possuir menos parâmetros, tornando sua otimização mais simples. Porém, para se obter resultados com a precisão próxima do FCI, é necessária a utilização de mais camadas do circuito com parâmetros independentes. Um diagrama representando as transições consideradas em cada caso está apresentado na Figura \ref{fig:transicoes}. Os operadores de \textit{cluster} utilizados e a forma do estado final em cada versão são:
\begin{itemize}
    \item UCC:
    \begin{equation*}
        \begin{split}
            & \hat{S}_U = (\hat{T}_1 - \hat{T}_1^{\dagger}) + (\hat{T}_2 - \hat{T}_2^{\dagger}), \\
            & \hat{T}_1=\sum_{a_1\in o}\sum_{r_1\in v} t_{a_1}^{r_1} \hat{a}_{r_1}^{\dagger}\hat{a}_{a_1}, \\
            & \hat{T}_2=\sum_{\underset{a_{1}<a_{2}}{a_{1},a_{2}\in o}}\sum_{\underset{r_{1}<r_{2}}{r_{1},r_{2}\in v}} t_{a_1,a_2}^{r_1,r_2} \hat{a}_{r_1}^{\dagger}\hat{a}_{r_2}^{\dagger}\hat{a}_{a_2}\hat{a}_{a_1},  \\
            & \ket{\psi_{\text{final}}} = e^{\hat{S}_U}\ket{\psi_0}.
        \end{split}
    \end{equation*}
    \item UCCG: 
    \begin{equation*}
        \begin{split}
            & \hat{S}_{UG} = (\hat{T}_{G,1} - \hat{T}_{G,1}^{\dagger}) + (\hat{T}_{G,2} - \hat{T}_{G,2}^{\dagger}), \\ 
            & \hat{T}_{G,1}=\sum_{p_1\in o \cup v}\sum_{q_1\in o \cup v} t_{p_1}^{q_1} \hat{a}_{q_1}^{\dagger}\hat{a}_{p_1},\\
            &  \hat{T}_{G,2}=\sum_{\underset{p_{1}<p_{2}}{p_{1},p_{2}\in o \cup v}}\sum_{\underset{q_{1}<q_{2}}{q_{1},q_{2}\in o \cup v}}t_{p_1,p_2}^{q_1,q_2} \hat{a}_{q_1}^{\dagger}\hat{a}_{q_2}^{\dagger}\hat{a}_{p_2}\hat{a}_{p_1}, \\
            & \ket{\psi_{\text{final}}} = e^{\hat{S}_{UG}}\ket{\psi_0}.
        \end{split}
    \end{equation*}

    \item k-UpCCG: 
    \begin{equation*}
        \begin{split}
            &\hat{S}_{UpG} = (\hat{T}_{pG,1} - \hat{T}_{pG,1}^{\dagger}) + (\hat{T}_{pG,2} - \hat{T}_{pG,2}^{\dagger}), \\
            &\hat{T}_{pG,1}=\sum_{p_1\in o \cup v}\sum_{q_1\in o \cup v} t_{p_1}^{q_1} \hat{a}_{q_1}^{\dagger}\hat{a}_{p_1}, \\
            &\hat{T}_{pG,2}=\sum_{p\in \tilde{o} \cup \tilde{v}}\sum_{q\in \tilde{o} \cup \tilde{v}} t_{p}^{q} \hat{a}_{q\uparrow}^{\dagger}\hat{a}_{q\downarrow}^{\dagger}\hat{a}_{p\uparrow}\hat{a}_{p\downarrow},\\
            & \ket{\psi_{\text{final}}} = \prod_{\alpha=1}^{k}e^{\hat{S}^{\alpha}_{UpG}}\ket{\psi_0}.
        \end{split}
    \end{equation*}
\end{itemize}
onde $o$ é o conjunto dos spin-orbitais ocupados, $v$ dos virtuais, $\Tilde{o}$ o conjunto dos orbitais ocupados e $\Tilde{v}$ dos orbitais virtuais e $\alpha$ indica a camada com parâmetros independentes.

\section{Complexidade temporal teórica do algoritmo}
\label{sec:complexity}

Conforme apresentado na Figura \ref{fig:VQA-diagram} a realização de um VQA geral pode ser dividida em cinco partes: (i) cálculos preliminares (pré-processamento) que antecedem a execução do algoritmo, realizado em computadores clássicos, como as integrais das equações \eqref{eq:integral1Corpo} e \eqref{eq:integral2Corpo} e a obtenção dos autoestados e autoenergias do operador de Fock na base trabalhada, (ii) a preparação do estado inicial, (iii) a execução do circuito quântico, (iv) o processo de medida e (v) a otimização de parâmetros.

A complexidade temporal das etapas são: (i) complexidade clássica $X$ do pré-processamento; (ii) a profundidade $I$ do circuito quântico não parametrizável utilizado para a preparação do estado inicial; (iii) a profundidade $P$ do circuito variacional; (iv) a quantidade $Q$ de medidas necessárias para o cálculo da função custo e (v) a quantidade $K$ de estados que o otimizador utiliza para gerar o estado seguinte e otimizar o circuito.

A partir dessa consideração, podemos construir uma expressão para a complexidade total do algoritmo a partir de uma sequência de argumentos lógicos: O pré-processamento clássico só é necessário ser realizado uma única vez, então ele soma uma complexidade $X$ ao algoritmo, enquanto a parte quântica e o otimizador clássico adicionam um segundo termo $f(I,P,Q,K)$ que depende das demais etapas, gerando uma complexidade total $X + f(I,P,Q,K)$. Fazer essa distinção é essencial, pois, para o algoritmo ser vantajoso para aplicações práticas, é necessário que ambas as partes tenham baixa complexidade. 

O circuito quântico total é formado pela junção da parte não variacional seguida pela aplicação do circuito variacional, assim, sua profundidade total é dada pela soma das duas profundidades: $I + P$. Para cada execução do circuito precisamos realizar uma medida, dessa maneira, se são necessárias $Q$ medidas para o cálculo da função custo, a complexidade por passo é dada por $(I+P)\cdot Q$. Por fim, como são necessários $K$ cálculos da função custo para a otimização, temos que $f(I,P,Q,K) = (I+P) \cdot Q \cdot K$. Assim, a complexidade total do algoritmo é:
\begin{equation}
    X + (I+P)\cdot Q \cdot K.
    \label{eq:totalTempComplex}
\end{equation}

Por outro lado, como comentado na Seção \ref{sec:recursos}, nosso foco será na complexidade para o cálculo da função custo por meio do circuito quântico, que pode ser feito de forma precisa e não depende de fatores imprevisíveis; dessa forma, não será calculado o termo $K$ que é afetado pela escolha do otimizador, a solução inicial e a superfície de energia em si. Na prática, geralmente os algoritmos são executados diversas vezes, utilizando diferentes otimizadores e soluções iniciais que, apesar de possuírem certo grau de aleatoriedade, são influenciadas por fatores como propriedades do sistema, da solução inicial e também da experiência do programador.  Para tornar essa discussão toda mais palpável, iremos apresentar o cálculo de $X +(I+P)\cdot Q$ para um exemplo do VQE onde pretende-se obter a energia do estado fundamental de um sistema com $N_e$ elétrons, considerando $N_f$ funções de base e o mapeamento de Jordan-Wigner. Também iremos considerar três circuitos variacionais: UCC, UCCG e k-UpCCG.

Devido ao foco deste trabalho ser no algoritmo variacional, a discussão sobre o pré-processamento e obtenção de $X$ será apresentada de forma mais sucinta, focando apenas em resultados já presentes na literatura. Para o exemplo em discussão, o pré-processamento necessário exige cálculos do método de Hartree-Fock. A obtenção das autofunções do operador de Fock é feita a partir de um método Self-Consistent Field (SCF), conforme descrito na Seção \ref{sec:HartreeFock}, no qual a função é atualizada iterativamente a partir de um chute inicial até a sua convergência. De forma geral, em um caso onde são utilizadas $N_f$ funções de base de $1$ elétron para representar um sistema de $N_e$ elétrons, a realização de cada passo da iteração escala com $O(N_f^4)$ devido às integrais de termos de dois corpos, que possuem quatro funções de base cada~\cite{whitfield2013computational,echenique2007mathematical}. Entretanto, conforme o número de funções aumenta, a contribuição de diversas das integrais se torna desprezível, ficando abaixo da precisão do método. O número de integrais necessárias diminui mais rapidamente do que a quantidade de integrais novas que surgem com o aumento da base e podemos reduzir a complexidade para $O(N_f^2)$ no limite de $N_f$ grande~\cite{whitfield2013computational}. Por outro lado, a convergência pode se tornar um problema e é demonstrado que, devido a isso, obter a solução de Hartree-Fock é um problema NP-Completo no pior caso~\cite{whitfield2013computational}. No final, a utilização de heurísticas e a desconsideração de integrais que contribuem pouco para a solução permitem que ele escale com $O(N_f^3)$ para diversos problemas de interesse prático, valor que iremos considerar para o nosso exemplo.

\subsection{Profundidade do circuito}
\label{sec:profundidadeCircuito}

\begin{figure}[H]
    \centering
    \vspace{-0.5cm}
    \hspace{-0.8cm}\includegraphics[width=0.7\columnwidth]{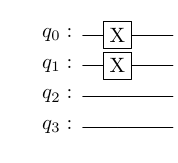} 
    \vspace{-0.8cm}
    \caption{Exemplo de circuito que gera o estado de Hartree-Fock para um problema com $4$ funções de base e dois spin-orbitais ocupados. Considera-se que cada qubit representa a ocupação de um autoestado do operador de Fock. Os qubits que representam orbitais ocupados são colocados em $\ket{1}$ pela atuação da porta lógica $X$.}
    \label{fig:XGate}
\end{figure}

A profundidade do circuito é dada pela soma das complexidades do termo necessário para a obtenção do estado inicial com o termo do circuito variacional. Para o exemplo apresentado, o estado inicial será o estado de Hartree-Fock de $N_e$ elétrons obtido a partir dos cálculos preliminares; além disso, por meio do mapeamento de Jordan-Wigner, consideramos que o qubit $0$ representa a ocupação do autoestado do operador de Fock de menor autovalor, o qubit $1$ representa o segundo de menor autovalor, e assim por diante. Como o estado de Hartree-Fock é aquele em que os $N_e$ autoestados de menor autovalor estão ocupados, então sua representação no espaço de qubits é dada simplesmente por
\[
\ket{\psi_{HF}} = \vert\underset{N_e}{\underbrace{1\cdots1}} \ \underset{N_f-N_e}{\underbrace{0\cdots0}\rangle},
\]
que pode ser preparado com a aplicação simultânea de portas $X$ nos $N_e$ primeiros qubits, assim, a profundidade do circuito não parametrizável que prepara o estado inicial é $I=1$, como mostra a Figura \ref{fig:XGate} para um problema com $N_e=2$ e $N_f=4$.

Como a complexidade da parte não-variacional é $1$, ela pode ser desconsiderada para o cálculo de complexidade de \textit{big} O, restando apenas a parte variacional que será feita para as três variações do \textit{Coupled Cluster} apresentados na Seção \ref{sec:VQEsteps}

Para obter a complexidade do circuito variacional, primeiro é necessário apresentar como converter o operador $e^{\hat{S}}$ em um circuito quântico. Relembremos que $\hat{S}$ de forma geral é dado por
\begin{equation*}
    \hat{S} = \sum_{i}t^{(1)}_i\left( \hat{T}_{1}^{(i)} - \hat{T}_1^{(i)\dagger}\right) + \sum_{i}t_i^{(2)}\left( \hat{T}_2^{(i)} - \hat{T}_2^{(i)\dagger}\right),
\end{equation*}
onde $\hat{T}_1^{i}$ e $\hat{T}_2^{i}$ representam diferentes operadores de excitação de um e dois corpos, respectivamente, para qualquer um dos três ansatz considerados (UCC, UCCG e k-UpCCG), com $\{t_i^{(j)}\}$ sendo os parâmetros a serem otimizados.

Por meio do mapeamento de Jordan-Wigner podemos converter $\hat{S}$ em uma somatória de \textit{Pauli strings}
\begin{equation}
    \hat{S} \to \sum_j-i\frac{\theta_j}{2}P_j,
    \label{eq:StoPauli}
\end{equation}
com $P_i$ sendo uma \textit{Pauli string}. Por exemplo, considerando $\hat{a}_1^{\dagger}\hat{a}_3 - h.c.$  em um sistema com $5$ funções de base, temos:
\begin{equation*}
    \begin{split}
        \hat{a}_1^{\dagger}\hat{a}_3 - h.c.\to & \left[Z_{0}\sigma_1^+I_{2}I_{3}I_{4}\right] \left[Z_{0}Z_{1}Z_{2}\sigma_3^-I_{4}\right] - h.c.\\
                           =  & I_0\sigma_1^+Z_2\sigma_3^-I_4 - h.c.\\
                           =  & \frac{1}{4}\left(I_0X_1Z_2X_3I_4 + iI_0X_1Z_2Y_3I_4 \right. - h.c.\\
                           -  & \left.iI_0Y_1Z_2X_3I_4 + I_0Y_1Z_2Y_3I_4\right) - h.c. \\
                           =  & \frac{i}{2} \left(I_0X_1Z_2Y_3I_4 - I_0Y_1Z_2X_3I_4 \right). 
    \end{split}
\end{equation*}
Dessa forma, exponenciando a somatória da equação \eqref{eq:StoPauli}, a unitária que será aplicada no estado de Hartree-Fock do espaço de qubits é:
\begin{equation}
    U\left( \boldsymbol{\theta} \right) = e^{\sum_j-i\frac{\theta_j}{2}P_j}.
    \label{eq:circFormulaExato}
\end{equation}

A profundidade do circuito depende da quantidade de termos que aparece na somatório da equação \eqref{eq:circFormulaExato}. Assim, faremos uma estimativa dessa quantidade antes prosseguir com o cálculo da complexidade. Para facilitar a análise, como na química geralmente utiliza-se $N_f$ e $N_e$ da mesma ordem de grandeza, ainda com $N_f>N_e$, vamos assumir que $N_f=2N_e=2N$ para a complexidade de \textit{big} O. Considere primeiro o ansatz UCC cujos operadores presentes são:
\begin{equation*}
    \begin{split}
            & \hat{T}_1=\sum_{a_1}^{o}\sum_{r_1}^{v} t_{a_1}^{r_1} \hat{a}_{r_1}^{\dagger}\hat{a}_{a_1}, \\
            & \hat{T}_2=\sum_{a_1<a_2}^{o}\sum_{r_1<r_2}^{v} t_{a_1,a_2}^{r_1,r_2} \hat{a}_{r_1}^{\dagger}\hat{a}_{r_2}^{\dagger}\hat{a}_{a_2}\hat{a}_{a_1}.
        \end{split}
\end{equation*}
Como a somatória em $o$ representa os orbitais ocupados cuja quantidade é igual ao número de elétrons $N_e$, enquanto a somatória em $v$ é sobre os orbitais virtuais, com quantidade igual a $N_f-N_e$, a quantidade de termos em $\hat{T}_1$ é igual à quantidade de formas que podemos escolher um orbital ocupado e um virtual. Como são $N_e$ formas de escolher o primeiro e $N_f-N_e$ de escolher o segundo, no total há $N_e(N_f-N_e)=O(N^2)$ termos em $\hat{T}_1$.

Podemos usar um raciocínio semelhante para $\hat{T}_2$, porém, agora são dois orbitais ocupados e dois virtuais diferentes entre si, além disso, como a soma é feita sobre $a_2>a_1$ e $r_2>r_1$ também é necessário dividir por quatro o resultado final para retirar somas repetidas em ambos os casos. Assim, a quantidade de termos em $\hat{T}_2$ é $\frac{1}{4}N_e(N_e-1)(N_f-N_e)(N_f-N_e-1)=O(N^4)$. 

Como os termos em $\hat{T}_j$ e $\hat{T}_j^{\dagger}$ são os mesmos, então podemos agrupá-los em uma mesma soma. Assim, a quantidade total de termos em $\hat{S}$ do UCC fica igual à soma da quantidade de termos de $\hat{T}_1$ e $\hat{T}_2$, que resulta da ordem de $O(N^4)$.

Para o UCCG não possuímos mais a restrição sobre orbitais ocupados e virtuais, porém, devido à subtração dos adjuntos de cada operador $\hat{T}$, os casos em que $q_1=p_1$ para $\hat{T}_{G,1}$ e $p_1=q_1$ e $p_2=q_2$ em $\hat{T}_{G,2}$ vão dar zero; assim, precisamos ainda considerar spin-orbitais diferentes.
Para $\hat{T}_{G,1}$ existem $N_f$ formas de escolher $q_1$ e $N_f-1$ formas de escolher $p_1$, gerando um total de $N_f(N_f - 1)$ termos, que contribui para a complexidade com um termo da ordem de $O(N^2)$. Já para $\hat{T}_{G,2}$, há $N_f$ formas de escolher $q_1$ e $q_2$, sobrando $N_f - 1$ formas de escolher $p_1$ e $p_2$, com um total de termos igual a $N_f^2(N_f-1)^2=O(N^4)$. Somando ambas, o número de termos totais para $\hat{S}_{UG}$ também é $O(N^4)$.

Por último, para o k-UpCCG temos que $\hat{T}_{pG,1}=\hat{T}_{G,1}$, possuindo, portanto, $O(N^2)$ termos. Por outro lado, para o $\hat{T}_{pG,2}$, como são feitas excitações apenas entre pares de elétrons com spins diferentes e mesmo orbital, então a escolha de $q\uparrow$ deve ser feita sobre os $N_f/2$ orbitais, em vez dos $N_f$ spin orbitais. Dessa forma, fazendo-se tal escolha, $q\downarrow$ fica automaticamente definido. Restam então $N_f/2 - 1$ orbitais para a escolha de $p\uparrow$, que uma vez escolhido, define $p\downarrow$, ficando com um total de $\frac{N_f}{2} \left(\frac{N_f}{2} - 1\right)$ termos que são da ordem de $O(N^2)$. Como resultado, a quantidade de termos em $\hat{S}_{UpG}$ é da ordem de $O(N^2)$. Entretanto, como é necessário implementar $\hat{S}_{UpG}$ sequencialmente $k$ vezes, sua complexidade se torna $O(kN^2)$.
A forma como essas quantidades impactam na complexidade ficará clara após a discussão seguinte. 

Apesar da unitária na equação \eqref{eq:circFormulaExato} definir um circuito quântico que pode ser decomposto diretamente em um conjunto de portas universais com precisão arbitrária, essa decomposição pode ser extremamente complicada e custosa. Para contornar esse problema, é possível encontrar uma aproximação da unitária que é mais simples de ser decomposta. 

\begin{figure*}[!htp]
    \centering
    \includegraphics[width=0.7\linewidth]{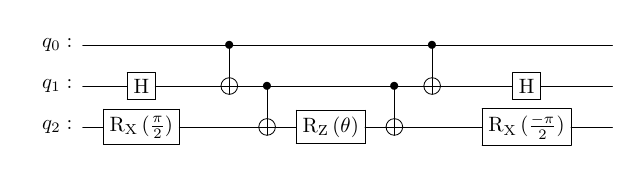} 
    \caption{Circuito que representa o operador $e^{-i\frac{\theta}{2}Z_0X_1Y_2}$. As portas controladas e a rotação em torno de $Z$ geram o efeito multqubit da exponencial $e^{-i\frac{\theta}{2}Z_0Z_1Z_2}$. Já a aplicação das portas Hadamard no começo e final da linha do qubit $1$ realizam, respectivamente, mudanças da base $Z$ para $X$ e de volta de $X$ para $Z$, enquanto as portas $R_X(\frac{\pi}{2})$ e $R_X(\frac{-\pi}{2})$ fazem o mesmo com as bases $Z$ e $Y$ para o qubit 2, garantindo que a exponencial desejada de diferentes matrizes de Pauli seja aplicada no sistema.}
    \label{fig:exp}
\end{figure*}

A dificuldade de decompor a unitária da equação \eqref{eq:circFormulaExato} ocorre devido à somatória de diferentes \textit{Pauli strings} presentes na exponencial. Por outro lado, é possível decompor a exponencial de uma única \textit{Pauli string} em portas C-NOT, Hadamard, $R_x(\frac{\pi}{2})$, $R_x(-\frac{\pi}{2})$ e $R_z(\theta)$,  com a profundidade do circuito crescendo linearmente com a quantidade de operadores não identidade presentes na \textit{Pauli strings}~\cite{anand2022quantum}, \textit{i. e.} o \textit{Pauli weight}. 
Apesar da demonstração dessa decomposição fugir do escopo do trabalho, um exemplo da implementação da exponencial $e^{-i\frac{\theta}{2}Z_0X_1Y_2}$ é apresentado na Figura \ref{fig:exp} e permite o entendimento de como essa decomposição é feita. Basicamente deve-se aplicar a porta Hadamard $H$ em todos os qubits que aparecem a matriz $X$ e a porta  $R_x(\frac{\pi}{2})$ nos que aparecem a matriz $Y$ na \textit{Pauli strings}. 
Em seguida aplica-se portas C-NOT entre todos os qubits que não têm uma matriz identidade na \textit{Pauli string} e depois a porta $R_z(\theta)$ no último qubit com o ângulo desejado. Por fim, inverte-se a aplicação das C-NOT, Hadamard e $R_x(\frac{\pi}{2})$. 

Para implementar a unitária da equação \eqref{eq:circFormulaExato} busca-se uma maneira de aproximá-la por um produto de exponenciais de uma única \textit{Pauli string}, pois, dessa forma, cada uma pode ser implementada sequencialmente. Infelizmente, diferentemente dos números complexos, para os quais sempre vale a relação $e^{z+w}=e^ze^w$ (com $z,w\in\mathbb{C}$), devido à não comutatividade de operadores em espaços de Hilbert, como acontece com as \textit{Pauli strings}, para dois operadores $\hat{A}$ e $\hat{B}$, vale a igualdade
\begin{equation}
    e^{\hat{A}+\hat{B}}=e^{\hat{A}}e^{\hat{B}}e^{-\frac{1}{2}[\hat{A},\hat{B}]}e^{\frac{1}{6}\left(2[\hat{B},[\hat{A},\hat{B}]]+[\hat{A},[\hat{A},\hat{B}]]\right)}\cdots,
    \label{eq:zassenhaus}
\end{equation}
obtida a partir do teorema de Baker-Hausdorff, de forma que apenas se $[\hat{A},\hat{B}]=0$ podemos escrever $e^{\hat{A}+\hat{B}}=e^{\hat{A}}e^{\hat{B}}$. A presença dos termos proporcionais ao comutador é um problema pois, além de serem infinitos, não são simples de serem implementados devido ao comutador ser uma somatória de \textit{Pauli strings}. Para contornar esse problema, podemos usar a seguinte igualdade deduzida a partir da equação \eqref{eq:zassenhaus}:
\begin{equation*}
    e^{\hat{A}+\hat{B}}=\left(e^{\frac{\hat{A}+\hat{B}}{l}}\right)^l = \left(e^{\frac{\hat{A}}{l}}e^{\frac{\hat{B}}{l}}\right)^l + O\left(\frac{1}{l}\right),
\end{equation*}
onde $ O\left(\frac{1}{l}\right)$ representa um termo da ordem de $\frac{1}{l}$ que é menor quanto maior for $l$. Graças a isso, é possível realizar a seguinte aproximação, conhecida como Trotterização~\cite{trotter1959product,suzuki1976generalized},
\begin{equation*}
    U\left( \boldsymbol{\theta} \right) = e^{\sum_j-i\frac{\theta_j}{2}P_j} \approx \left(\prod_{j} e^{\frac{-i\theta_j}{2l}P_j} \right)^l
\end{equation*}
onde o termo mais a direita é a versão trotterizada de $U\left( \boldsymbol{\theta} \right)$ que é uma aproximação melhor quanto maior for $l$, e cada termo no parênteses é chamado de \textit{Trotter step}. É possível absorver o termo $\frac{1}{l}$ em $\theta_j$ a ser otimizável e o circuito se torna uma sequência de exponenciais de \textit{Pauli strings} com cada uma podendo ser implementada em um hardware real. Dessa forma, a complexidade do circuito variacional é dada pelo número de termos sequenciais que precisam ser aplicados e pela profundidade de cada termo.

\begin{figure}[H]
\captionsetup[subfigure]{justification=centering}
     \centering
     \begin{subfigure}[b]{\linewidth}
         \centering
         \hspace{-0.8cm}\includegraphics[width=0.5\linewidth]{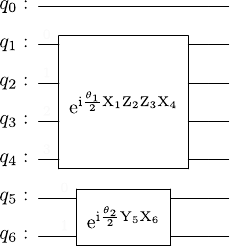}
         \caption{}
         \label{fig:Parallel}
     \end{subfigure}
     
     \vspace{0.5cm} 

     \begin{subfigure}[b]{\linewidth}
         \centering
         \hspace{-0.8cm}\includegraphics[width=0.8\linewidth]{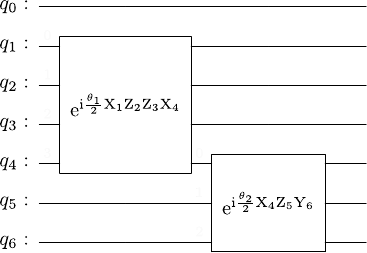}
         \caption{}
         \label{fig:nonParallel}
     \end{subfigure}
     
     \caption{(a) Circuito de seis qubits onde são aplicadas as operações $e^{i\frac{\theta}{2}X_1Z_2Z_3X_4}$ e $e^{i\frac{\theta}{2}Y_5X_6}$. Como as exponenciais atuam em qubits diferentes, elas podem ser realizadas paralelamente. (b) Circuito de seis qubits onde são aplicadas as operações $e^{i\frac{\theta}{2}X_1Z_2Z_3X_4}$ e $e^{i\frac{\theta}{2}X_4Z_5Y_6}$. Como as exponenciais atuam em um qubit em comum, elas não podem ser realizadas paralelamente.}
\end{figure}

Enfatiza-se que a quantidade de termos sequenciais é diferente da simples quantidade de termos porque é possível aplicar exponenciais diferentes em paralelo: considere as exponenciais das \textit{Pauli strings} $X_1Z_2Z_3X_4$, $Y_5X_6$ e $X_4Z_5Y_6$. As duas primeiras atuam, respectivamente, nos qubits $(1,2,3,4)$ e $(5,6)$, sendo todos distintos. Dessa maneira, caso queira-se aplicar ambas as exponenciais no sistema, não é necessário esperar a aplicação de uma para se realizar a outra, podendo ser realizadas em paralelo. Por outro lado, a terceira exponencial atua em qubits comuns às duas e, assim, caso queira-se implementar ela no mesmo circuito que a primeira ou a segunda, é necessário realizar uma e depois a outra. As Figuras \ref{fig:Parallel} e \ref{fig:nonParallel} exemplificam essa situação.

Essa possibilidade de paralelismo pode ser bastante útil para codificações com \textit{Pauli-weight} baixo, como a de Bravye-Kitaev, pois, nesses casos, as exponenciais afetam poucos qubits e assim há uma maior chance de existirem diversas exponenciais que atuam em qubits diferentes. Esse não é o caso do mapeamento de Jorda-Wigner, que necessita de muitos operadores $Z$ para manter a anti-simetria dos estados do espaço de Fock. Quando se mapeia operadores de criação e aniquilação de spin-orbitais com índices baixos o \textit{Pauli-weight} será baixo, porém, para índices altos ele será alto, de forma que, em média, o \textit{Pauli-weight} da codificação resulta em $\frac{N}{2}$ e a paralelização tem pouco efeito na quantidade de termos sequenciais. Graças a isso, o número de termos sequenciais é da mesma ordem da quantidade de termos nos operadores $\hat{S}$, \textit{i. e.} $O(N^4)$ para UCC e UCCG e $O(kN^2)$ para o k-UpCCG. Aos interessados, no trabalho de O'Gorman \textit{et al.}~\cite{o2019generalized} é proposta uma forma de tornar o mapeamento de JW mais paralelizável a partir de portas FSwap~\cite{verstraete2009quantum, kivlichan2018quantum}, que são capazes de reduzir a quantidade de termos sequenciais necessários.

Considerando que é necessário implementar, para cada \textit{Trotter step}, $O(N^4)$ termos sequenciais para o UCC e o UCCG e $O(kN^2)$ para o k-UpCCG, com profundidade média de $O(N)$, a profundidade total do circuito $P$ que utiliza cada um dos diferentes ansatz são, respectivamente, $O(N^5)$, $O(N^5)$, $O(kN^3)$.

\subsection{Escalonamento do número de medições em VQE}
\label{sec:measurement}

Um dos desafios fundamentais na implementação do VQE é o número de medições necessárias para se obter resultados precisos, uma vez que é necessário decompor o Hamiltoniano da molécula em \textit{Pauli strings}, e realizar medições em cada um desses termos, várias vezes, para estimar com precisão o valor esperado da energia. O número de medições aumenta rapidamente com o número de termos na decomposição, que depende do tamanho e da complexidade da molécula. À medida que esses requisitos aumentam, também aumenta-se a quantidade de medições necessárias, limitando a eficiência do VQE em simulações moleculares complexas. Essa necessidade de um grande número de medições representa um desafio significativo para a viabilidade do VQE em aplicações químicas práticas.

Para obter a chamada ``precisão química'' nas simulações de moléculas via VQE, é necessário que as medições dos termos Hamiltonianos forneçam uma estimativa de energia com uma precisão da ordem de $1,6$ milihartrees (mHa) ($1$Ha  $\approx4,4\cdot10^{-18}$J)~\cite{pople1999nobel}. Essa precisão permite a obtenção de valores de energia suficientemente próximos aos observados experimentalmente, o que é fundamental para cálculos de afinidades eletrônicas, energias de dissociação e outras propriedades moleculares relevantes. Isso implica reduzir a incerteza no valor esperado da energia $\langle H \rangle$, que é calculado como a soma ponderada dos valores esperados dos operadores de Pauli na decomposição do Hamiltoniano. Se o Hamiltoniano da molécula é representado como uma combinação linear de \(M\) \textit{Pauli strings} \(P_i\):

\begin{equation}
H = \sum_{i=1}^{M} c_i P_i,
\end{equation}
onde cada \(c_i\) é um coeficiente real, a estimativa de $\langle \hat{H} \rangle$ exige medir cada \(P_i\) um número suficiente de vezes para minimizar o erro da estimativa total. O número de medições \(S_i\) necessário para alcançar uma precisão de \( \epsilon \) em $\langle \hat{H} \rangle$ depende da variância $\text{Var}(P_i)$ e pode ser aproximado por

\begin{equation}
S_i \approx \frac{\text{Var}(P_i)}{\epsilon^2}.
\end{equation}

Para um conjunto de \(M\) termos, o número total de medições $S_{\text{total}}$ necessário pode ser aproximado somando-se as medições requeridas para cada \(P_i\):

\begin{equation}
S_{\text{total}} \approx \sum_{i=1}^{M} \frac{\text{Var}(P_i)}{\epsilon^2}.
\end{equation}

Dado que o número de termos \(M\) aumenta com a complexidade do sistema, o custo experimental em tempo e recursos de hardware também cresce, dificultando a obtenção da exatidão química em moléculas grandes ou complexas.

Quanto ao número de \textit{Pauli strings} no Hamiltoniano de uma molécula, isso depende do mapeamento usado para traduzir o problema molecular ao formalismo de qubits. No caso dos mapeamentos de Jordan-Wigner, paridade e Bravyi-Kitaev, a conversão do Hamiltoniano em termos de operadores eletrônicos de segunda quantização em uma soma de \textit{Pauli strings}  resulta em um número de termos que cresce com $O(N^4)$. Isso ocorre principalmente devido às interações de dois corpos que fazem parte do Hamiltoniano molecular, necessários para representar todas as interações elétron-elétron, cuja quantidade cresce com $O(N^4)$ pelos mesmos argumentos de análise combinatória usados na Seção \ref{sec:profundidadeCircuito}.
Esse crescimento de ordem $O(N^4)$ no número de \textit{Pauli strings} implica um aumento significativo no número de medições necessárias para calcular $\langle H \rangle$ com precisão.

Portanto, para um nível específico de precisão $\epsilon$ em cada \textit{Pauli string} medida separadamente, o escalonamento total do número de \textit{shots} necessários para estimar a energia é:
\begin{equation}
   O\left(\frac{N^{4}}{\epsilon^{2}}\right).
\end{equation}

Em especial, devido ao fenômeno do \textit{Barren plateau}~\cite{mcclean2018barren, larocca2025barren}, torna-se necessária uma alta precisão nos valores esperados. Esse fenômeno é caracterizado pelo espaço de soluções apresentar, em média, diferenças exponencialmente pequenas no valor de derivadas da função custo ou na própria função com o aumento do sistema. Isso faz com que surjam ``grandes regiões'' na superfície praticamente planas podendo haver pequenas regiões onde essas diferenças são significativamente maiores, que são exatamente as regiões onde encontram-se as soluções de interesse. Contudo, como a região de grandes variações é pequena, normalmente a solução inicial do algoritmo estará na região de pequenas variações, o que tornará extremamente custosa a otimização, uma vez que será necessária uma alta precisão para se obter essas diferenças (seja das derivadas ou da função) que são essenciais nos métodos de otimização. Em particular, para os métodos de otimização por gradiente, valores pequenos de gradiente podem tornar o processo de otimização extremamente longo mesmo que o gradiente fosse conhecido exatamente. Dessa forma, o \textit{Barren plateau} implica tanto em uma dificuldade de otimização quanto no processo de medida.

Sumarizando os resultados de nosso exemplo, o tempo necessário para os cálculos de pré-processamento do Hartree-Fock cresce com $O(N^4)$, enquanto o cálculo da função custo total, levando em conta a profundidade do circuito e a quantidade de medidas, dado por $\left(I+P\right)\cdot Q$, é da ordem de
\begin{equation}
        O\left(N^5\cdot\frac{N^4}{\epsilon^2}\right)=O\left(\frac{N^9}{\epsilon^2}\right)
\end{equation}
para os ansatz UCC e UCCG, enquanto para o k-UpCCG é
\begin{equation}
    O\left(kN^3\cdot\frac{N^4}{\epsilon^2}\right)=O\left(\frac{kN^7}{\epsilon^2}\right)
\end{equation}
onde a precisão necessária pode variar com base em fatores como a presença ou não de \textit{Barren plateau} no problema de interesse.

Essa complexidade dificulta a viabilidade do VQE como um todo. Em primeiro lugar, a profundidade do circuito crescendo com $O(kN^3)$ faz com que mesmo circuitos com algumas dezenas de funções de base possam se tornar extremamente profundos e irrealizáveis em computadores da era NISQ devido ao tempo de coerência dos qubits. Por outro lado, a escalabilidade total regida por um termo de ordem $N^7$ poderia deixar o cálculo extremamente demorado para moléculas de interesse prático, mesmo que existissem computadores ideais, com portas lógicas que não introduzissem erro e tempo de coerência infinito. Entretanto, métodos para diminuir a profundidade do circuito e a quantidade de medidas necessárias vêm sendo fortemente estudados na literatura~\cite{seeley2012bravyi,o2019generalized,lee2018generalized, BMrm,GC, QW,CS}, sendo que alguns desses são comprovadamente capazes de reduzir as complexidades apresentadas nesse trabalho.

Entretanto, a maior dificuldade se localiza no \textit{Barren plateau}, pois, é argumentado que uma ampla gama de circuitos variacionais que geram funções custo livre de \textit{Barren plateau} são possíveis de serem simulados de forma eficiente em computadores clássicos, gerando um questionamento se é possível o desenvolvimento de algum algoritmo variacional que apresente vantagem com relação aos computadores clássicos~\cite{cerezo2023does}. Isso abriu espaço para o desenvolvimento e estudo de estratégias diversas que buscam lidar com ambos os problemas (possuir \textit{Barren plateau} e ser classicamente simuladas), bem como de novas formas para se obter vantagem quântica na era NISQ~\cite{larocca2025barren, cerezo2023does, ramoa2025reducing}.

\section{Conclusão}
\label{sec:conclusoes}

Ao longo deste trabalho mostramos como o \textit{Variational Quantum Eigensolver} pode ser utilizado como simulador quântico para a obtenção da energia do estado fundamental de moléculas para configurações nucleares quaisquer, que são essenciais para a obtenção de propriedades úteis para desenvolvimentos científicos e tecnológicos, especialmente em áreas como a química e a farmácia. Também foi discutido em detalhes como obter a complexidade computacional para o cálculo da função custo do algoritmo, separando as diferentes etapas e exemplificando a implementação de cada uma delas. 

O resultado obtido para a profundidade dos circuitos, chegando no melhor cenário à ordem $O(kN^3)$, revela que o método poderia encontrar dificuldades em ser realizado em computadores da era NISQ com curtos tempos de coerência, enquanto a complexidade total de $O\left( \frac{kN^7}{\epsilon^2}\right)$, principalmente devido ao \textit{Barren plateau}, mas também influenciada pela quantidade de \textit{Pauli Strings} a serem medidas, poderia ser um impeditivo à realização do método como um todo. Porém, a forma como apresentamos o algoritmo pode ser entendida como a ``mais simples'', por propósitos didáticos e diversas técnicas sofisticadas que buscam simplificar/melhorar as etapas do VQE vêm sendo propostas, sendo hoje um tópico ativo de pesquisa. Por exemplo buscando-se a  redução do número de medidas necessárias, propostas de novos circuitos parametrizáveis, novas técnicas de otimização e formas de evitar o problema do \textit{Barren plateau}.

Em conclusão, o VQE é um dos principais candidatos para a obtenção de uma vantagem quântica na era NISQ na área de simulações quânticas devido à necessidade de circuitos menos profundos em comparação aos algoritmos puramente quânticos. Por outro lado, o VQE ainda apresenta diversos desafios que precisam ser superados, que têm sido fortemente investigados pela comunidade científica atualmente.

\bibliographystyle{unsrt}
\bibliography{bibliografy.bib}

\begin{thebibliography}{10}

\bibitem{Steane1998}
Andrew Steane.
\newblock Quantum computing.
\newblock {\em Reports on Progress in Physics}, 61:117--173, 2 1998.

\bibitem{Valiev2005}
Kamil'~A Valiev.
\newblock Quantum computers and quantum computations.
\newblock {\em Physics-Uspekhi}, 48:1--36, 1 2005.

\bibitem{Zoller2005}
P.~Zoller, Th. Beth, D.~Binosi, R.~Blatt, H.~Briegel, D.~Bruss, T.~Calarco, J.~I. Cirac, D.~Deutsch, J.~Eisert, A.~Ekert, C.~Fabre, N.~Gisin, P.~Grangiere, M.~Grassl, S.~Haroche, A.~Imamoglu, A.~Karlson, J.~Kempe, L.~Kouwenhoven, S.~Kröll, G.~Leuchs, M.~Lewenstein, D.~Loss, N.~Lütkenhaus, S.~Massar, J.~E. Mooij, M.~B. Plenio, E.~Polzik, S.~Popescu, G.~Rempe, A.~Sergienko, D.~Suter, J.~Twamley, G.~Wendin, R.~Werner, A.~Winter, J.~Wrachtrup, and A.~Zeilinger.
\newblock Quantum information processing and communication.
\newblock {\em The European Physical Journal D}, 36:203--228, 11 2005.

\bibitem{Feynman1982}
Richard~P. Feynman.
\newblock Simulating physics with computers.
\newblock {\em International Journal of Theoretical Physics}, 21:467--488, 1982.

\bibitem{Preskill2021}
J.~Preskill.
\newblock Quantum computing and the entanglement frontier.
\newblock {\em arXiv}, 2021.

\bibitem{cirac1995quantum}
Juan~I Cirac and Peter Zoller.
\newblock Quantum computations with cold trapped ions.
\newblock {\em Physical review letters}, 74(20):4091--4094, 1995.

\bibitem{Lloyd1996}
Seth Lloyd.
\newblock Universal quantum simulators.
\newblock {\em Science}, 273:1073--1078, 8 1996.

\bibitem{quantummanifest}
Aymard de~Touzalin, Charles Marcus, Freeke Heijman, Ignacio Cirac, Richard Murray, and Tommaso Calarco.
\newblock Quantum manifesto for quantum technologies, 2016.

\bibitem{blattquantum}
R.~Blatt and D.~Wineland.
\newblock Entangled states of trapped atomic ions.
\newblock In T.~Beth and G.~Leuchs, editors, {\em Quantum Information Processing}, pages 143--160. Wiley-VCH, Weinheim, Germany, 2005.

\bibitem{nielsen2010}
M.~A. Nielsen and I.~L. Chuang.
\newblock {\em Quantum Computation and Quantum Information}.
\newblock Cambridge University Press, Cambridge, England, 2nd edition, 2010.
\newblock Cited pages: 5, 6, 13--15, 20, 111--118, 175.

\bibitem{Qing2023}
Maomin Qing and Wei Xie.
\newblock Use vqe to calculate the ground energy of hydrogen molecules on ibm quantum.
\newblock {\em arXiv}, 5 2023.

\bibitem{shor1997}
Peter~W. Shor.
\newblock Polynomial-time algorithms for prime factorization and discrete logarithms on a quantum computer.
\newblock {\em SIAM Journal on Computing}, 26:1484--1509, 10 1997.

\bibitem{Grover1996}
Lov~K. Grover.
\newblock A fast quantum mechanical algorithm for database search.
\newblock {\em arXiv}, 5 1996.

\bibitem{Peruzzo2014}
Alberto Peruzzo, Jarrod McClean, Peter Shadbolt, Man-Hong Yung, Xiao-Qi Zhou, Peter~J. Love, Alán Aspuru-Guzik, and Jeremy~L. O’Brien.
\newblock A variational eigenvalue solver on a photonic quantum processor.
\newblock {\em Nature Communications}, 5:4213, 7 2014.

\bibitem{Li2019}
Yifan Li, Jiaqi Hu, Xiao‐Ming Zhang, Zhigang Song, and Man‐Hong Yung.
\newblock Variational quantum simulation for quantum chemistry.
\newblock {\em Advanced Theory and Simulations}, 2, 4 2019.

\bibitem{tilly2022variational}
Jules Tilly, Hongxiang Chen, Shuxiang Cao, Dario Picozzi, Kanav Setia, Ying Li, Edward Grant, Leonard Wossnig, Ivan Rungger, George~H Booth, et~al.
\newblock The variational quantum eigensolver: a review of methods and best practices.
\newblock {\em Physics Reports}, 986:1--128, 2022.

\bibitem{Frey2016}
Jann~A. Frey, Christof Holzer, Wim Klopper, and Samuel Leutwyler.
\newblock Experimental and theoretical determination of dissociation energies of dispersion-dominated aromatic molecular complexes.
\newblock {\em Chemical Reviews}, 116(9):5614–5641, April 2016.

\bibitem{Bao2017}
Junwei~Lucas Bao and Donald~G. Truhlar.
\newblock Variational transition state theory: theoretical framework and recent developments.
\newblock {\em Chemical Society Reviews}, 46(24):7548–7596, 2017.

\bibitem{Levine2013-pg}
Ira~N Levine.
\newblock {\em Quantum chemistry}.
\newblock Pearson, Upper Saddle River, NJ, USA, 7 edition, February 2013.

\bibitem{Porto2020}
Caio~M. Porto and Nelson~H. Morgon.
\newblock Analytical approach for the tunneling process in double well potentials using irc calculations.
\newblock {\em Computational and Theoretical Chemistry}, 1187:112917, October 2020.

\bibitem{Born1927}
M~Born and R.~Oppenheimer.
\newblock Zur quantentheorie der molekeln.
\newblock {\em Annalen der Physik}, 389:457--484, 1927.

\bibitem{piza2003mecanica}
Antonio Fernando Ribeiro De~Toledo Piza.
\newblock {\em Mec{\^a}nica Qu{\^a}ntica Vol. 51}.
\newblock EdUSP, São Paulo, Brazil, 2003.

\bibitem{morgon2001funccoes}
Nelson~Henrique Morgon and Rog{\'e}rio Cust{\'o}dio.
\newblock Fun{\c{c}}{\~o}es de base: o ajuste variacional.
\newblock {\em Revista Chemkeys}, (2):1--11, 2001.

\bibitem{hartree1928wave}
Douglas~R Hartree.
\newblock The wave mechanics of an atom with a non-coulomb central field. part i. theory and methods.
\newblock {\em Mathematical Proceedings of the Cambridge Philosophical Society}, 24(1):89--110, 1928.

\bibitem{fock1930naherungsmethode}
Vladimir Fock.
\newblock N{\"a}herungsmethode zur l{\"o}sung des quantenmechanischen mehrk{\"o}rperproblems.
\newblock {\em Zeitschrift f{\"u}r Physik}, 61:126--148, 1930.

\bibitem{szabo1996modern}
Attila Szabo and Neil~S. Ostlund.
\newblock {\em Modern Quantum Chemistry: Introduction to Advanced Electronic Structure Theory}.
\newblock Dover Publications, Mineola, NY, USA, 1996.

\bibitem{Jensen2007}
Frank Jensen.
\newblock {\em Introduction to Computational Chemistry}.
\newblock John Wiley \& Sons, Nashville, TN, USA, 2 edition, January 2007.

\bibitem{Morgon2018}
Nelson~Henrique Morgon and Rogério Custódio.
\newblock Fun\c{c}ões de base: o ajuste variacional.
\newblock {\em Revista Chemkeys}, (2):1–11, September 2018.

\bibitem{Hehre1969}
W.~J. Hehre, R.~F. Stewart, and J.~A. Pople.
\newblock Self-consistent molecular-orbital methods. i. use of gaussian expansions of slater-type atomic orbitals.
\newblock {\em The Journal of Chemical Physics}, 51(6):2657–2664, September 1969.

\bibitem{Binkley1980}
J.~Stephen Binkley, John~A. Pople, and Warren~J. Hehre.
\newblock Self-consistent molecular orbital methods. 21. small split-valence basis sets for first-row elements.
\newblock {\em Journal of the American Chemical Society}, 102(3):939–947, January 1980.

\bibitem{Ditchfield1971}
R.~Ditchfield, W.~J. Hehre, and J.~A. Pople.
\newblock Self-consistent molecular-orbital methods. ix. an extended gaussian-type basis for molecular-orbital studies of organic molecules.
\newblock {\em The Journal of Chemical Physics}, 54(2):724–728, January 1971.

\bibitem{Gordon1982}
Mark~S. Gordon, J.~Stephen Binkley, John~A. Pople, William~J. Pietro, and Warren~J. Hehre.
\newblock Self-consistent molecular-orbital methods. 22. small split-valence basis sets for second-row elements.
\newblock {\em Journal of the American Chemical Society}, 104(10):2797–2803, May 1982.

\bibitem{Krishnan1980}
R.~Krishnan, J.~S. Binkley, R.~Seeger, and J.~A. Pople.
\newblock Self-consistent molecular orbital methods. xx. a basis set for correlated wave functions.
\newblock {\em The Journal of Chemical Physics}, 72(1):650–654, January 1980.

\bibitem{Hehre_1972}
W.~J. Hehre, R.~Ditchfield, and J.~A. Pople.
\newblock Self—consistent molecular orbital methods. xii. further extensions of gaussian—type basis sets for use in molecular orbital studies of organic molecules.
\newblock {\em The Journal of Chemical Physics}, 56(5):2257–2261, March 1972.

\bibitem{Nagy2017}
Balazs Nagy and Frank Jensen.
\newblock Basis sets in quantum chemistry.
\newblock In Abby~L. Parrill and Kenny~B. Lipkowitz, editors, {\em Reviews in Computational Chemistry}, chapter~3, pages 93--149. Wiley, Nova Jersey, USA, 2017.

\bibitem{Johnson_2013}
Erin~R. Johnson, Alberto Otero-de-la Roza, Stephen~G. Dale, and Gino~A. DiLabio.
\newblock Efficient basis sets for non-covalent interactions in xdm-corrected density-functional theory.
\newblock {\em The Journal of Chemical Physics}, 139(21), December 2013.

\bibitem{kirschner2020performance}
Karl~N. Kirschner, Dirk Reith, and Wolfgang Heiden.
\newblock The performance of dunning, jensen, and karlsruhe basis sets on computing relative energies and geometries.
\newblock {\em Soft Materials}, 18(2-3):200--214, 2020.

\bibitem{Pitman2023}
Samuel~J. Pitman, Alicia~K. Evans, Robbie~T. Ireland, Felix Lempriere, and Laura~K. McKemmish.
\newblock Benchmarking basis sets for density functional theory thermochemistry calculations: Why unpolarized basis sets and the polarized 6-311g family should be avoided.
\newblock {\em The Journal of Physical Chemistry A}, 127(48):10295–10306, November 2023.

\bibitem{Weigend2005}
Florian Weigend and Reinhart Ahlrichs.
\newblock Balanced basis sets of split valence, triple zeta valence and quadruple zeta valence quality for h to rn: Design and assessment of accuracy.
\newblock {\em Physical Chemistry Chemical Physics}, 7(18):3297--3305, 2005.

\bibitem{Dunning1989}
Thom~H. Dunning.
\newblock Gaussian basis sets for use in correlated molecular calculations. i. the atoms boron through neon and hydrogen.
\newblock {\em The Journal of Chemical Physics}, 90(2):1007–1023, January 1989.

\bibitem{Crawford2000}
T.~Daniel Crawford and Henry~F. Schaefer.
\newblock An introduction to coupled cluster theory for computational chemists.
\newblock In Kenneth~B. Lipkowitz and Donald~B. Boyd, editors, {\em Reviews in Computational Chemistry}, chapter~2, pages 33--136. Wiley, 2000.

\bibitem{Bartlett2011}
Rodney~J. Bartlett.
\newblock Coupled‐cluster theory and its equation‐of‐motion extensions.
\newblock {\em WIREs Computational Molecular Science}, 2(1):126–138, July 2011.

\bibitem{Bartlett2007}
Rodney~J. Bartlett and Monika Musiał.
\newblock Coupled-cluster theory in quantum chemistry.
\newblock {\em Reviews of Modern Physics}, 79(1):291–352, February 2007.

\bibitem{Lyakh2011}
Dmitry~I. Lyakh, Monika Musiał, Victor~F. Lotrich, and Rodney~J. Bartlett.
\newblock Multireference nature of chemistry: The coupled-cluster view.
\newblock {\em Chemical Reviews}, 112(1):182–243, December 2011.

\bibitem{tanenbaum2015}
Andrew~S. Tanenbaum.
\newblock {\em Organização Estruturada de Computadores}.
\newblock Pearson, London, England, 2015.
\newblock Cited pages: 2--5, 28--30.

\bibitem{patterson2017}
David~A. Patterson and John~L. Hennessy.
\newblock {\em Organização e Projeto de Computadores: A Interface Hardware/Software}.
\newblock Elsevier, Rio de Janeiro, Brazil, 2017.

\bibitem{stallings2016}
William Stallings.
\newblock {\em Organização e Arquitetura de Computadores: Projetando para o Desempenho}.
\newblock Pearson, London, England, 2016.

\bibitem{moore1965}
Gordon~E. Moore.
\newblock Cramming more components onto integrated circuits.
\newblock {\em Electronics}, 38(8):82--85, 1965.

\bibitem{ladd2010}
Thaddeus~D Ladd, Fedor Jelezko, Raymond Laflamme, Yasunobu Nakamura, Christopher Monroe, and Jeremy~Lloyd O’Brien.
\newblock Quantum computers.
\newblock {\em nature}, 464(7285):45--53, 2010.

\bibitem{preskill2018}
John Preskill.
\newblock Quantum computing in the nisq era and beyond.
\newblock {\em Quantum}, 2:79, 2018.

\bibitem{bennett1984}
Charles~H Bennett and Gilles Brassard.
\newblock Quantum cryptography: Public key distribution and coin tossing.
\newblock In {\em Proceedings of IEEE International Conference on Computers, Systems and Signal Processing}, pages 175--179, 1984.

\bibitem{james2001measurement}
Daniel~FV James, Paul~G Kwiat, William~J Munro, and Andrew~G White.
\newblock Measurement of qubits.
\newblock {\em Physical Review A}, 64(5):052312, 2001.

\bibitem{einstein1935can}
Albert Einstein, Boris Podolsky, and Nathan Rosen.
\newblock Can quantum-mechanical description of physical reality be considered complete?
\newblock {\em Physical review}, 47(10):777, 1935.

\bibitem{bell1964einstein}
John~S Bell.
\newblock On the einstein podolsky rosen paradox.
\newblock {\em Physics Physique Fizika}, 1(3):195, 1964.

\bibitem{deutsch1992rapid}
David Deutsch and Richard Jozsa.
\newblock Rapid solution of problems by quantum computation.
\newblock {\em Proceedings of the Royal Society of London. Series A: Mathematical and Physical Sciences}, 439(1907):553--558, 1992.

\bibitem{bennett1993teleporting}
Charles~H Bennett, Gilles Brassard, Claude Cr{\'e}peau, Richard Jozsa, Asher Peres, and William~K Wootters.
\newblock Teleporting an unknown quantum state via dual classical and einstein-podolsky-rosen channels.
\newblock {\em Physical review letters}, 70(13):1895, 1993.

\bibitem{deutsch1989quantum}
David~Elieser Deutsch.
\newblock Quantum computational networks.
\newblock {\em Proceedings of the royal society of London. A. mathematical and physical sciences}, 425(1868):73--90, 1989.

\bibitem{barenco1995elementary}
Adriano Barenco, Charles~H Bennett, Richard Cleve, David~P DiVincenzo, Norman Margolus, Peter Shor, Tycho Sleator, John~A Smolin, and Harald Weinfurter.
\newblock Elementary gates for quantum computation.
\newblock {\em Physical review A}, 52(5):3457, 1995.

\bibitem{acharya2024quantum}
Rajeev Acharya, Laleh Aghababaie-Beni, Igor Aleiner, Trond~I Andersen, Markus Ansmann, Frank Arute, Kunal Arya, Abraham Asfaw, Nikita Astrakhantsev, Juan Atalaya, et~al.
\newblock Quantum error correction below the surface code threshold.
\newblock {\em arXiv preprint arXiv:2408.13687}, 2024.

\bibitem{fernandes2022ions}
Gabriel~PLM Fernandes, Alexandre~C Ricardo, Fernando~R Cardoso, and Celso~J Villas-Boas.
\newblock {\'I}ons aprisionados como arquitetura para computa{\c{c}}{\~a}o qu{\^a}ntica.
\newblock {\em Revista Brasileira de Ensino de F{\'\i}sica}, 45:e20220218, 2022.

\bibitem{aghaee2025scaling}
H~Aghaee~Rad, T~Ainsworth, RN~Alexander, B~Altieri, MF~Askarani, R~Baby, L~Banchi, BQ~Baragiola, JE~Bourassa, RS~Chadwick, et~al.
\newblock Scaling and networking a modular photonic quantum computer.
\newblock {\em Nature}, pages 1--8, 2025.

\bibitem{jones2001nmr}
Jonathan~A Jones.
\newblock Nmr quantum computation.
\newblock {\em Progress in Nuclear Magnetic Resonance Spectroscopy}, 38(4):325--360, 2001.

\bibitem{bloch2008quantum}
Immanuel Bloch.
\newblock Quantum coherence and entanglement with ultracold atoms in optical lattices.
\newblock {\em Nature}, 453(7198):1016--1022, 2008.

\bibitem{cormen2022introduction}
Thomas~H Cormen, Charles~E Leiserson, Ronald~L Rivest, and Clifford Stein.
\newblock {\em Introduction to algorithms}.
\newblock MIT press, Cambridge, MA, USA, 2022.

\bibitem{cerezo2021variational}
Marco Cerezo, Andrew Arrasmith, Ryan Babbush, Simon~C Benjamin, Suguru Endo, Keisuke Fujii, Jarrod~R McClean, Kosuke Mitarai, Xiao Yuan, Lukasz Cincio, et~al.
\newblock Variational quantum algorithms.
\newblock {\em Nature Reviews Physics}, 3(9):625--644, 2021.

\bibitem{cohen2005history}
Leon Cohen.
\newblock The history of noise [on the 100th anniversary of its birth].
\newblock {\em IEEE Signal Processing Magazine}, 22(6):20--45, 2005.

\bibitem{shannon1948mathematical}
Claude~Elwood Shannon.
\newblock A mathematical theory of communication.
\newblock {\em The Bell system technical journal}, 27(3):379--423, 1948.

\bibitem{hamming1950error}
Richard~W Hamming.
\newblock Error detecting and error correcting codes.
\newblock {\em The Bell system technical journal}, 29(2):147--160, 1950.

\bibitem{krinner2022realizing}
Sebastian Krinner, Nathan Lacroix, Ants Remm, Agustin Di~Paolo, Elie Genois, Catherine Leroux, Christoph Hellings, Stefania Lazar, Francois Swiadek, Johannes Herrmann, et~al.
\newblock Realizing repeated quantum error correction in a distance-three surface code.
\newblock {\em Nature}, 605(7911):669--674, 2022.

\bibitem{postler2022demonstration}
Lukas Postler, Sascha Heu$\beta$en, Ivan Pogorelov, Manuel Rispler, Thomas Feldker, Michael Meth, Christian~D Marciniak, Roman Stricker, Martin Ringbauer, Rainer Blatt, et~al.
\newblock Demonstration of fault-tolerant universal quantum gate operations.
\newblock {\em Nature}, 605(7911):675--680, 2022.

\bibitem{ai2024quantum}
Google~Quantum AI et~al.
\newblock Quantum error correction below the surface code threshold.
\newblock {\em Nature}, 638(8052):920--926, 2024.

\bibitem{frisch2024trapped}
Albert Frisch, Alexander Erhard, Thomas Feldker, Florian Girtler, Max Hettrich, Wilfried Huss, Georg Jacob, Christine Maier, Gregor Mayramhof, Daniel Nigg, et~al.
\newblock Trapped-ion quantum computing.
\newblock In {\em Quantum Software: Aspects of Theory and System Design}, pages 251--283. Springer, Cham, Switzerland, 2024.

\bibitem{jordan1928paulische}
Pascual Jordan and Eugene~Paul Wigner.
\newblock {\"U}ber das paulische {\"a}quivalenzverbot.
\newblock {\em Zeitschrift f{\"u}r Physik}, 47:631--651, 1928.

\bibitem{seeley2012bravyi}
Jacob~T Seeley, Martin~J Richard, and Peter~J Love.
\newblock The bravyi-kitaev transformation for quantum computation of electronic structure.
\newblock {\em The Journal of chemical physics}, 137(22), 2012.

\bibitem{mcclean2018barren}
Jarrod~R McClean, Sergio Boixo, Vadim~N Smelyanskiy, Ryan Babbush, and Hartmut Neven.
\newblock Barren plateaus in quantum neural network training landscapes.
\newblock {\em Nature communications}, 9(1):4812, 2018.

\bibitem{larocca2025barren}
Mart{\'\i}n Larocca, Supanut Thanasilp, Samson Wang, Kunal Sharma, Jacob Biamonte, Patrick~J Coles, Lukasz Cincio, Jarrod~R McClean, Zo{\"e} Holmes, and M~Cerezo.
\newblock Barren plateaus in variational quantum computing.
\newblock {\em Nature Reviews Physics}, pages 1--16, 2025.

\bibitem{taube2006new}
Andrew~G Taube and Rodney~J Bartlett.
\newblock New perspectives on unitary coupled-cluster theory.
\newblock {\em International journal of quantum chemistry}, 106(15):3393--3401, 2006.

\bibitem{kutzelnigg2010unconventional}
Werner Kutzelnigg.
\newblock Unconventional aspects of coupled-cluster theory.
\newblock In Jirí~Pittner Petr~Cársky, Josef~Paldus, editor, {\em Recent Progress in Coupled Cluster Methods: Theory and Applications}, chapter~12, pages 299--356. Springer, Dordrecht, Netherlands, 2010.

\bibitem{nooijen2000can}
Marcel Nooijen.
\newblock Can the eigenstates of a many-body hamiltonian be represented exactly using a general two-body cluster expansion?
\newblock {\em Physical review letters}, 84(10):2108--2111, 2000.

\bibitem{mukherjee2004some}
Debashis Mukherjee and Werner Kutzelnigg.
\newblock Some comments on the coupled cluster with generalized singles and doubles (ccgsd) ansatz.
\newblock {\em Chemical physics letters}, 397(1-3):174--179, 2004.

\bibitem{lee2018generalized}
Joonho Lee, William~J Huggins, Martin Head-Gordon, and K~Birgitta Whaley.
\newblock Generalized unitary coupled cluster wave functions for quantum computation.
\newblock {\em Journal of chemical theory and computation}, 15(1):311--324, 2018.

\bibitem{whitfield2013computational}
James~Daniel Whitfield, Peter~John Love, and Al{\'a}n Aspuru-Guzik.
\newblock Computational complexity in electronic structure.
\newblock {\em Physical Chemistry Chemical Physics}, 15(2):397--411, 2013.

\bibitem{echenique2007mathematical}
Pablo Echenique and Jos{\'e}~Luis Alonso.
\newblock A mathematical and computational review of hartree--fock scf methods in quantum chemistry.
\newblock {\em Molecular Physics}, 105(23-24):3057--3098, 2007.

\bibitem{anand2022quantum}
Abhinav Anand, Philipp Schleich, Sumner Alperin-Lea, Phillip~WK Jensen, Sukin Sim, Manuel D{\'\i}az-Tinoco, Jakob~S Kottmann, Matthias Degroote, Artur~F Izmaylov, and Al{\'a}n Aspuru-Guzik.
\newblock A quantum computing view on unitary coupled cluster theory.
\newblock {\em Chemical Society Reviews}, 51(5):1659--1684, 2022.

\bibitem{trotter1959product}
Hale~F Trotter.
\newblock On the product of semi-groups of operators.
\newblock {\em Proceedings of the American Mathematical Society}, 10(4):545--551, 1959.

\bibitem{suzuki1976generalized}
Masuo Suzuki.
\newblock Generalized trotter's formula and systematic approximants of exponential operators and inner derivations with applications to many-body problems.
\newblock {\em Communications in Mathematical Physics}, 51(2):183--190, 1976.

\bibitem{o2019generalized}
Bryan O'Gorman, William~J Huggins, Eleanor~G Rieffel, and K~Birgitta Whaley.
\newblock Generalized swap networks for near-term quantum computing.
\newblock {\em arXiv preprint arXiv:1905.05118}, 2019.

\bibitem{verstraete2009quantum}
Frank Verstraete, J~Ignacio Cirac, and Jos{\'e}~I Latorre.
\newblock Quantum circuits for strongly correlated quantum systems.
\newblock {\em Physical Review A}, 79(3):032316, 2009.

\bibitem{kivlichan2018quantum}
Ian~D Kivlichan, Jarrod McClean, Nathan Wiebe, Craig Gidney, Al{\'a}n Aspuru-Guzik, Garnet Kin-Lic Chan, and Ryan Babbush.
\newblock Quantum simulation of electronic structure with linear depth and connectivity.
\newblock {\em Physical review letters}, 120(11):110501, 2018.

\bibitem{pople1999nobel}
John~A Pople.
\newblock Nobel lecture: Quantum chemical models.
\newblock {\em Reviews of Modern Physics}, 71(5):1267, 1999.

\bibitem{BMrm}
Giacomo Torlai, Guglielmo Mazzola, Giuseppe Carleo, and Antonio Mezzacapo.
\newblock Precise measurement of quantum observables with neural-network estimators.
\newblock {\em Phys. Rev. Res.}, 2:022060, Jun 2020.

\bibitem{GC}
Pranav Gokhale, Olivia Angiuli, Yongshan Ding, Kaiwen Gui, Teague Tomesh, Martin Suchara, Margaret Martonosi, and Frederic~T. Chong.
\newblock Minimizing state preparations in variational quantum eigensolver by partitioning into commuting families.
\newblock {\em arXiv preprint arXiv:1907.13623}, 2019.

\bibitem{QW}
Vladyslav Verteletskyi, Tzu-Ching Yen, and Artur~F. Izmaylov.
\newblock Measurement optimization in the variational quantum eigensolver using a minimum clique cover.
\newblock {\em The Journal of Chemical Physics}, 152(12):124114, 03 2020.

\bibitem{CS}
Charles Hadfield, Sergey Bravyi, Roger Raymond, et~al.
\newblock Measurements of quantum hamiltonians with locally-biased classical shadows.
\newblock {\em Communications in Mathematical Physics}, 391:951--967, 2022.

\bibitem{cerezo2023does}
Marco Cerezo, Martin Larocca, Diego Garc{\'\i}a-Mart{\'\i}n, Nelson~L Diaz, Paolo Braccia, Enrico Fontana, Manuel~S Rudolph, Pablo Bermejo, Aroosa Ijaz, Supanut Thanasilp, et~al.
\newblock Does provable absence of barren plateaus imply classical simulability? or, why we need to rethink variational quantum computing.
\newblock {\em arXiv preprint arXiv:2312.09121}, 2023.

\bibitem{ramoa2025reducing}
Mafalda Ram{\^o}a, Panagiotis~G Anastasiou, Luis~Paulo Santos, Nicholas~J Mayhall, Edwin Barnes, and Sophia~E Economou.
\newblock Reducing the resources required by adapt-vqe using coupled exchange operators and improved subroutines.
\newblock {\em npj Quantum Information}, 11(1):1--19, 2025.

\end{thebibliography}

\end{multicols}

\end{document}